\documentclass[lettersize,journal]{IEEEtran}
\usepackage{amsmath,amsfonts}
\usepackage{algorithmic}
\usepackage{algorithm}
\usepackage{array}
\usepackage[caption=false,font=normalsize,labelfont=sf,textfont=sf]{subfig}
\usepackage{textcomp}
\usepackage{stfloats}
\usepackage{url}
\usepackage{verbatim}
\usepackage{graphicx}
\usepackage{cite}
\usepackage[utf8]{inputenc}
\usepackage[normalem]{ulem}
\usepackage{amsmath}

\usepackage{soul}
\usepackage{color}
\usepackage{multirow}
\usepackage{makecell}

\newcommand{\Revise}[1]{#1}

\usepackage{tikz}
\usetikzlibrary{calc}

\newcommand*\blackcircled[1]{\tikz[baseline=(char.base)]{
        \node[shape=circle,fill={rgb,255:red,0;green,0;blue,0}, text=white, font=\small, inner sep=0.6pt] (char) {#1};}}

\begin{document}

%\title{A Comprehensive Review on Distributed Training of Graph Neural Networks}
\title{A Comprehensive Survey on \\ Distributed Training of Graph Neural Networks}

\author{By Haiyang Lin, Mingyu Yan, Xiaochun Ye, Dongrui Fan, \textit{Senior Member, IEEE}, \\Shirui Pan, Wenguang Chen, and Yuan Xie, \textit{Fellow, IEEE}
\thanks{Haiyang Lin, Mingyu Yan, Xiaochun Ye, and Dongrui Fan are with State Key Lab of Processors (SKLP), Institute of Computing Technology (ICT), Chinese Academy of Sciences (CAS), Beijing 100864, China and also with the University of Chinese Academy of Sciences (UCAS), Beijing 100049, China (e-mail: linhaiyang18z@ict.ac.cn; yanmingyu@ict.ac.cn; yexiaochun@ict.ac.cn; fandr@ict.ac.cn).
}% <-this % stops a space
\thanks{Shirui Pan is with the School of Information and Communication Technology, Griffith University, Queensland 4215, Australia (e-mail: s.pan@griffith.edu.au).
}
\thanks{Wenguang Chen is with the Department of Computer Science and Technology, Tsinghua University, Beijing 100084, China (e-mail: cwg@tsinghua.edu.cn).
}
%\thanks{Yuan Xie is with the Department of Electrical and Computer Engineering, University of California at Santa Barbara, Santa Barbara, CA 93106 USA (e-mail: yuanxie@ucsb.edu).
%}
\thanks{Yuan Xie is with the Department of Electronic and Computer Engineering, The Hong Kong University of Science and Technology (e-mail: yuanxie@ust.hk).
}
\thanks{Corresponding author: Mingyu Yan}
}

%\author{IEEE Publication Technology,~\IEEEmembership{Staff,~IEEE,}
        % <-this % stops a space
%\thanks{This paper was produced by the IEEE Publication Technology Group. They are in Piscataway, NJ.}% <-this % stops a space
%\thanks{Manuscript received April 19, 2021; revised August 16, 2021.}}

\markboth{To Appear in Proceedings of the IEEE}%
{H. Lin \MakeLowercase{\textit{et al.}}: A Comprehensive Survey on Distributed Training of Graph Neural Networks}
%\markboth{Proceedings of the IEEE}%
%{H. Lin \MakeLowercase{\textit{et al.}}: A Comprehensive Survey on Distributed Training of Graph Neural Networks}


\maketitle

\begin{abstract}
% \todo{Final check:
% %图表要离描述的内容近
% %格式，问号等等
% }
%Graph neural networks (GNNs) have been demonstrated to be a powerful algorithmic model in broad application fields for its effectiveness in learning over graphs. 
Graph neural networks (GNNs) have been demonstrated to be a powerful algorithmic model in broad application fields for their effectiveness in learning over graphs. 
To scale GNN training up for large-scale and ever-growing graphs, the most promising solution is distributed training which distributes the workload of training across multiple computing nodes.
%\todo{R1 Q10} 
\Revise{
At present, the volume of related research on distributed GNN training is exceptionally vast, accompanied by an extraordinarily rapid pace of publication. Moreover, the approaches reported in these studies exhibit significant divergence. This situation poses a considerable challenge for newcomers, hindering their ability to grasp a comprehensive understanding of the workflows, computational patterns, communication strategies, and optimization techniques employed in distributed GNN training. As a result, there is a pressing need for a survey to provide correct recognition, analysis, and comparisons in this field.
}
In this paper, we provide a comprehensive survey of distributed GNN training by investigating various optimization techniques used in distributed GNN training.
First, distributed GNN training is classified into several categories according to their workflows. In addition, their computational patterns and communication patterns, as well as the optimization techniques proposed by recent work are introduced.
Second, the software frameworks and hardware platforms of distributed GNN training are also introduced for a deeper understanding. 
Third, distributed GNN training is compared with distributed training of deep neural networks, emphasizing the uniqueness of distributed GNN training.
Finally, interesting issues and opportunities in this field are discussed.

\end{abstract}


\section{Introduction} \label{sec:introduction}

\begin{figure*}[!t]
    %\vspace{-15pt}
    \centering
    \includegraphics[page=1, width=0.99\textwidth]{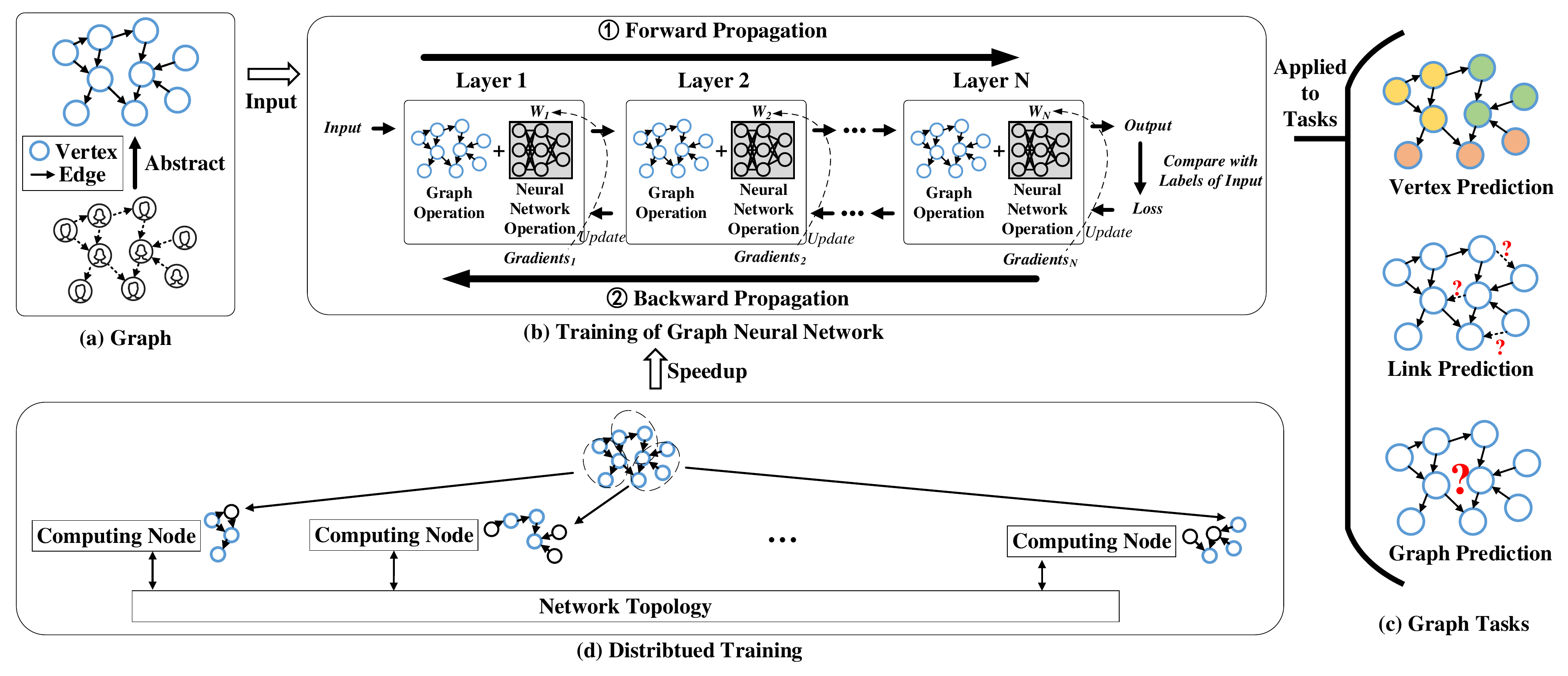}
    %\vspace{-15pt}
    \caption{
    Distributed GNN training: (a) Illustration of graphs; \Revise{(b) Illustration of a GNN and its training procedure ($W_N$ means the weights of GNN $N_{th}$ layer);} (c) Illustration of graph tasks; (d) Illustration of distributed training.}
    %\todo{R1 Q1}
    \label{fig:01ley_gnndt}
    %\vspace{-20pt}
\end{figure*}

\IEEEPARstart{G}{raph} is a well-known data structure widely used in many critical application fields due to its powerful representation capability of data, especially in expressing the associations between objects \cite{introduction_to_graph_theory, graph_applications}.
Many real-world data can be naturally represented as graphs which consist of a set of vertices and edges.
Take social networks as an example~\cite{social_network_analysis,What_is_Twitter}, the vertices in the graph represent people and the edges represent interactions between people on Facebook \cite{DBLP:conf/nips/learning_to_discover/facebook_dataset_ego/}.
An illustration of graphs for social networks is illustrated in Fig. \ref{fig:01ley_gnndt} (a), where the circles represent the vertices, and the arrows represent the edges.
Another well-known example is knowledge graphs \cite{Freebase_knowledge_graph, ashburner2000gene}, in which the vertices represent entities while the edges represent relations between the entities \cite{WordNet_knowledge_graph}.

Graph neural networks (GNNs) demonstrate superior performance compared to other algorithmic models in learning over graphs \cite{tkde_deep_learning_on_graphs_gnngood, tnnls_a_comprehensive_survey_gnngood, aiopen_graph_neural_networks_a_review_gnngood}.
Deep neural networks (DNNs) have been widely used to analyze Euclidean data such as images~\cite{Nature_Deep_Learning}. 
\Revise{
However, they encounter challenges when dealing with non-Euclidean data, characterized by arbitrary size and complex topological structures that can be efficiently represented using graphs~\cite{nips_inductive_representation_ley_graphsage_minibatch}.
}
Besides, a major weakness of deep learning paradigms identified by industry is that they cannot effectively carry out causal reasoning, which greatly reduces the cognitive ability of intelligent systems~\cite{geometric_learning}. To this end, GNNs have emerged as the premier paradigm for graph learning and endowed intelligent systems with cognitive ability.

%\todo{R1 Q1}
\Revise{
An illustration of a GNN and its training procedure is shown in Fig.~\ref{fig:01ley_gnndt} (b).
GNNs consist of multiple layers. Each layer comprises two types of operations corresponding to two distinct steps: Aggregation and Combination.
The Aggregation step performs graph operations, often referred to as aggregation operations. For instance, it gathers neighboring features for each vertex.
The Combination step executes neural network operations, which are commonly known as combination operations. For example, it uses a multi-layer perceptron to update the features of each vertex.
The training process for GNNs involves forward propagation and backward propagation. 
During forward propagation, the input data undergoes layer-by-layer processing to produce the final output. This final output is then compared to the true labels of the input data to calculate the loss.
Subsequently, the loss is utilized in the backward propagation to calculate gradients, which are then used to update the model weights ($W$).
This iterative process continues until the loss converges.
}
Finally, the trained model can be applied to graph tasks, including vertex prediction \cite{iclr_semi_supervised_classification_with_nodeprediction} (predicting the properties of specific vertices), link prediction \cite{nips_link_prediction_based_on_linkprediction} (predicting the existence of an edge between two vertices), and graph prediction \cite{aaai_an_end_to_end_deep_learing_graphprediction} (predicting the properties of the whole graph), as shown in Fig. \ref{fig:01ley_gnndt} (c). 

Thanks to the superiority of GNNs, they have been widely used in various real-world applications in many critical fields. 
These real-world applications include 
knowledge inference~\cite{gcn_knowledge_graph}, natural language processing \cite{DBLP:journals/cim/YoungHPC18/gnn_nlp/, DBLP:conf/icdm/WuPZZP19/pan_gnn03}, machine translation \cite{gcn_machine_translation}, recommendation systems \cite{kdd_graph_convolutional_ley_pinsage,DBLP:conf/www/Fan0LHZTY19/gnn_recommend01/,DBLP:conf/aaai/MaMZSLC20/gnn_recommend02}, visual reasoning \cite{gcn_visual_reasoning}, chip design \cite{google_ai_chip_design, GCN-RL_circuit_designer,GCN_EDA}, traffic prediction \cite{google_map_gnn, DBLP:conf/aaai/ChenLTZWWZ19/gnn_traffic_prediction01/, DBLP:conf/aaai/LiZ21/gnn_traffic_prediction02/}, ride-hailing demand forecasting~\cite{gcn_ride-hailing_demand_forecasting}, spam review detection~\cite{xianyu_spam_review_detection}, molecule property prediction \cite{DBLP:conf/nips/FoutBSB17/gnn_protein/}, and so forth. 
GNNs enhance the machine intelligence when processing a broad range of real-world applications, such as giving $>$50\% accuracy improvement for real-time ETAs in Google Maps~\cite{google_map_gnn}, generating $>$40\% higher-quality recommendations in Pinterest~\cite{kdd_graph_convolutional_ley_pinsage}, achieving $>$10\% improvement of ride-hailing demand forecasting in Didi~\cite{gcn_ride-hailing_demand_forecasting}, improving $>$66.90\% of \textit{recall at 90\% precision} for spam review detection in Alibaba~\cite{xianyu_spam_review_detection}.

However, both industry and academia are still eagerly expecting the acceleration of GNN training for the following reasons \cite{pvldb_aligraph, pvldb_agl, eurosys_flexgraph, ia3_distdgl}:
\begin{itemize}
    \item \textit{The scale of graph data is rapidly expanding, consuming a great deal of time for GNN training.}
    With the explosion of information on the Internet, new graph data are constantly being generated and changed, such as the establishment and demise of interpersonal relationships in social communication and the changes in people's preferences for goods in online shopping.
    The scales of vertices and edges in graphs are approaching or even outnumbering the order of billions and trillions, respectively \cite{nips_open_graph_benchmark_ogbdataset, sc_reducint_communication_in_graph_cagnet, DBLP:journals/tjs/HuangQWC20/chen_largegraph02/, DBLP:journals/tpds/HuangWFWC22/chen_largegraph03/}.
    The growth rate of graph scales is also astonishing.
    For example, the number of vertices (i.e., users) in Facebook's social network is growing at a rate of 17\% per year \cite{/facebook/17percent/}.
    Consequently, GNN training time dramatically increases due to the ever-growing scale of graph data.
    \item \textit{Swift development and deployment of novel GNN models involves repeated training, in which a large amount of training time is inevitable.}
    Much experimental work is required to develop a highly-accurate GNN model since repeated training is needed \cite{tnnls_a_comprehensive_survey_gnngood, aiopen_graph_neural_networks_a_review_gnngood, tkde_deep_learning_on_graphs_gnngood}.
    Moreover, expanding the usage of GNN models to new application fields also requires much time to train the model with real-life data.
    Such a sizeable computational burden calls for faster methods of training.
\end{itemize}

Distributed training is a popular solution to speed up GNN training \cite{atc_neugraph, mlsys_roc, sc_reducint_communication_in_graph_cagnet, eurosys_dgcl, sc_distgnn, eurosys_flexgraph, arxiv_mg_gcn, osdi_dorylus, arxiv_sar, pvldb_aligraph, ia3_distdgl, pvldb_agl, socc_pagraph, cluster_2pgraph, arxiv_llcg, arxiv_distdglv2, arxiv_graphtheta, arxiv_salient, /osdi/Gandhi/p3/distributed_deep_graph/, DBLP:conf/infocom/LuoBW22/new02/DGTP}.
It tries to accelerate the entire computing process by adding more computing resources, or ``nodes", to the computing system with parallel execution strategies, as shown in Fig. \ref{fig:01ley_gnndt} (d). 
NeuGraph \cite{atc_neugraph}, proposed in 2019, is the first published work of distributed GNN training.
Since then, there has been a steady stream of attempts to improve the efficiency of distributed GNN training in recent years with significantly varied optimization techniques, including workload partitioning \cite{atc_neugraph, mlsys_roc, eurosys_dgcl, sc_distgnn}, transmission planning \cite{mlsys_roc, atc_neugraph, eurosys_dgcl, eurosys_flexgraph}, caching strategy \cite{pvldb_aligraph, socc_pagraph, cluster_2pgraph}, etc.

\begin{table*}[!htbp]
\centering
\caption{Content guidance of this paper.} \label{table:organization}
\renewcommand\arraystretch{1.5} 
\setlength{\tabcolsep} {1.5mm}{
\begin{tabular}{|c|cc|}
\hline 
\textbf{Section} &  \textbf{Subsection} & \textbf{Index} \\ \hline \hline
\multirow{3}{*}{Background} 
&  Graphs  &  Section \ref{sec:background:graphs}\\ 
&  Graph Neural Networks  &  Section \ref{sec:background:graph_neural_network} \\ 
%&  General Programming Model for Graph Neural Network  &   Section \ref{sec:background:programming_model}\\ 
&  Training Methods for Graph Neural Networks  &   Section \ref{sec:background:training_method}\\ \hline
\multirow{7}{*}{\makecell[c]{Taxonomy of \\Distributed GNN Training}} 
&  Distributed Full-batch Training  & Section \ref{sec:taxonomy:dft_intro} \\ 
&  Distributed Mini-batch Training  & Section \ref{sec:taxonomy:dmt_intro} \\ 
%&  Comparison Between Distributed Full-batch Training and Distributed Mini-batch Training  & Section \ref{sec:taxonomy:dft_dmt_comparison} \\ 
& \multirow{2}{*}{\begin{tabular}[c]{@{}c@{}}  Comparison Between Distributed Full-batch Training \\and Distributed Mini-batch Training \end{tabular}}  
& \multirow{2}{*}{\begin{tabular}[c]{@{}c@{}}  Section \ref{sec:taxonomy:dft_dmt_comparison} \end{tabular}}   \\ 
&   &  \\
&  Other Information of Taxonomy  & Section \ref{sec:taxonomy:others_information} \\ 
& \multirow{2}{*}{\begin{tabular}[c]{@{}c@{}}  \Revise{Comparison Between Non-distributed GNN Training} \\\Revise{and Distributed GNN Training} \end{tabular}}  
& \multirow{2}{*}{\begin{tabular}[c]{@{}c@{}}  Section \ref{sec:taxonomy:non_dt_and_dt_comparison} \end{tabular}}   \\ 
&   &  \\
\hline
\multirow{2}{*}{Distributed Full-batch Training} 
&  Dispatch-workload-based Execution  & Section \ref{sec:dft:chunk_execution}  \\ 
&  Preset-workload-based Execution  & Section \ref{sec:dft:partition_execution} \\ \hline
\multirow{2}{*}{Distributed Mini-batch Training} 
&  Individual-sample-based Execution  & Section \ref{sec:dmt:sample_individual_execution} \\
&  Joint-sample-based Execution  & Section \ref{sec:dmt:sample_joint_execution} \\ 
\hline
%\makecell[c]{Distributed Full-batch Training V.S. \\Distributed Mini-batch Training} &  (Comparison of the Computational Pattern) & Section \ref{sec:comparison_full_mini_distributed_training} \\ \hline
%\multirow{2}{*}{\makecell[c]{Distributed Full-batch Training V.S. \\Distributed Mini-batch Training}} 
%&  Joint-sample-based Execution  & Section \ref{sec:dmt:sample_individual_execution} \\ 
%&  Individual-sample-based Execution  & Section \ref{sec:dmt:sample_joint_execution} \\ \hline
\makecell[c]{\multirow{2}{*}{\begin{tabular}[c]{@{}c@{}}Software Frameworks for \\Distributed GNN Training\end{tabular}}} & \multirow{2}{*}{(Introduction of the Software Frameworks)}  & \multirow{2}{*}{Section \ref{sec:software_framework}} \\ 
\makecell[c]{} &  &  \\ \hline
\multirow{3}{*}{\makecell[c]{Hardware Platforms for \\Distributed GNN Training}} 
&  Multi-CPU Hardware Platform  &  Section \ref{sec:hp:multi_cpu_hardware_platform} \\ 
&  Multi-GPU Hardware Platform  &  Section \ref{sec:hp:multi_gpu_hardware_platform} \\ 
&  Multi-CPU Hardware Platform V.S. Multi-GPU Hardware Platform  &  Section \ref{sec:hp:comparison_cpu_or_cpu} \\ \hline
\multirow{3}{*}{\makecell[c]{Comparison to \\Distributed DNN Training}} 
&  Brief Introduction to Distributed DNN Training  &  Section \ref{sec:comparison_to_dnn:brief_intro_dnn} \\ 
&  Distributed Full-batch Training V.S. DNN Model Parallelism  & Section \ref{sec:comparison_to_dnn:full_vs_model}  \\ 
&  Distributed Mini-batch Training V.S. DNN Data Parallelism  &  Section \ref{sec:comparison_to_dnn:mini_vs_data} \\ \hline
\multirow{7}{*}{\makecell[c]{Summary and Discussion}} 
&Quantitative Analysis of Performance Bottleneck   &Section \ref{sec:summary:quantitative}   \\
&Performance Benchmark   & Section \ref{sec:summary:performance_benchmark} \\
&Distributed Training on Extreme-scale Hardware Platform   & Section \ref{sec:summary:distributed_training_on_extreme_scale} \\
&Domain-specific Distributed Hardware Platform   & Section \ref{sec:summary:domain_specific} \\
&General Communication Library for Distributed GNN Training   & Section \ref{sec:summary:general_communication} \\ 
&\Revise{Distributed Training for Dynamic GNNs}   & Section \ref{sec:summary:dynamic_gnn} \\ 
&\Revise{Distributed Training for Deep GNNs}   & Section \ref{sec:summary:performance_deterioration} \\
\hline
%Trends & (Summary of the Optimization Trends)  & Section \ref{sec:trends} \\ \hline
Conclusion &  (Conclusion of the Survey) & Section \ref{sec:conclusion} \\ \hline
%Background & \multicolumn{2}{c}{Related Concepts and Training Methods of GNN} &  Section \ref{sec:BACKGROUND}       \\ \hline 
%GNN Training Methods            &  Full-batch Training and Mini-batch Training    &          &  \\ \hline 
%\multirow{2}{*}{Distributed Full-batch Training of GNN}
% &  Dispatch-workload-based Execution             &          &  \\
% &  Preset-workload-based Execution         &          &  \\ \hline 
%\multirow{2}{*}{Distributed Mini-batch Training of GNN}
% &  Individual-sample-based Execution  &          &  \\
% &  Joint-sample-based Execution        &          &  \\ \hline 
%\multirow{2}{*}{Software Frameworks for distributed GNN training}
% &  Characteristics of Software Frameworks &          &  \\ 
% &          &          &  \\ \hline 
% \multirow{2}{*}{Hardware Platforms for distributed GNN training}
% &  Multi-CPU Hardware Platform  &          &  \\
% &  Multi-GPU Hardware Platform  &          &  \\ \hline 
%Uniqueness of Distributed GNN Training &  Comparison between Distributed GNN Training and %Distributed DNN Training       &  \\ \hline 
%Optimization Trends for Distributed GNN Training  &          &  \\ \hline 

\end{tabular}
}
\end{table*}

% 问题
Despite the aforementioned efforts, there is still a dearth of review of distributed GNN training.
The need for management and cooperation among multiple computing nodes leads to a different workflow, resulting in complex computational and communication patterns, and making it a challenge to optimize distributed GNN training. 
However, regardless of plenty of efforts on this aspect have been or are being made, there are hardly any surveys on these challenges and solutions.
Current surveys mainly focus on GNN models and hardware accelerators \cite{tkde_deep_learning_on_graphs_gnngood, aiopen_graph_neural_networks_a_review_gnngood, tnnls_a_comprehensive_survey_gnngood, computing_survey_computing_graph_neural_networks_gnnsurvey1, arxiv_relational_inductive_gnnsurvey2, arxiv_machine_learning_gnnsurvey3, ijcai_graph_neural_networks_gnnsurvey4}, but they are not intended to provide a careful taxonomy and a general overview of distributed training of GNNs, especially from the perspective of the workflows, computational patterns, communication patterns, and optimization techniques.

%\cite{ia3_distdgl, atc_neugraph, socc_pagraph, pvldb_aligraph, arxiv_salient}. %{\color{red}not the first one, there are some review on distributed gnn training yet!!}
%As far as we know, this is the first comprehensive review on distributed GNN training.
%The purpose is to provide scholars with an intuitive presentation about distributed GNN training.

This paper presents a comprehensive review of distributed training of GNNs by investigating its various optimization techniques.
First, we summarize distributed GNN training into several categories according to their workflows. In addition, we introduce their computational patterns, communication patterns, and various optimization techniques proposed in recent works to facilitate scholars to quickly understand the principle of recent optimization techniques and the current research status.
Second, we introduce the prevalent software frameworks and hardware platforms for distributed GNN training and their respective characteristics.
Third, we emphasize the uniqueness of distributed GNN training by comparing it with distributed DNN training.
Finally, we discuss interesting issues and opportunities in this field.

Our main goals are as follows:
\begin{itemize}
    \item Introducing the basic concepts of distributed GNN training.
    %\item Summarizing and categorizing the techniques used in distributed GNN training.
    \item Analyzing the workflows, computational patterns, and communication patterns of distributed GNN training and summarizing the optimization techniques.
    %\item Highlighting the differences of workflow, computational pattern, and communication pattern between distributed GNN training and distributed DNN training \\
    \item Highlighting the differences between distributed GNN training and distributed DNN training
    \item Discussing interesting issues and opportunities in the field of distributed GNN training.
\end{itemize}

%This paper is arranged in accordance with the above-mentioned rules.
%To accomplish the above goals, we organize the rest of paper as follows.

%\todo{Must check description and Table \ref{table:organization} in final.}

The rest of this paper is described in accordance with these goals. Its organization is shown in Table \ref{table:organization} and summarized as follows:

\begin{itemize}
    \item Section \ref{sec:background} introduces the basic concepts of graphs and GNNs as well as two training methods of GNNs: full-batch training and mini-batch training. 
    
    \item Section \ref{sec:taxonomy} introduces the taxonomy of distributed GNN training and makes a comparison between them. \Revise{A comparison of non-distributed GNN training and distributed GNN training is also included.}
    
    \item Section \ref{sec:dft} introduces distributed full-batch training of GNNs and further categorizes it into two types, of which the workflows, computational patterns, communication patterns, and optimization techniques are introduced in detail.
    
    \item Section \ref{sec:dmt} introduces distributed mini-batch training and classifies it into two types. We also present the workflow, computational pattern, communication pattern, and optimization techniques for each type.
 
    \item Section \ref{sec:software_framework} introduces software frameworks currently supporting distributed GNN training and describes their characteristics.
    
    \item Section \ref{sec:hardware_platform} introduces hardware platforms for distributed GNN training.
    
    \item Section \ref{sec:comparison_to_dnn} highlights the uniqueness of distributed GNN training by comparing it with distributed DNN training.
    
    \item Section \ref{sec:summary} summarizes the distributed full-batch training and distributed mini-batch training, and discusses several interesting issues and opportunities in this field. 
    
    \item Section \ref{sec:conclusion} is the conclusion of this paper.

\end{itemize}

\section{Background} \label{sec:background} 

This section provides some background concepts of graphs as well as GNNs, and introduces the two training methods applied in GNN training: full-batch and mini-batch.

% \todo{Update Fig.~\ref{fig:background:graphs}
% %保持字体与其他图大小一致
% }
\begin{figure}[hbtp]
    \vspace{-10pt}
    \centering
    \includegraphics[page=9, width=0.35\textwidth]{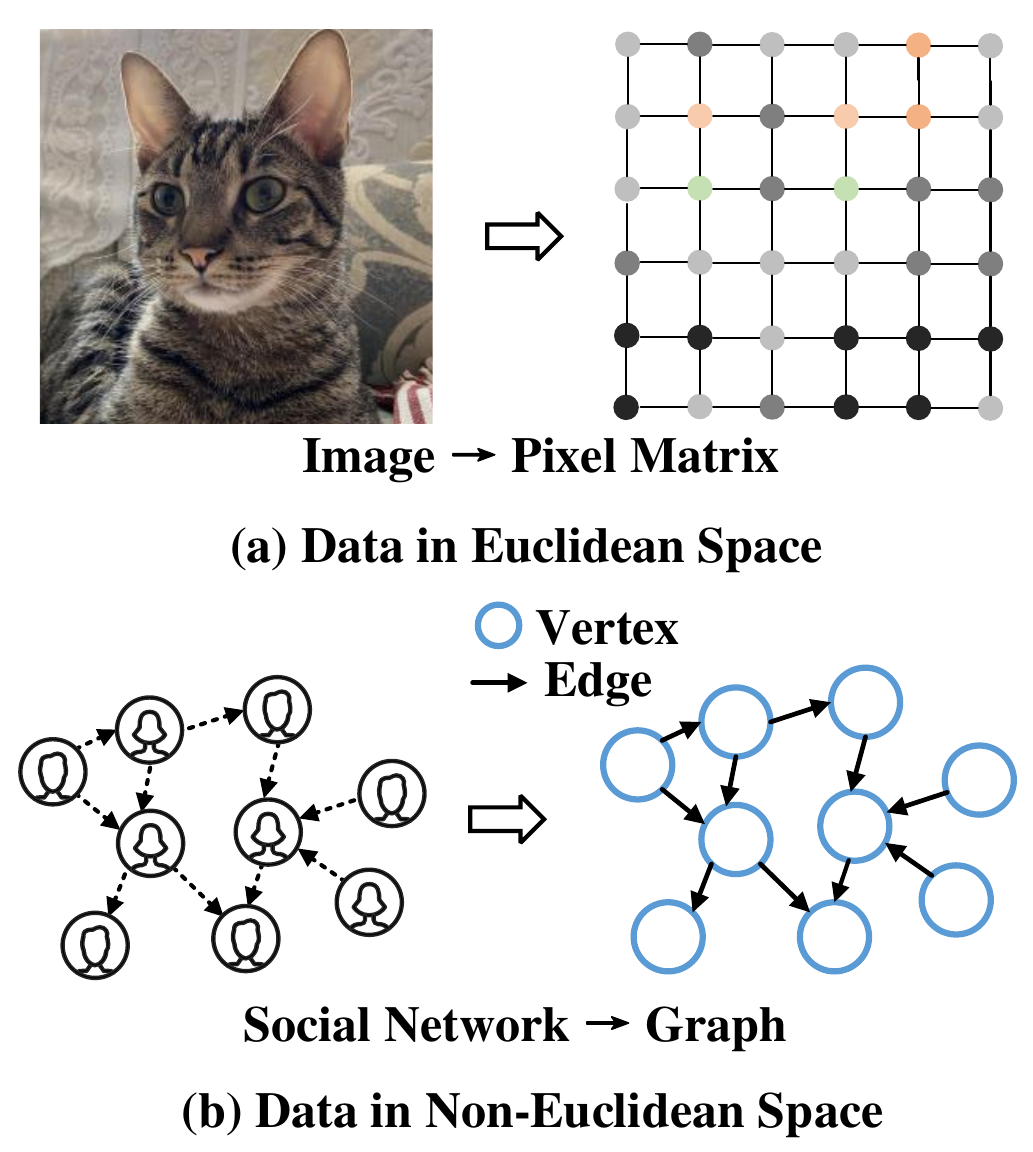}
    \vspace{-5pt}
    \caption{\Revise{Illustrations: (a) Data in Euclidean space; (b) Data in non-Euclidean space.}}
    %\todo{R1 Q2}
    \label{fig:background:graphs}
    %\vspace{-20pt}
\end{figure}

\subsection{Graphs} \label{sec:background:graphs}

%\todo{R1 Q2}
% \todo{Need to revise and check by Prof. Pan.
% %目前一个问题是：图是表示形式，还是就是数据。两个概念，Graph是表示形式，Graph data是表示之后的数据。欧基里德和非欧几里得数据都能用图表示。
% %Sound和文本都不是欧几里得数据，其次图只是一种数据的表示形式，图数据是图表示后的数据。
% %修改一下图{fig:background:graphs}，分成上下两个子图。第一个子图的左半部分是图片，右半部分是格子；第二个子图的左半部分是social network data，右半部分是抽象成的graph。
% }
\Revise{

Graphs' efficient representation of non-Euclidean data has positioned them as a popular choice for various learning tasks involving diverse sets of non-Euclidean data~\cite{iclr_graph_attention_ley_gat, RGCN, GIN, spatial_temporal_GCN, nips_inductive_representation_ley_graphsage_minibatch}.
%The efficient representation of non-Euclidean data using graphs has positioned them as a popular choice for various learning tasks involving diverse sets of non-Euclidean data~\cite{iclr_graph_attention_ley_gat, RGCN, GIN, spatial_temporal_GCN, nips_inductive_representation_ley_graphsage_minibatch}.

Euclidean data refers to data represented in Euclidean space where the Euclidean metric is used to measure distances and spatial relationships. This type of data includes various modalities such as images. Images are typically composed of pixels, where each pixel contains color information (such as red, green, and blue channels). These color values can be represented as points in Euclidean space. In a two-dimensional image, the coordinates of each pixel can be represented as points in two-dimensional Euclidean space. Fig.~\ref{fig:background:graphs} (a) illustrates an image and its corresponding pixel matrix.

Conversely, non-Euclidean data refers to data that cannot be represented in Euclidean space. This type of data may exist in spaces with curvature or non-linear metrics, such as Riemannian manifolds~\cite{lee2006riemannian}.
Graphs are widely used to represent non-Euclidean data because they provide a versatile and intuitive way to model complex relationships and structures. 
In a graph, the connections between vertices can be arbitrary and are not constrained by spatial positions. The connection between two vertices can represent entirely different relationships, which may be immeasurable or highly complex. For example, in a social network, the connection between two people could represent friendship, collaboration, and so on – relationships that are difficult to quantify using Euclidean distance in space. Fig.~\ref{fig:background:graphs} (b) shows a social network and its corresponding graph.

Graphs are commonly defined as $\mathcal{G} = (\mathcal{V}, \mathcal{E})$, where $\mathcal{V} = \{v_i | i = 1, 2, \ldots, N\}$ denotes vertices, $N$ signifies the number of vertices, and $\mathcal{E} = \{e_{ij} | v_i, v_j \in V\}$, with $|\mathcal{E}| \leq N^2$, represents edges connecting vertices. Common representation methods also include utilizing an adjacency matrix $A$ to denote the connections between vertices. 
The adjacency matrix $A$ is a $N \times N$ matrix with Boolean values, where $A_{ij}$ is 1 if $e_{ij}$ exists in $\mathcal{E}$, otherwise, it is 0. 
Through such methods, graphs can effectively capture non-Euclidean data and be harnessed for various analytical and learning tasks, especially in the context of GNNs. 
}

There are mainly three taxonomies of graphs:

\begin{itemize}
    \item \textbf{Directed/Undirected Graphs:}
    Every edge in a directed graph has a fixed direction, indicating that the connection is only from a source vertex to a destination vertex.
    %\todo{R1 Q3}
    \Revise{
    In directed graphs, a vertex's neighbors are typically classified as in-neighbors, referring to vertices connected by edges pointing towards the target vertex, and out-neighbors, referring to vertices connected by edges pointing away from the target vertex.
    However, in an undirected graph, the connection represented by an edge is bi-directional between the two vertices.
    An undirected graph can be transformed into a directed one, in which two edges in the opposite direction represent an undirected edge in the original graph.}
    
    \item \textbf{Homogeneous/Heterogeneous Graphs:}
    Homogeneous graphs contain a single type of vertex and edge, while heterogeneous graphs contain multiple types of vertices and multiple types of edges. 
    Thus, heterogeneous graphs are more powerful in expressing the relationships between different objects.
    
    \item \textbf{Static/Dynamic Graphs:}
    The structure and the features of static graphs are always unchanged, while those of dynamic graphs can change over time.
    A dynamic graph can be represented by a series of static graphs with different timestamps.
    
\end{itemize}

\subsection{Graph Neural Networks}\label{sec:background:graph_neural_network}

\begin{figure*}[hbtp]
    %\vspace{-15pt}
    \centering
    \includegraphics[page=8, width=0.60\textwidth]{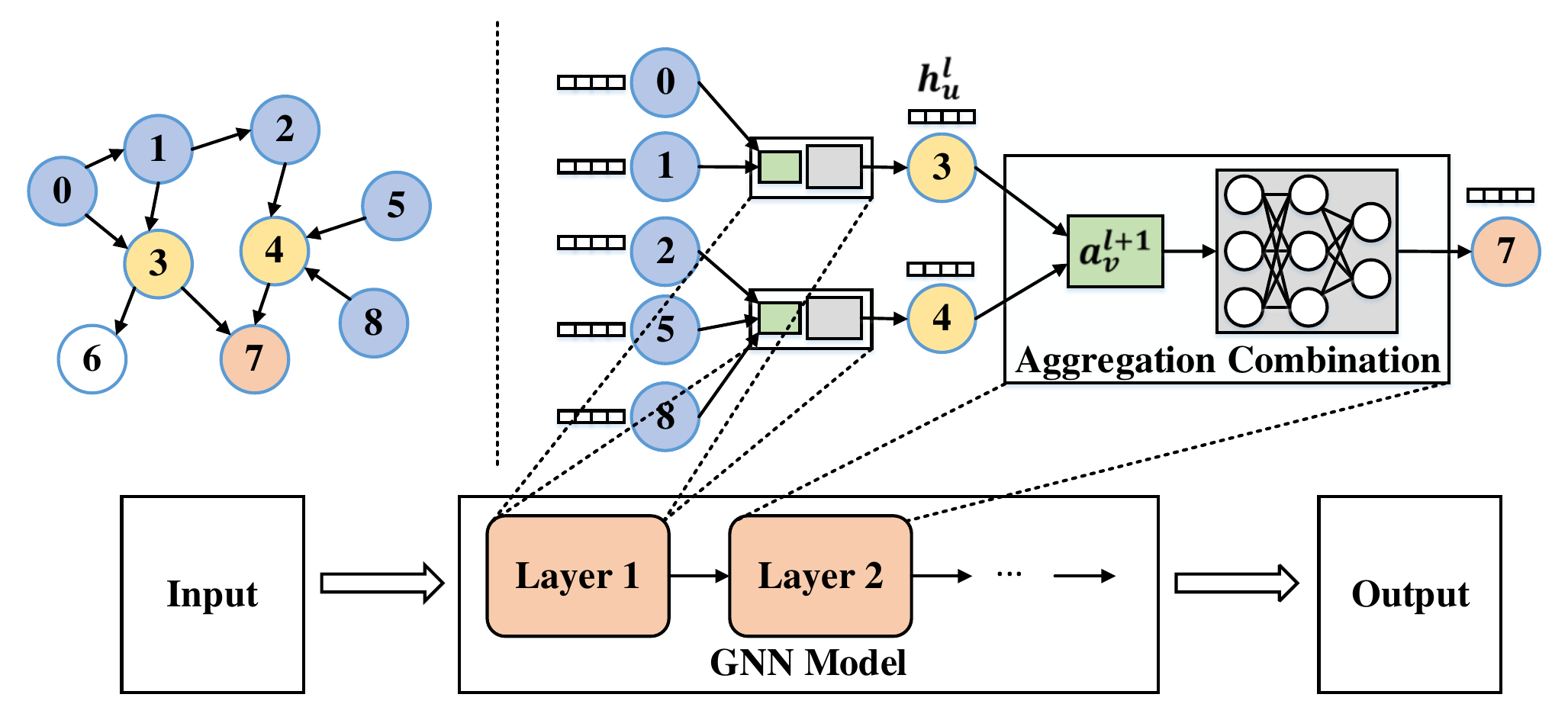}
    %\vspace{-15pt}
    \caption{\Revise{
    %Model architecture of GNNs ($h_v^l$ represents the features of vertex $v$ at the $l$-th layer and $a_v^l$ denotes the aggregation result of vertex $v$ at the $l$-th layer.).
    Model architecture of GNNs ($h_u^l$ represents the features of vertex $u$ at the $l$-th layer and $a_v^{l+1}$ denotes the aggregation result of vertex $v$ at the ($l$+1)-th layer.).
    } 
    }
    %\todo{R1 Q3}
    \label{fig:01_gnn}
    %\vspace{-20pt}
\end{figure*}

GNNs have been dominated to be a promising algorithmic model for learning knowledge from graph data~\cite{RGCN,GIN,spatial_temporal_GCN, DBLP:conf/kdd/WuPL0CZ20/pan_gnn01, DBLP:journals/tkdd/WuPDZ21/pan_gnn02, DBLP:conf/icdm/ZhuZPZW19/pan_gnn04}.
%\Revise{to do: add pan article ``GNN, GCN''  here}
It takes the graph data as input and learns a representation vector for each vertex in the graph. The learned representation can be used for down-stream tasks such as vertex prediction \cite{iclr_semi_supervised_classification_with_nodeprediction}, 
link prediction \cite{nips_link_prediction_based_on_linkprediction}, and
graph prediction \cite{aaai_an_end_to_end_deep_learing_graphprediction}.

\Revise{
A GNN model consists of one or multiple layers, each of which includes the Aggregation step and the Combination step. An illustration is provided in Fig. \ref{fig:01_gnn}.
In the Aggregation step, the Aggregate function $Aggregate(~)$ is used to aggregate the feature vectors of in-neighbors from the previous layer or input layer for each target vertex.
For example, in Fig. \ref{fig:01_gnn}, vertex 4 gathers the feature vectors of itself and its in-neighbors (i.e., vertex 2, 5, 8) using graph operations.
In the Combination step, the Combine function $Combine(~)$ transforms the aggregated feature vector of each vertex using neural network operations.
}
To sum up, the aforementioned computation on a graph $\mathcal{G(V,E)}$ can be formally expressed by
\begin{align}   % 这个包可以用&表示不同行公式之间的对齐，在多行公式上会更美观
    a_v^l & = Aggregate(\{h_u^{l-1}|u \in N(v) \cup \{v\}\}) \label{eq1} \\
    h_v^l & = Combine(a_v^l) \label{eq2}
\end{align}
where $h_v^l$ denotes the feature vector of vertex $v$ at the $l$-th layer,  $a_v^l$ represents the aggregation result of vertex $v$ at the $l$-th layer, and $N(v)$ represents the in-neighbors of vertex $v$.
Specifically, the input features of vertex $v \in \mathcal{V}$ is denoted as $h_v^0$. 
%\todo{R1 Q3}
\Revise{
Note that in GNNs, every vertex in a directed graph gathers information from its in-neighbors. Throughout this paper, the term ``the vertex's neighbors'' specifically refers to its in-neighbors. In the context of an undirected graph, it can be converted into a directed graph by splitting each edge into two directed edges, each with opposite directions.
}

\Revise{
To better grasp the computation process, we explain the process using an example of a two-layer GNN as shown in Fig.~\ref{fig:01_gnn}.
We first select vertex 7 as our target vertex. 
This is the vertex from which we aim to obtain its final features.
Vertex 7 has 1-hop in-neighbors, namely vertex 3 and 4, and 2-hop in-neighbors, which include vertex 0, 1, 2, 5, and 8. 
The term ``$k$-hop'' indicates vertices that are reachable within exactly $k$ edges from a given vertex.
In the first layer, vertex 3 and 4 collect features from vertex 0, 1, 2, 5, and 8 and perform a combination operation. 
Subsequently, in the second layer, vertex 7 gathers features from vertex 3 and 4 and performs a combination operation to generate the final features.
}

\Revise{
In addition, it's crucial to clarify that the existence of cycles in a graph does not influence the computation of GNNs. In the Aggregation step of each GNN layer, every vertex gathers information from its in-neighbors. This entire computation process doesn't require consideration of whether the graph contains cycles.
However, when cycles are present in the graph, they can potentially impact the model accuracy.
The presence of cycles may result in the repetition and mixing of information after multiple rounds of GNN propagation, leading to what is known as the over-smoothing phenomenon. Over-smoothing occurs when vertex features become excessively similar, causing a loss of discriminative information between vertices~\cite{DBLP:conf/aistats/LiuQWL23/EEGNN/pd07_review, DBLP:conf/iclr/0002Y21/bottleneck_of_gnn/pd08}.
To address the issue, techniques such as DropEdge~\cite{DBLP:conf/iclr/RongHXH20/dropedge/pd02} and DropNode~\cite{DBLP:journals/corr/abs-2008-09864/pd09/dropnode} are employed. These techniques involve selectively removing edges and vertices. It will be further discussed in Section~\ref{sec:summary:performance_deterioration}.
}

\begin{figure*}[hbtp]
    %\vspace{-15pt}
    \centering
    \includegraphics[page=2, width=0.99\textwidth]{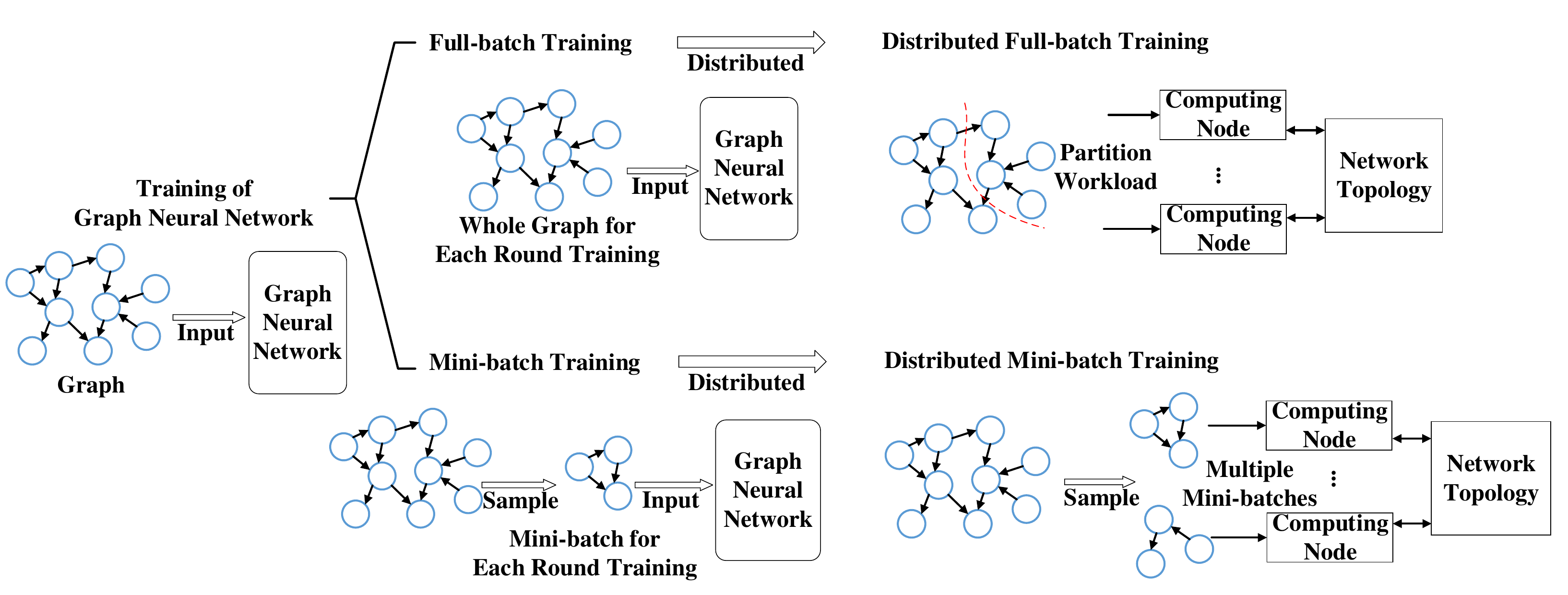}
    %\vspace{-15pt}
    \caption{Training methods for GNNs and their distributed implementations.}
    \label{fig:overall_training_method}
    %\vspace{-20pt}
\end{figure*}

%Due to that, the distributed training of GNNs can also be divided into distributed full-batch training and distributed mini-batch training.
%For brevity, in the rest of the paper, we refer to distributed full-batch training as DFT, and refer to distributed mini-batch training as DMT.

\begin{figure*}[hbtp]
    %\vspace{-15pt}
    \centering
    \includegraphics[page=11, width=0.75\textwidth]{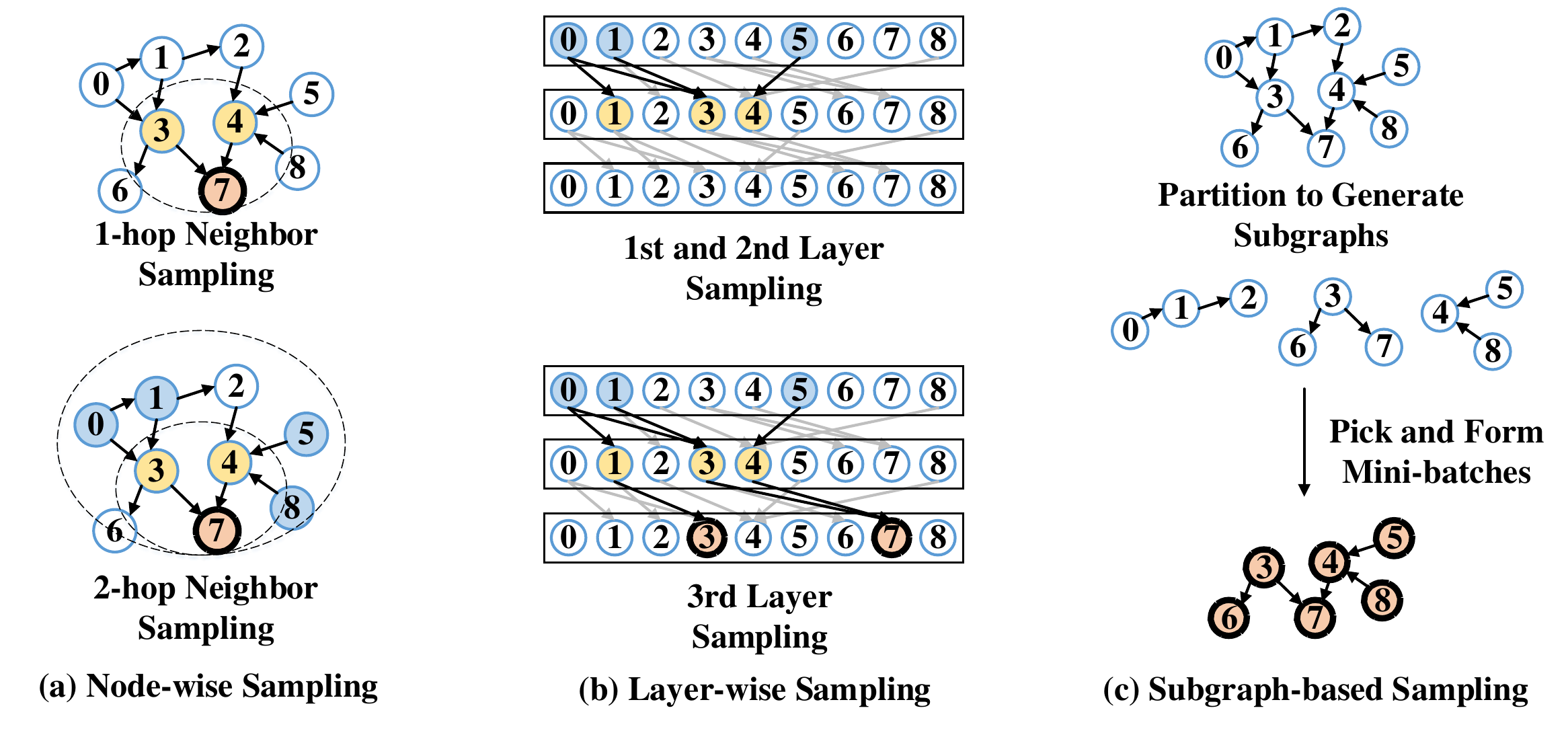}
    %\vspace{-15pt}
    %\caption{\Revise{Illustrations of three groups of sampling methods: (a) Node-wise sampling (Vertex 7 is the target vertex.); (b) Layer-wise sampling; (c) Subgraph-based sampling.}}
    \caption{\Revise{Illustrations of three groups of sampling methods: (a) Node-wise sampling (vertex 7 is the target vertex, vertices colored in yellow represent sampled 1-hop in-neighbors, and vertices colored in blue represent sampled 2-hop in-neighbors.); (b) Layer-wise sampling (vertices colored in blue, yellow, and orange respectively represent the vertices sampled from the first, second, and third layers.); (c) Subgraph-based sampling.}}
    %\todo{R1 Q5}
    \label{fig:background_sampling_methods}
    %\vspace{-20pt}
\end{figure*}

\subsection{Training Methods for Graph Neural Networks} \label{sec:background:training_method}

In this subsection, we introduce the training methods for GNNs, which are approached in two ways including full-batch training \cite{iclr_semi_supervised_classification_ley_gcn_fullbatch, iclr_graph_attention_ley_gat} and mini-batch training \cite{iclr_fastgcn_ley_minibatch, nips/adaptive_sampling/layerwise2, nips_inductive_representation_ley_graphsage_minibatch, iclr_graphsaint, kdd_gns}.

A typical training procedure of neural networks, including GNNs, includes forward propagation and backward propagation.
In forward propagation, the input data is passed through the layers of neural networks towards the output.
Neural networks generate differences of the output of forward propagation by comparing it to the predefined labels.
Then in backward propagation, these differences are propagated through the layers of neural networks in the opposite direction, generating gradients to update the model parameters.

%As illustrated in Fig. \ref{fig:overall_training_method}, training methods of GNNs can be classified into full-batch training \cite{iclr_semi_supervised_classification_ley_gcn_fullbatch, iclr_graph_attention_ley_gat} and mini-batch training \cite{iclr_fastgcn_ley_minibatch, nips/adaptive_sampling/layerwise2, nips_inductive_representation_ley_graphsage_minibatch, iclr_graphsaint, kdd_gns}, depending on whether the whole graph is involved.
As illustrated in Fig. \ref{fig:overall_training_method}, training methods of GNNs can be classified into full-batch training \cite{iclr_semi_supervised_classification_ley_gcn_fullbatch, iclr_graph_attention_ley_gat} and mini-batch training \cite{iclr_fastgcn_ley_minibatch, nips/adaptive_sampling/layerwise2, nips_inductive_representation_ley_graphsage_minibatch, iclr_graphsaint, kdd_gns}, depending on whether the whole graph is involved in each \textbf{round}.
% Tip: 既然下面已经下了定义，就没必要再非要强调一遍round包括参数更新了
Here, we define a \textbf{round} of full-batch training consisting of a model computation phase, including forward and backward propagation, and a parameter update phase.
On the other hand, a \textbf{round} in mini-batch training additionally contains a sampling phase, which samples a small-sized workload required for the subsequent model computation and thus locates prior to the other two phases.
Thus, an \textbf{epoch}, which is defined as an entire pass of the data, is equivalent to a round of full-batch training, while that in mini-batch training usually contains several rounds. % TODO: Check whether this is accurate
Details of these two methods are introduced below.

%In full-batch training, a round represents  one epoch, where an epoch means to train the model parameters once using all the data in the training set.
%The model parameters are updated only when all the raw graph data has been processed once.
%In mini-batch training, the steps in a round include sampling raw graph data to generate one mini-batch and performing forward propagation as well as backward propagation on this mini-batch. 

\subsubsection{Full-batch Training} \label{sec:background:full_batch_training}
Full-batch training utilizes the whole graph to update model parameters in each round.

%\todo{R1 Q4}
\Revise{
%V 全部数据集，从中抽取出 Vt 训练数据集，Vval 验证数据集，Vtest测试数据集
A complete dataset $\mathcal{V}$ is typically divided into a training set $\mathcal{V}_t$, a validation set $\mathcal{V}_{val}$, and a test set $\mathcal{V}_{test}$.
Using the training set $\mathcal{V}_t \subseteq \mathcal{V}$, the loss function of full-batch training is
}
% TODO: 需再次确认这里应该用哪种包含符号（subset、subseteq还是subsetneqq）
% 好像\subseteq更合适，可以等于  --林
\begin{equation}\label{eq3_full_loss}
    \mathcal{L} = \frac{1}{|\mathcal{V}_t|}\sum_{v_i \in \mathcal{V}_t} \nabla l(y_i, z_i)
\end{equation}
where \Revise{$\mathcal{L}$ is the loss,} $\nabla l(~)$ is a loss function, $y_i$ is the known label of vertex $v_i$, and $z_i$ is the output of GNN model when inputting features $x_i$ of $v_i$.
In each epoch, GNN model needs to aggregate representations of all neighboring vertices for each vertex in $\mathcal{V}_t$ all at once. % TODO: Check whether this statement is accurate.
As a result, the model parameters are updated only once at each epoch.

\subsubsection{Mini-batch Training}\label{sec:background:mini_batch_training}

%{\color{red}{a graph include 4 execution phase of mini-batch training}}

Mini-batch training utilizes part of the vertices and edges in the graph to update model parameters in every forward propagation and backward propagation. 
%It aims to reduce the number of vertices involved in the computation of one round to reduce the requirement of computing resource and memory resource.
It aims to reduce the number of vertices involved in the computation of one round to reduce the computing and memory  resource requirements.

Before each round of training, a mini-batch $\mathcal{V}_s$ is sampled from the training dataset $\mathcal{V}_t$.
By replacing the full training dataset $\mathcal{V}_t$ in equation (\ref{eq3_full_loss}) with the sampled mini-batch $\mathcal{V}_s$, we obtain the loss function of mini-batch training:
\begin{equation}\label{eq3_mini_loss}
    \mathcal{L} = \frac{1}{|\mathcal{V}_s|}\sum_{v_i \in \mathcal{V}_s} \nabla l(y_i, z_i)
\end{equation}
It indicates that for mini-batch training, the model parameters are updated multiple times at each epoch, since numerous mini-batches are needed to have an entire pass of the training dataset, resulting in many rounds in an epoch.
Stochastic Gradient Descent (SGD) \cite{compstat_sgd}, a variant of gradient descent which applies to mini-batch, is used to update the model parameters according to the loss $\mathcal{L}$.

\textbf{Sampling:}
Mini-batch training requires a sampling phase to generate the mini-batches.
The sampling phase first samples a set of vertices, called target vertices, from the training set according to a specific sampling strategy, and then it samples the neighboring vertices of these target vertices to generate a complete mini-batch.
%as shown in Fig. \ref{fig:background_sampling_methods}.
The sampling method can be generally categorized into three groups: Node-wise sampling, Layer-wise sampling, and Subgraph-based sampling \cite{liuxin_sampling_survey, arxiv_salient}.
%\todo{R1 Q5}
\begin{itemize}
    \item \textbf{Node-wise sampling} \cite{nips_inductive_representation_ley_graphsage_minibatch, kdd_graph_convolutional_ley_pinsage, icml_learing_steady_states_ley_nodewise1, icml_stochastic_training_ley_nodewise2} is directly applied to the neighbors of a vertex: the algorithm selects a subset of each vertex's neighbors, \Revise{as depicted in Fig. \ref{fig:background_sampling_methods} (a).} 
    It is typical to specify a different sampling size for each layer. 
    For example, in GraphSAGE \cite{nips_inductive_representation_ley_graphsage_minibatch}, it samples at most 25 neighbors for each vertex in the first layer and at most 10 neighbors in the second layer.
    \item \textbf{Layer-wise sampling} \cite{iclr_fastgcn_ley_minibatch, nips_layer_dependent_ley_layerwise1, nips/adaptive_sampling/layerwise2} enhances Node-wise sampling. 
    %It selects multiple vertices at a time and then proceeds recursively layer by layer.
    It selects multiple vertices in each layer and then proceeds recursively layer by layer, \Revise{as shown in Fig. \ref{fig:background_sampling_methods} (b).}
    \item \textbf{Subgraph-based sampling} \cite{iclr_graphsaint, kdd/cluster_gcn/subgraph_based1, ijcnn/ripple_walk_training/subgraph_based2, ipdps/accurate_efficient/subgraph_based3} first partition the original graph into multiple subgraphs, and then samples the mini-batches from one or a certain number of them, \Revise{as illustrated in Fig. \ref{fig:background_sampling_methods} (c).}
\end{itemize}

%\tableofcontents

\section{Taxonomy of Distributed GNN Training}\label{sec:taxonomy}

\begin{figure*}[hbtp]
    %\vspace{-15pt}
    \centering
    \includegraphics[page=10, width=0.75\textwidth]{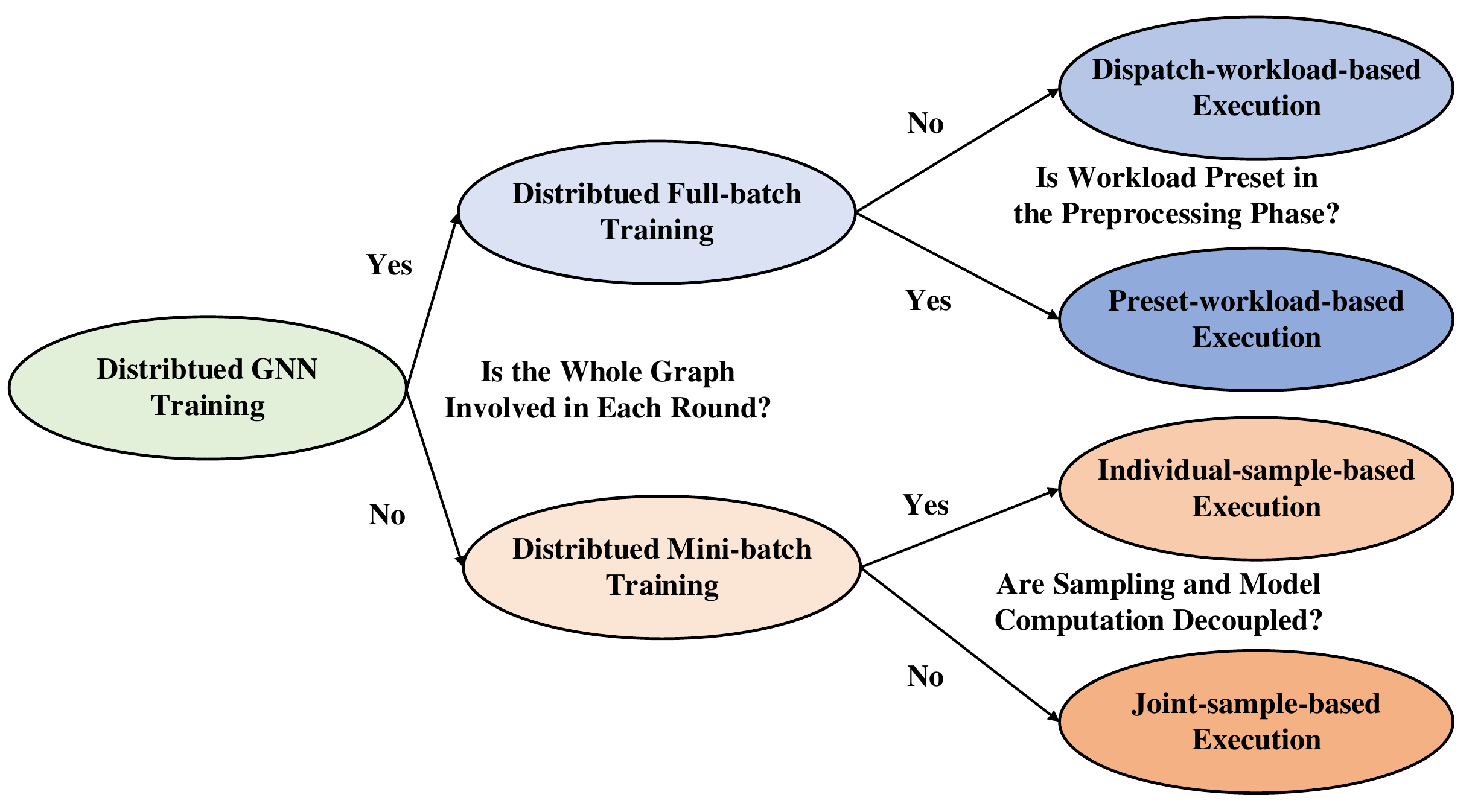}
    %\vspace{-15pt}
    \caption{Taxonomy of distributed GNN training.}
    \label{fig:taxonomy:tax_of_cate}
    %\vspace{10pt}
\end{figure*}

This section introduces the taxonomy of distributed GNN training.
As shown in Fig. \ref{fig:taxonomy:tax_of_cate}, we firstly categorize it into distributed full-batch training and distributed mini-batch training, according to the training method introduced in Section \ref{sec:background:training_method}, i.e., whether the whole graph is involved in each round, and show the key differences between the two types.
Each of the two types is classified further into two detailed types respectively by analyzing their workflows. This section introduces the first-level category, that is, distributed full-batch training and distributed mini-batch training, and makes a comparison between them.
The second-level category of the two types is introduced in Section \ref{sec:dft} and Section \ref{sec:dmt}, respectively.

\subsection{Distributed Full-batch Training} \label{sec:taxonomy:dft_intro}

%Distributed GNN training can be divided into distributed full-batch training and distributed mini-batch training according to the classification strategy shown in Fig. \ref{fig:overall_training_method}, which is also a currently recognized classification \cite{ia3_distdgl, arxiv_salient, arxiv_distdglv2}.

%\todo{
% workflow 特征
% communication pattern

%加维度
%1. 传输数据谁占主导：   节点feature      mini-batch传输
%2. 节点计算独立性:        low              high
%3. 通讯量大小：           多               少

%4. 是否涉及到全图的计算

%5.
% advan/disadvan
%改成 mmeomry capacity requirement, bandwidth requirements， communication volume

%6.irregularity of communication   high   low

%精度别写了

%在comparison里， high bandwidth引用下 distgnn
%}

Distributed full-batch training is the distributed implementation of GNN full-batch training, as illustrated in Fig. \ref{fig:overall_training_method}.
Except for graph partition, a major difference is that multiple computing nodes need to synchronize gradients before updating model parameters, so that the models across the computing nodes remain unified.
Thus, a round of distributed full-batch training includes two phases: model computation (forward propagation + backward propagation) and gradient synchronization. The model parameter update is included in the gradient synchronization phase.

Since each round involves the entire raw graph data, a considerable amount of computation and a large memory footprint are required in each round \cite{arxiv_sar,eurosys_flexgraph,sc_distgnn}.
To deal with it, distributed full-batch training mainly adopts the workload partitioning method \cite{atc_neugraph,mlsys_roc}: split the graph to generate small workloads, and hand them over to different computing nodes.

Such a workflow leads to a lot of irregular inter-node communications in each round, mainly for transferring the features of vertices along the graph structure.
This is because the graph data is partitioned and consequently stored in a distributed manner, and the irregular connection pattern in a graph, such as the arbitrary number and location of a vertices' neighbors.
Therefore, many uncertainties exist in the communication of distributed full-batch training, including the uncertainty of the communication content, target, time, and latency, leading to challenges in the optimization of distributed full-batch training.

As shown in Fig. \ref{fig:taxonomy:tax_of_cate}, we further classify distributed full-batch training more specifically into two categories according to whether the workload is preset in the preprocessing phase, namely dispatch-workload-based execution and preset-workload-based execution, as shown in the second column of Table \ref{table:category}.
Their detailed introduction and analysis are presented in Section \ref{sec:dft:chunk_execution} and Section \ref{sec:dft:partition_execution}.

\subsection{Distributed Mini-batch Training} \label{sec:taxonomy:dmt_intro}

Similar to distributed full-batch training, distributed mini-batch training is the distributed implementation of GNN mini-batch training as in Fig. \ref{fig:overall_training_method}.
It also needs to synchronize gradients prior to model parameter update, so a round of distributed mini-batch training includes three phases: sampling, model computation, and gradient synchronization. The model parameter update is included in the gradient synchronization phase.

Distributed mini-batch training parallelizes the training process by processing several mini-batches simultaneously, one for each computing node.
The mini-batches can be sampled either by the computing node itself or by other devices, such as another node specifically for sampling.
Each computing node performs forward propagation and backward propagation on its own mini-batch.
Then, the nodes synchronize and accumulate the gradients, and update the model parameters accordingly.
Such a process can be formulated by 
\begin{equation}\label{eq01_ley_minibatch_gradient} 
    \mathcal{W}_{i+1} = \mathcal{W}_i + \sum^n_{j = 1} \nabla g_{i,j}
\end{equation}
where $\mathcal{W}_i$ is the weight parameters of model in the $i^{\text{th}}$ round of computation, $\nabla g_{i,j}$ is the gradients generated in the backward propagation of the computing node $j$ in the $i^{\text{th}}$ round of computation, and the $n$ is the number of the computing nodes.

As shown in Fig. \ref{fig:taxonomy:tax_of_cate}, we further classify it more specifically into two categories according to whether the sampling and model computation are decoupled, namely individual-sample-based execution and joint-sample-based execution, as shown in the second column of Table \ref{table:category}.
Their detailed introduction and analysis are presented in Section \ref{sec:dmt:sample_individual_execution} and Section \ref{sec:dmt:sample_joint_execution}.

\begin{table*}[!t]
    %\vspace{-12pt}
 \caption{Comparison between distributed full-batch training and distributed mini-batch training.} \label{table:overall_comparison_distrbuted_training}
 %\vspace{-10pt}
 \centering
\renewcommand\arraystretch{1.8} 
\resizebox{1\textwidth}{!}{
\begin{tabular}{|c|cc|}
\hline
 & \textbf{Distributed Full-batch Training}   & \textbf{Distributed Mini-batch Training}  \\ \hline \hline
 \textbf{Workflow Characteristic} & Collaborative Computation with Workload Partition & Independent Computation with Periodic Synchronization \\ \hline
 \textbf{Independence of Computation}   & Low   & High  \\ \hline
 \textbf{Involvement of Entire Graph?}  & Yes   & No  \\ \hline
 \textbf{Memory Capacity Requirement}   & High & Low \\ \hline
  \textbf{Communication Volume}          & Large & Small \\ \hline
 \textbf{Prime Communication Content}   & Features of Neighboring Vertices & Mini-batches \\ \hline
  \textbf{Communication Irregularity}   & High & Low \\ \hline
 \textbf{Communication Uncertainty}     & High & Low \\ \hline
 \textbf{Communication Bandwidth Requirement} & High & Low \\ \hline
 \textbf{Additional Overhead}           & -    & An Extra Phase (Sampling Phase) \\ \hline
 \textbf{Primary Challenges}            & Workload Imbalance and Massive Transmissions & Insufficient Sampling Performance \\ \hline

\end{tabular}
    }
\end{table*}

\subsection{Comparison Between Distributed Full-batch Training and Distributed Mini-batch Training} \label{sec:taxonomy:dft_dmt_comparison}

%The comparison between distributed full-batch training and distributed mini-batch training is summarized in Table. \ref{table:overall_comparison_distrbuted_training} in terms of workflow, communication uncertainty, communication bandwidth requirement, advantage, and disadvantage.

%Its communication characteristics include large communication volume, high communication irregularity, and high communication uncertainty.
%Next we explain them separately.
%Therefore, it requires high communication bandwidth \cite{sc_distgnn}.

%This section compares distributed full-batch training with distributed mini-batch training of GNNs. The major differences are also summarized in Table \ref{table:overall_comparison_distrbuted_training}.
This subsection compares distributed full-batch training with distributed mini-batch training of GNNs. The major differences are also summarized in Table \ref{table:overall_comparison_distrbuted_training}.

The workflow of distributed full-batch training is summarized as collaborative computation with workload partition.
Since the computation in each round involves the entire graph, the computing nodes need to cache it locally, leading to a high memory capacity requirement \cite{atc_neugraph}.
Also, the communication volume of distributed full-batch training is large \cite{sc_distgnn, sc_reducint_communication_in_graph_cagnet}.
In every round, the Aggregate function needs to collect the features of neighbors for each vertex, causing a large quantity of inter-node communication requests since the graph is partitioned and stored on different nodes.
Considering that the communication is based on the irregular graph structure, the communication irregularity of distributed full-batch training is high \cite{eurosys_dgcl}.
Another characteristic of communication is high uncertainty.
The time of generating the communication request is indeterminate, since each computing node sends communication requests according to the currently involved vertices in its own computing process.
As a result, the main challenges of distributed full batch training are workload imbalance and massive transmissions \cite{eurosys_dgcl, sc_distgnn, mlsys_roc}.

In contrast, the workflow of distributed mini-batch training is summarized as independent computation with periodic synchronization.
%The major transmission content are mini-batches, which are sent from the sampling node (or component) to the computing node (or component) responsible for the current mini-batch.
The major transmission content is the mini-batches, sent from the sampling node (or component) to the computing node (or component) responsible for the current mini-batch \cite{socc_pagraph, cluster_2pgraph, /cal22/linhaiyang/characterizing_and_gnn/}.
As a result, these transmissions have low irregularity and low uncertainty, as the direction and content of transmission are deterministic.
Since the computation of each round only involves the mini-batches, it triggers less communication volume and requires less memory capacity \cite{socc_pagraph}.
However, the extra sampling phase may cause some new challenges.
Since the computation of the sampling phase is irregular and requires access to the whole graph for neighbor information of a given vertex, it is likely to encounter the problem of insufficient sampling performance, causing the subsequent computing nodes (or components) to stall due to lack of input, resulting in a performance penalty \cite{/cal22/linhaiyang/characterizing_and_gnn/, arxiv_salient}.

\begin{table*}[hbtp]
    %\vspace{-12pt}
 \caption{Summary of recent studies on distributed GNN training.} \label{table:category}
 %\vspace{-10pt}
 \centering
 \renewcommand\arraystretch{1.5} 
    \resizebox{0.99\textwidth}{!}{
\begin{tabular}{|c|rrrrrr|}
\hline
\textbf{Name}     & \textbf{Taxonomy}   & \textbf{Software Framework} & \textbf{Hardware Platform} &\textbf{GPU Types}  &\textbf{Year} &\textbf{Code Available}\\ \hline \hline
\multicolumn{7}{|c|}{\begin{tabular}[c]{@{}c@{}} \textbf{Distributed Full-batch Training} \end{tabular}} \\ \hline
             NeuGraph \cite{atc_neugraph}          &Dispatch-workload-based Execution        &NeuGraph   &multi-GPU  &Tesla P100 &2019 &-\\
             Roc \cite{mlsys_roc}                  &Dispatch-workload-based Execution        &Roc        &multi-GPU  &Tesla P100 &2020 &Yes\\
             CAGNET \cite{sc_reducint_communication_in_graph_cagnet}              &Preset-workload-based Execution        &PyTorch    &multi-GPU  &Tesla V100  &2020 &Yes\\
             DGCL \cite{eurosys_dgcl}              &Preset-workload-based Execution    &PyTorch       &multi-GPU  &Tesla V100  &2021 &-\\
             DistGNN \cite{sc_distgnn}             &Preset-workload-based Execution    &DGL        &multi-CPU  &-   &2021   &Yes\\
             FlexGraph \cite{eurosys_flexgraph}    &Preset-workload-based Execution    &FlexGraph  &multi-CPU  &-   &2021   &-\\
             MG-GCN \cite{arxiv_mg_gcn}            &Preset-workload-based Execution        &MG-GCN     &multi-GPU  &Tesla V100/A100 &2021 &Yes\\
             Dorylus \cite{osdi_dorylus}           &Preset-workload-based Execution        &Dorylus    &multi-CPU  &- &2021 &Yes\\
             SAR\cite{arxiv_sar}                   &Preset-workload-based Execution    &DGL        &multi-CPU  &- &2021 &-\\ 
\hline 
\multicolumn{7}{|c|}{\begin{tabular}[c]{@{}c@{}} \textbf{Distributed Mini-batch Training} \end{tabular}} \\ \hline
            AliGraph \cite{pvldb_aligraph}     &Joint-sample-based Execution     &AliGraph   &multi-CPU  &- &2019 &-\\
            DistDGL \cite{ia3_distdgl}         &Joint-sample-based Execution     &DGL    &multi-CPU  &-       &2020       &-\\
            AGL \cite{pvldb_agl}               &Individual-sample-based Execution          &AGL    &multi-CPU  &- &2020 &- \\
            PaGraph \cite{socc_pagraph}        &Joint-sample-based Execution     &DGL    &multi-GPU  &GTX 1080Ti  &2020   &-\\
            2PGraph \cite{cluster_2pgraph}     &Joint-sample-based Execution     &PyTorch    &multi-GPU  &RTX 3090 &2021  &-\\
            LLCG \cite{arxiv_llcg}             &Joint-sample-based Execution     &PyG    &multi-GPU  &RTX 8000    &2021   &Yes\\
            %DistDGLv2 \cite{arxiv_distdglv2}       &Joint-sample-based Execution &DistDGLv2  &multi-GPU  &Tesla T4  &2021 &-\\
            GraphTheta \cite{arxiv_graphtheta}     &Joint-sample-based Execution       &GraphTheta &multi-CPU  &- &2021 &-\\
            SALIENT \cite{arxiv_salient}       &Individual-sample-based Execution           &PyG    &multi-GPU  &Tesla V100    &2021   &-\\
            P3 \cite{/osdi/Gandhi/p3/distributed_deep_graph/}  &Joint-sample-based Execution &DGL &multi-GPU &Tesla P100 &2021 &-\\
            DistDGLv2 \cite{arxiv_distdglv2}       &Joint-sample-based Execution &DistDGLv2  &multi-GPU  &Tesla T4  &2022 &-\\
            DGTP \cite{DBLP:conf/infocom/LuoBW22/new02/DGTP} &Individual-sample-based Execution & DGL &multi-GPU &GTX 1080Ti &2022 &-\\
\hline 
\end{tabular}
    }
\end{table*}

\subsection{Other Information of Taxonomy}\label{sec:taxonomy:others_information}

%In order to give a comprehensive review for recent studies, we supplement following information in Table \ref{table:category} to show more details about recent studies. 

Table \ref{table:category} provides a summary of the current studies on distributed GNN training using our proposed taxonomy.
Except for the aforementioned classifications, we also add some supplemental information in the table to provide a comprehensive review of them.

\textbf{Software frameworks.}
The software frameworks used by the various studies are shown in the third column of Table \ref{table:category}.
PyTorch Geometric (PyG) \cite{arxiv_pyg} and Deep Graph Library (DGL) are the most popular among them.
In addition, there are many newly proposed software frameworks aiming at distributed training of GNNs, and many of them are the optimization version of PyG \cite{arxiv_pyg} or DGL \cite{arxiv_dgl}.
%However, most of them are not available now.
A detailed introduction to the software frameworks of distributed GNN training is presented in Section \ref{sec:software_framework}.

\textbf{Hardware platforms.}
Multi-CPU platform and multi-GPU platform are the most common hardware platforms of distributed GNN training, as shown in the fourth column of Table \ref{table:category}.
Multi-CPU platform usually refers to a network with multiple servers, which uses CPUs as the only computing component. % TODO: Check correctness
On the contrary, in multi-GPU platforms, GPUs are responsible for the major computing work, while CPU(s) handle some computationally complex tasks, such as workload partition and sampling.
A detailed introduction to the hardware platforms is presented in Section \ref{sec:hardware_platform}.

\textbf{Year.} 
The contribution of distributed GNN training began to emerge in 2019 and is now showing a rapid growth trend.
This is because more attention is paid to it due to the high demand from industry and academia to shorten the training time of GNN model.

\textbf{Code available.}
The last column of Table \ref{table:category} simply records the open source status of the corresponding study on distributed GNN training for the convenience of readers.

\begin{table*}[hbtp]
    %\vspace{-12pt}
 \caption{\Revise{Comparison Between Non-distributed GNN Training and Distributed GNN Training.}} \label{table:comparison_between_distrbuted_and_non_distributed}
 %\vspace{-10pt}
 \centering
\renewcommand\arraystretch{1.8} 
\resizebox{0.65\textwidth}{!}{
\begin{tabular}{|c|cc|}
\hline
 & \textbf{Non-distributed GNN Training}   & \textbf{Distributed GNN Training}  \\ \hline \hline
\textbf{Scalability} & Weak & Strong \\ \hline
\textbf{Training Speed} & Slow & Fast \\ \hline
\textbf{Data Processing and Storage} & Local File System & Distributed File System \\ \hline
\textbf{Communication Overhead} & Low & High \\ \hline
%\textbf{Fault Tolerance and Stability} & High & Low \\ \hline
\textbf{Data Security} & High & Low \\ \hline
\textbf{Model Accuracy} & --- & May Decrease \\ \hline

\end{tabular}
    }
\end{table*}

%\todo{R2 Q1}
\Revise{
\subsection{Comparison Between Non-distributed GNN Training and Distributed GNN Training}\label{sec:taxonomy:non_dt_and_dt_comparison}

To clearly distinguish distributed GNN training, a detailed comparison between non-distributed and distributed GNN training is presented below. A comprehensive summary of this comparison is also provided in Table \ref{table:comparison_between_distrbuted_and_non_distributed}.

In the context of non-distributed GNN training, its scalability is limited, rendering it challenging to accommodate the demands of larger graph datasets and larger model training requirements. 
The training speed is significantly constrained by the finite computational and storage resources available on a single machine. 
Furthermore, data preparation and storage predominantly rely on local data file systems. This approach is characterized by low communication overhead and is well-suited for computing systems that lack high-speed network infrastructure support.
%In terms of fault tolerance and stability, non-distributed training exhibits a high degree of robustness, primarily due to its single-machine computation, simplifying system maintenance. 
Moreover, non-distributed training demonstrates high data security compared with distributed training, as it exclusively involves local data processing, thereby enhancing its resistance to external security threats.

In contrast, distributed GNN training offers enhanced scalability, permitting the dynamic addition of computing nodes to accommodate larger graph datasets and models as needed. 
The availability of greater computational and storage resources contributes to substantially faster training speeds. 
Data processing and storage primarily rely on distributed file systems. 
However, this distributed approach results in higher communication overhead, necessitating frequent data and model parameter transfers across computing nodes. 
%In addition, distributed training presents relatively lower fault tolerance and stability due to the inherent complexities of multi-machine computation. 
%The failure of a single machine can potentially disrupt the entire computation, warranting the implementation of protective measures. 
Furthermore, distributed training exhibits lower data security due to network transmission and multi-machine computation.

Transitioning from non-distributed training to distributed training can enhance training speed by utilizing additional computing nodes, but it may also impact the model's accuracy. 
For commonly used distributed training methods, they typically retain the same fundamental computational processes, merely transitioning from single-machine computation to multi-machine computation. 
For instance, in synchronous training methods within distributed mini-batch training~\cite{ia3_distdgl}, the computational process remains essentially equivalent to that of single-machine computation.
Consequently, the convergence behavior and final accuracy of the model remain consistent with single-machine training.

Subsequently, we delve into the reasons why some distributed training might affect model accuracy. Firstly, distributed training often involves the use of larger batch sizes, where the batch size refers to the number of samples used for a single model parameter update during training. 
Larger batch sizes can potentially lead to overfitting, wherein the model excessively fits the training data, thereby affecting accuracy. To address this issue, a common approach is to adjust the learning rate. 
Secondly, in the case of some other distributed training methods, changes in model accuracy often stem from the adoption of aggressive computational strategies, such as delayed aggregation in distributed full-batch training and asynchronous training in distributed mini-batch training. 
These aggressive methods typically enhance parallel computing efficiency and reduce node stagnation waiting time by allowing computing nodes to utilize partial or outdated information, including vertex features or model parameters.
The impact of these techniques on accuracy is discussed in corresponding sections, i.e., Section \ref{sec:dft:preset:tech} and Section \ref{sec:dmt:joint:tech}.
}

\section{Distributed Full-batch Training}\label{sec:dft}

\begin{figure*}[hbtp]
    %\vspace{-15pt}
    \centering
    \includegraphics[page=3, width=0.99\textwidth]{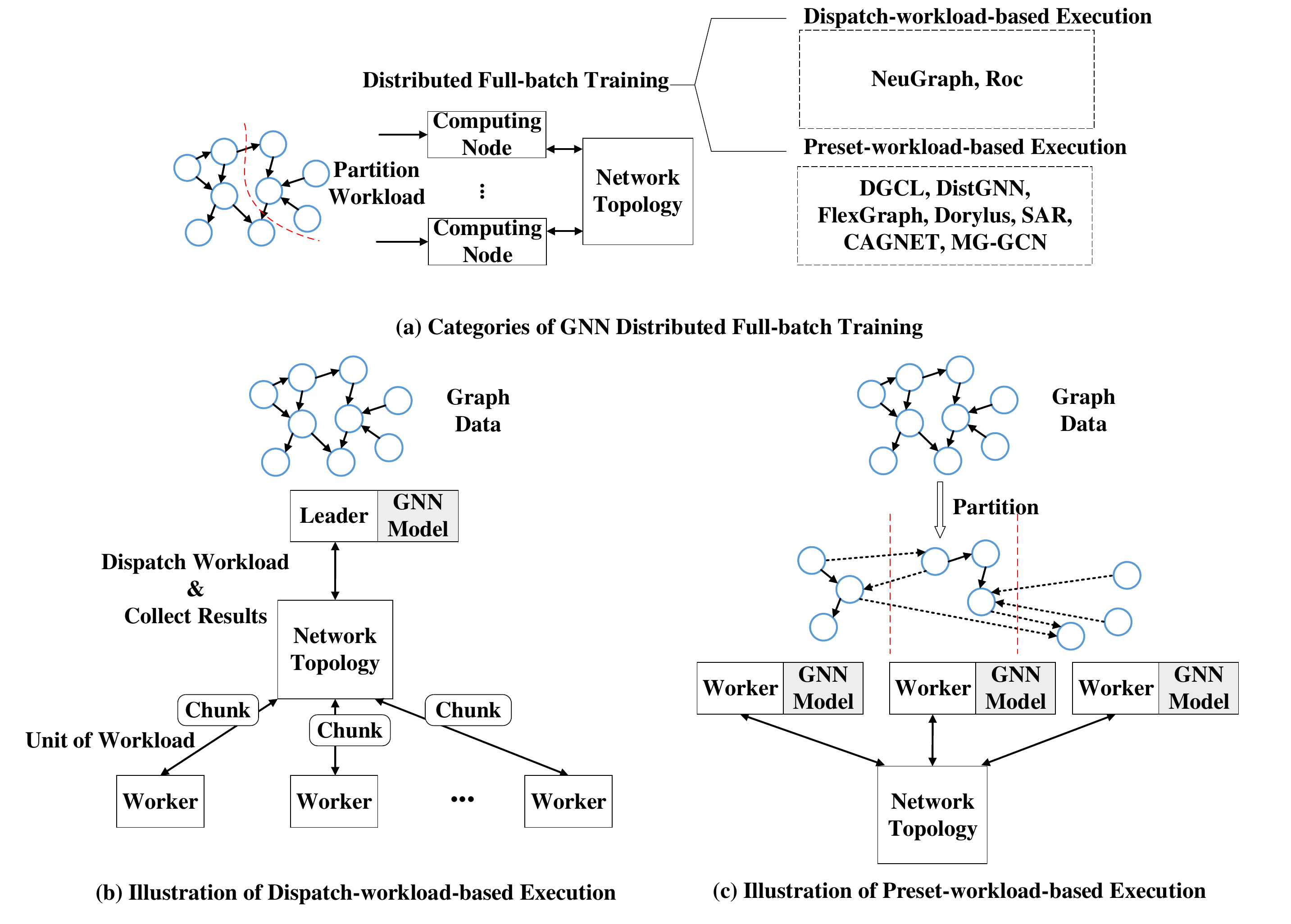}
    %\vspace{-15pt}
    \caption{GNN distributed full-batch training: (a) Categories of GNN distributed full-batch training; (b) Illustration of dispatch-workload-based execution; (c) Illustration of preset-workload-based execution.}
    \label{fig:02_dft}
    %\vspace{-20pt}
\end{figure*}

%In this section, the distributed full-batch training is described in detail.
%Its taxonomy is illustrated in Fig. \ref{fig:02_dft} (a).
%Its categories include \textbf{dispatch-workload-based execution} and \textbf{preset-workload-based execution}, which are introduced in detail as follows.

This section describes GNN distributed full-batch training in detail.
Our taxonomy classifies it into two categories according to whether the workload is preset in the preprocessing phase, namely dispatch-workload-based execution and preset-workload-based execution, as shown in Fig. \ref{fig:02_dft} (a).

% \todo{R1 general}
% \leyc{
% %根据师兄意见：

% %此处需要去 看看新的文章，补充补充。   然后解释给review1  为什么full-batch部分内容少。（因为本身文章就少，现在新增一些文章，补充上）    

% %-----没去看... 10-04
% % 先目前这样，等师兄您意见
% }

\subsection{Dispatch-workload-based Execution}\label{sec:dft:chunk_execution}

The dispatch-workload-based execution of distributed full-batch training is illustrated in Fig. \ref{fig:02_dft} (b). Its workflow, computational pattern, communication pattern, and optimization techniques are introduced in detail as follows.

\subsubsection{Workflow}
%工作流程和工作分配

%The dispatch-workload-based execution of distributed full-batch training is demonstrated in Fig. \ref{fig:02_dft} (b).
%The computational pattern of dispatch-workload-based execution type is demonstrated in Fig. \ref{fig:02_dft} (b).

%--
%In the dispatch-workload-based execution, a leader and multiple workers are used to perform training.
%The leader holds the model parameters and accesses to the graph data, as it is responsible for workload scheduling. 
%It splits the computing workloads into workload chunks and distributes them to the workers.
%It also collects the intermediate results from workers to continue the computation.
%The worker is responsible for completing the computing workloads received from the leader and sending the immediate intermediate results back to the leader. Note that, the \textit{chunk} we use here is as a unit of workload.

In the dispatch-workload-based execution, a leader and multiple workers are used to perform training.
The leader stores the model parameters and the graph data, and is also responsible for scheduling:
it splits the computing workloads into chunks, distributes them to the workers, and collects the intermediate results sent from the workers.
%It also processes these results and advances the computing.
It also processes these results and advances the computation.
Note that, the \textit{chunk} we use here is as a unit of workload.

\subsubsection{Computational Pattern} 
The computational patterns of forward propagation and backward propagation are similar in dispatch-workload-based execution: the latter can be seen as the reversed version of the former.
As a result, we only introduce forward propagation's computational pattern here for simplicity.
The patterns of the two functions in forward propagation (Aggregate and Combine) differ a lot and are introduced below respectively.
% 前面用的是Aggregate和Combine，不是Aggregation和Combination

\textbf{Aggregate function.} The computational pattern of the Aggregate function is dynamic and irregular, making workload partition for this step a challenge.
In the Aggregation step, each vertex needs to aggregate the features of its own neighbors.
As a result, the computation of the Aggregation step relies heavily on the graph structure, which is irregular or even changeable.
Thus, the number and memory location of neighbors vary significantly among vertices, resulting in the dynamic and irregular computational pattern \cite{hpca/hygcn/gnn_asic_acc}, causing the poor workload predictability and aggravating the difficulty of workload partition.

\textbf{Combine function.} The computational pattern of the Combine function is static and regular, thus the workload partition for it is simple.
The computation of the Combination step is to perform neural network operations on each vertex.
Since the structure of neural networks is regular and these operations share the same weight parameters, the Combination step enjoys a regular computational pattern.
Consequently, a simple partitioning method is sufficient to maintain workload balance, so it is relatively not a major consideration in dispatch-workload-based execution of GNN distributed full-batch training.

\subsubsection{Communication Pattern}
%在计算节点间通信上体现的行为，同时描述challenge或者concerns

%\todo{
%这个你不用管
%整体用Program Behavior表示程序的特征，Execution Pattern表示程序在硬件上的行为模式，用Computational Pattern表示处理节点内的行为模型，Communication表示处理节点间的行为模式
%改成communication pattern在前的描述，统一到Computational Pattern这个章节核心上。
%整理之后，把concerns重点描述。逻辑是，behavior-》pattern-》concern-》optimization
%简化版本：execution flow-》concerns/problems/computational pattern-》optimization
%划分其实本质是按照execution flow，而不是computational pattern吧？
%Computational pattern作为需要优化问题的基础描述，引出问题即可。流程是程序自身的行为，与硬件无关。计算pattern，是流程在硬件上的体现
%}

%The \textbf{main communication} is the transmission of input data and output results between the leader and the workers. 
%Since the leader is responsible for workload distribution and intermediate result collection, it needs to communicate with multiple workers.
%Such a communication structure results in a one-to-many communication pattern, which makes the leader's communication path easy to become a bottleneck.
%The good news is that these communications are relatively regular.
%Because the distribution of tasks is controlled by the leader.
%Therefore, the communication path congestion can be avoided as much as possible by reasonably scheduling computing tasks.

The majority of communication is the transmission of input data and output results between the leader and the workers \cite{atc_neugraph}. 
Since the leader is responsible for workload distribution and intermediate result collection, it needs to communicate with multiple workers.
Such a communication structure results in a one-to-many communication pattern, resulting in a possible bottleneck in the leader's communication path \cite{atc_neugraph}.
However, these communications are relatively regular since the distribution of tasks is controlled by the leader.
Therefore, the communication path congestion can be avoided to some extent by optimized scheduling techniques.

%优化的问题
%优化技术的核心
%引用具体工作详细描述优化技术或者优化技术变体

%\todo{
%你先想想怎么定义这两个Transmission Reduction 和Transmission Overhead Reduction。在他们第一次出现的时候，解释一下这两个概念。
%}
%\Revise{revised
%看了下， transmission overhead reduction只出现在几个表格中，就3次。
%我就统一改成。transmission reduction 吧。
%再第一次出现的表格的正文处，加了transmission reduction的解释
%}

\subsubsection{Optimization Techniques}\label{sec:dft:chunk_execution:opti_tech}
Next, we introduce the optimization techniques used to partition workload, balance workload, reduce transmission, and exploit parallelism for dispatch-workload-based execution. We classify them into four categories, including 
\blackcircled{1} Vertex-centric Workload Partition, 
\blackcircled{2} Balanced Workload Generation, 
\blackcircled{3} Transmission Planning, and 
\blackcircled{4} Feature-dimension Workload Partition. A summary for these categories is shown in Table \ref{table:summary_of_chunk_execution}.
%\Revise{Note that, transmission reduction here is used to refer to the reduction in the amount of transmission data or transmission time.}
Note that the transmission reduction in this paper refers to a reduction in the amount of transmitted data or in transmission time.

%\todo{ley
% 4个表 调整到 跟正文字体差不多
%检查下里面的 内容
%}
\begin{table*}[hbtp]
    %\vspace{-12pt}
 \caption{Summary of various categories of optimization techniques used in dispatch-workload-based execution of distributed full-batch training.} \label{table:summary_of_chunk_execution}
 %\vspace{-10pt}
 \centering
\renewcommand\arraystretch{1.5} 
\resizebox{1\textwidth}{!}{
\begin{tabular}{|l|cccc|}
\hline
\textbf{Optimization Technique}     
& \textbf{Function}  &  \textbf{Operation Object} & \textbf{Focus}  & \textbf{Related Work}\\ \hline \hline
\multirow{2}{*}{\begin{tabular}[c]{@{}c@{}}  \blackcircled{1} Vertex-centric Workload Partition   \end{tabular}} 
&\multirow{2}{*}{\begin{tabular}[c]{@{}c@{}}   Workload Partition  \end{tabular}}  
&\multirow{2}{*}{\begin{tabular}[c]{@{}c@{}}   Graph or Matrix \end{tabular}} 
&\multirow{2}{*}{\begin{tabular}[c]{@{}c@{}}   Graph or Matrix \\Partition Strategy \end{tabular}}  
&\multirow{2}{*}{\begin{tabular}[c]{@{}c@{}}   NeuGraph \cite{atc_neugraph}, Roc \cite{mlsys_roc} \end{tabular}}  \\
    &   &   &   &   \\ \hline
\multirow{2}{*}{\begin{tabular}[c]{@{}c@{}}    \blackcircled{2} Balanced Workload Generation \end{tabular}} 
&\multirow{2}{*}{\begin{tabular}[c]{@{}c@{}}   Workload Balance \end{tabular}}  
&\multirow{2}{*}{\begin{tabular}[c]{@{}c@{}}   Partition Operation \end{tabular}} 
&\multirow{2}{*}{\begin{tabular}[c]{@{}c@{}}   Partition Strategy \end{tabular}}  
&\multirow{2}{*}{\begin{tabular}[c]{@{}c@{}}    Roc \cite{mlsys_roc} \end{tabular}}  \\
    &   &   &   &   \\ \hline
\multirow{2}{*}{\begin{tabular}[c]{@{}c@{}}   \blackcircled{3} Transmission Planning  \end{tabular}} 
&\multirow{2}{*}{\begin{tabular}[c]{@{}c@{}}  Transmission Reduction  \end{tabular}}  
&\multirow{2}{*}{\begin{tabular}[c]{@{}c@{}}   Transmission Operation  \end{tabular}} 
&\multirow{2}{*}{\begin{tabular}[c]{@{}c@{}}  Transmission Strategy  \end{tabular}}  
&\multirow{2}{*}{\begin{tabular}[c]{@{}c@{}}  Roc \cite{mlsys_roc}, NeuGraph \cite{atc_neugraph} \end{tabular}}  \\
    &   &   &   &   \\ \hline
\multirow{2}{*}{\begin{tabular}[c]{@{}c@{}}  \blackcircled{4} Feature-dimension Workload Partition   \end{tabular}} 
&\multirow{2}{*}{\begin{tabular}[c]{@{}c@{}}  Parallelism Exploitation   \end{tabular}}  
&\multirow{2}{*}{\begin{tabular}[c]{@{}c@{}}  Vertices' Features  \end{tabular}} 
&\multirow{2}{*}{\begin{tabular}[c]{@{}c@{}}  Feature Partition Strategy  \end{tabular}}  
&\multirow{2}{*}{\begin{tabular}[c]{@{}c@{}}  NeuGraph \cite{atc_neugraph}  \end{tabular}}  \\
    &   &   &   &   \\ \hline

%\blackcircled{1} Vertex-centric Workload Partition   &  Workload Partition   & Graph or Matrix  & Graph or Matrix Partition Strategy       &  NeuGraph \cite{atc_neugraph}, Roc \cite{mlsys_roc}, CAGNET \cite{sc_reducint_communication_in_graph_cagnet}, MG-GCN \cite{arxiv_mg_gcn} \\ \hline
%\blackcircled{2} Balanced Workload Generation     &  Workload Balance         & Partition Operation   &  Partition Strategy    & Roc \cite{mlsys_roc}   \\ \hline   
%\blackcircled{3} Transmission Planning    &  Transmission Reduction   & Transmission Operation   &  Transmission Strategy & Roc \cite{mlsys_roc}, NeuGraph \cite{atc_neugraph}, CAGNET \cite{sc_reducint_communication_in_graph_cagnet}  \\ \hline  
%\blackcircled{4} Feature-dimension Workload Partition       &  Parallelism Exploitation   & Vertices' Feature   &   Feature Partition Strategy     & NeuGraph \cite{atc_neugraph}  \\ \hline  

\end{tabular}
    }
\end{table*}

\textbf{\blackcircled{1} Vertex-centric Workload Partition.}
Vertex-centric workload partition refers to the technique of generating workload chunks by partitioning the graph or matrix from the perspective of vertices. 
Specifically, the graph is partitioned into a list of subgraphs according to the source vertex $u$ and the destination vertex $v$ of edge $(u, v)$. % TODO: 这里是怎么个according法？
Then the subgraphs are taken as the workload chunk, and the leader distributes them to each computing node for computation.
%This is a very common partitioning method for processing large-scale graphs in the traditional graph analytics \cite{micro/graphicionado/graph_processing_asic_acc, micro/alleviating_irregularity_in/yan/graphdyns/graph_processing_asic_acc, pvldb/a_distributed_multi_gpu/roc_graph_partition}.
This is a very common partitioning method for processing large-scale graphs in the traditional graph application~\cite{micro/graphicionado/graph_processing_asic_acc, micro/alleviating_irregularity_in/yan/graphdyns/graph_processing_asic_acc, pvldb/a_distributed_multi_gpu/roc_graph_partition}.
%According to the partition object, vertex-level workload partition can be further classified into graph-based partition and matrix-based partition.

\begin{figure}[hbtp]
    %\vspace{-15pt}
    \centering
    \includegraphics[page=6, width=0.45\textwidth]{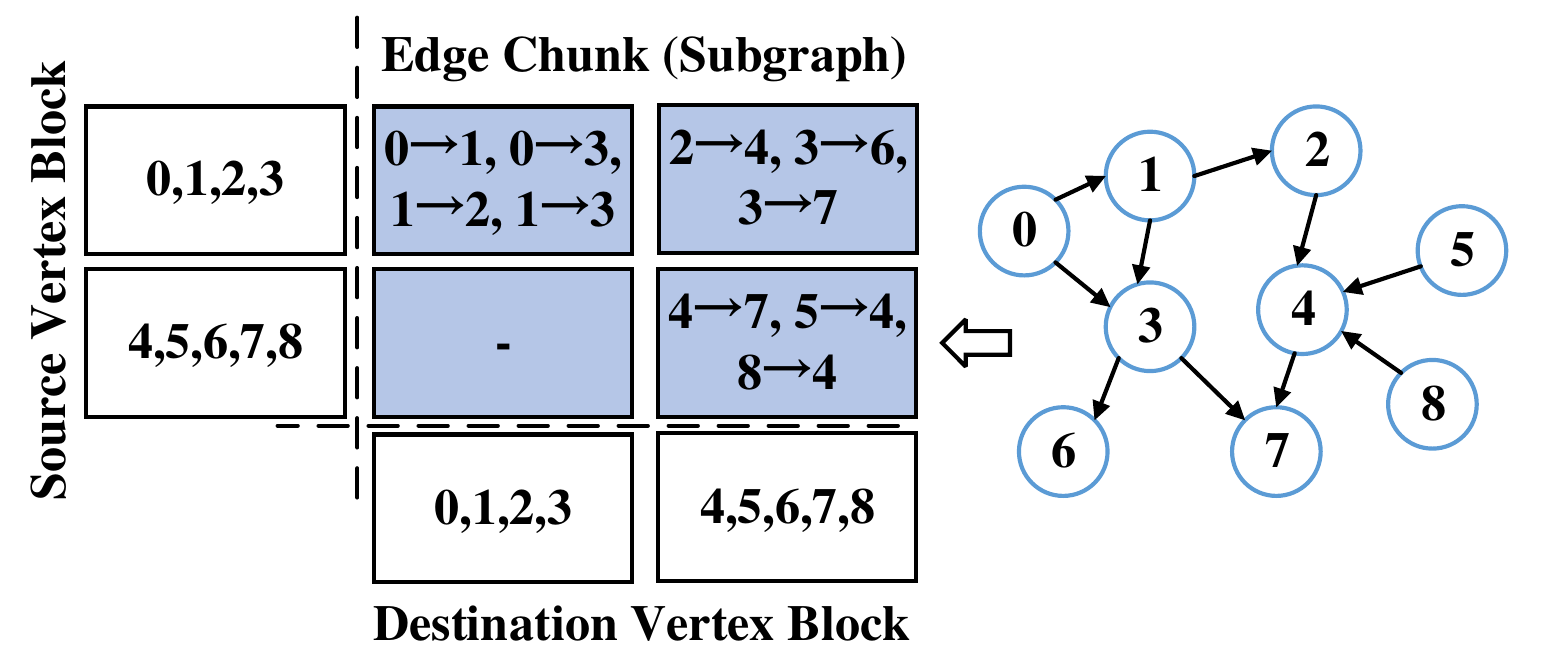}
    %\vspace{-15pt}
    \caption{Illustration of 2D graph partition.}
    \label{fig:05_chunk_vertex_partition}
    %\vspace{-10pt}
\end{figure}

Fig. \ref{fig:05_chunk_vertex_partition} illustrates 2D graph partition, an typical example of graph partition.
The vertices are firstly partitioned into $P$ disjoint blocks.
Then, we tile the edges into $P \times P$ chunks according to their source and destination vertices: in the $y^{\text{th}}$ chunk of the $x^{\text{th}}$ row, the source vertices of the edges all belong to $x^{\text{th}}$ source vertex block, and the destination vertices all belong to $y^{\text{th}}$ destination vertex block.
This partitioning method works well for the Aggregation step, which needs to transfer the information of the source vertex to the destination vertex along the edge.

%NeuGraph \cite{atc_neugraph} adopts a similar 2D graph-based partitioning method. 
NeuGraph \cite{atc_neugraph} adopts the 2D graph partitioning method. 
It chooses $P$ as the minimum integer satisfying the requirement to fit each chunk in the device memory of GPUs.
Roc \cite{mlsys_roc}, on the other hand, uses the graph partition strategy proposed in \cite{pvldb/a_distributed_multi_gpu/roc_graph_partition}, which can also be mentioned as 1D graph partition.
%\todo{R1 Q6}
\Revise{
In this approach, the graph, comprising $T$ vertices, is divided into $n$ subgraphs. To achieve this, $n-1$ numbers, ranging from 1 to $T$, are selected to split the vertices into $n$ parts. Consequently, each subgraph contains consecutively numbered vertices stored in adjacent locations, along with their respective in-edges. This arrangement maximizes coalesced access to device memory during subsequent computations. It's important to note that the 1D graph partitioning approach can be readily applied to directed graphs. For other graph structures, such as undirected graphs, a conversion to a directed graph format is necessary. For instance, in the case of undirected graphs, each edge must be represented as two opposing directed edges to facilitate the application of this strategy.
}
%\Revise{Suppose the graph is divided into $n$ subgraphs and the number of vertices in the graph is $T$. $n$-1 numbers, ranging from 1 to $T$, are selected to split the vertices into $n$ parts, that is, $n$ subgraphs. This approach ensures that each subgraph contains vertices numbered consecutively and stored in adjacent locations, along with their in-edges, maximizing coalesced access to device memory during subsequent computations. The 1D graph partitioning approach can be directly applied to directed graphs. In the case of other graph structures, conversion to a directed graph is required. For example, with undirected graphs, the transformation involves representing each edge as two opposing directed edges.}

\textbf{\blackcircled{2} Balanced Workload Generation.}
Workload balance is an extremely important optimization goal for dispatch-workload-based execution.
Since the workloads are split and distributed to multiple computing nodes, the prerequisite of the continuation of computing is that all the computing nodes have already returned their intermediate results.
If the workload is not evenly partitioned, the consequent lag of waiting will stall the training process.
Therefore, it is necessary to carefully adjust the workload partition, so as to make the workload of each computing node as balanced as possible.

In response to that, Roc \cite{mlsys_roc} proposes a linear regression cost model to produce balanced workloads in each round.
The cost model is used to predict the computation time of a GNN layer on an arbitrary input, which could be the whole or any subset of an input graph.

\textbf{\blackcircled{3} Transmission Planning.}
Planning data transmission is helpful to make full use of reusable data across computing components and nodes, thereby reducing data transmission.
In dispatch-workload-based execution, the major transmission overhead is caused by the requirement to transmit the input data and intermediate results.
Recent work focuses on harvesting two following optimization opportunities to reduce this overhead.

\textit{Avoid repeated transmission of overlapped data.}
The input data required by different computing tasks may overlap.
As a result, caching these overlapped parts on computing nodes or components is a reasonable way to reduce transmission.
Roc \cite{mlsys_roc} formulates GPU memory management as a cost minimization problem. 
It uses a recursive dynamic programming algorithm to find the global optimal solution to decide which part of the data should be cached in GPU memory for reuse, according to the input graph, the GNN model, and device information.
This minimizes data transmission between CPUs and GPUs.

\textit{Rationally select the source of the transmission.}
The overhead of data transmission can be reduced by allowing computing components or nodes to send data, so every component or node can rationally select the source of the transmission.
As a result, the design can reduce the overhead by decreasing the transmission distance.
For instance, NeuGraph \cite{atc_neugraph} employs a chain-based streaming scheduling scheme.
The idea is to have one GPU (which already holds the data chunk) forward the data chunk to its neighbor GPU under the same PCIe switch, which can eliminate the bandwidth contention on the upper-level shared inter-connection link.

\textbf{\blackcircled{4} Feature-dimension Workload Partition.}
Feature-dimension partition refers to the finer partition of the workload from the dimension of the vertex feature, to make full use of the parallel computing hardware in the computing node.
%\todo{R1 Q7}
\Revise{
The scales of vertices and edges in graphs are approaching or even outnumbering the order of billions and trillions, respectively \cite{nips_open_graph_benchmark_ogbdataset}. %Importantly, both vertices and edges can be processed independently. Consequently, in terms of parallelism in traditional graph applications, especially graph analytics, the primary focus is on vertex and edge parallel computation~\cite{micro/alleviating_irregularity_in/yan/graphdyns/graph_processing_asic_acc}.
Importantly, both vertices and edges can be processed independently. Consequently, in terms of parallelism in traditional graph applications, especially graph traversal (e.g. Single-Source Shortest Path), the primary focus is on vertex and edge parallel computation~\cite{micro/alleviating_irregularity_in/yan/graphdyns/graph_processing_asic_acc}.
However, GNNs stand apart from traditional graph applications due to their representation of vertex features as extensive vectors or even large tensors, rather than mere scalars. This distinction provides an opportunity for more nuanced parallel computation, specifically focusing on the dimensions of vertex features.

Modern computing hardware has advanced to facilitate large-scale parallel computations, including vector operations, through the implementation of hardware architectures that support single instruction multiple data (SIMD) parallel execution. For instance, in GPUs, each set of 32 consecutive threads forms a warp~\cite{nvidia_nd}. 
Threads within a single warp operate in parallel, following the SIMD parallel execution, ensuring efficient utilization of the underlying computing units~\cite{DBLP:conf/hpdc/FuJH22/tlpgnn}. Considering that vertex features in GNNs are high-dimensional vectors, often reaching dimensions of 1024 or more \cite{/cal22/linhaiyang/characterizing_and_gnn/, atc_neugraph}, it is intuitive to harness the parallel capabilities of the hardware. This can be achieved by leveraging parallelism at the feature-dimension level, distributing the computation of each element of the vertex features across threads.

To capitalize on this opportunity, NeuGraph \cite{atc_neugraph} adopts a clever strategy where the computations of each vertex are subdivided according to its feature dimensions. This approach effectively distributes the computations across different threads within a GPU warp. These threads duplicate neighboring vertex information, facilitating coarse-grained access to neighboring features while concurrently processing the vertex's computations at the feature-dimension level. This strategy ensures efficient utilization of the substantial parallel computing resources offered by GPUs, helping to achieve significant improvement in performance.

}

\subsection{Preset-workload-based Execution}\label{sec:dft:partition_execution}

The preset-workload-based execution of distributed full-batch training is illustrated in Fig. \ref{fig:02_dft} (c).
Its workflow, computational pattern, communication pattern, and optimization techniques are introduced in detail as follows.

\subsubsection{Workflow}

Preset-workload-based execution involves multiple collaborative workers to perform training.
The graph is firstly split into subgraphs through the partition operation in the preprocessing phase.
Then, each worker holds one subgraph and a replica of the model parameters.
During training, each worker is responsible to complete the computing tasks of all the vertices in its subgraph.
As a result, a worker needs to query the information from other workers to gather the information of neighboring vertices.
However, the Combine function can be performed directly locally, as the model parameters are replicated, but this also means that gradient synchronization is required for each round to ensure the consistency of model parameters across the nodes.

%优势：  把数据都存memory

%劣势：  通讯
%由于这样的特点，导致其主要因素就在通讯
%对此有两个方向   1 划分图   2 减少，overlap 开销

\subsubsection{Computational Pattern}
%在计算节点内计算上体现的行为，同时描述challenge

The computational pattern of preset-workload-based execution differs significantly in different steps.
In the Aggregation step, a node needs to query the neighbor information of vertices from other nodes, so only when each computing node cooperates efficiently can the data be supplied to the target computing node in a timely manner.
However, in the Combination step, each computing node conducts the operation on the vertices in its own subgraph independently due to the local replica of model parameters.
As a result, the Aggregation step is more prone to inefficiencies such as computational stagnation, and it is the key optimization point for preset-workload-based execution.

In addition, the preset workload has the benefit that the whole graph can be loaded at a time, as it is partitioned into small subgraphs which fit in the memory of computing nodes in the preprocessing phase.
This means that it is more scalable than dispatch-workload-based execution when either adding more computing resources or increasing the size of the dataset.
This avoids frequent accesses to graph data from low-speed storage such as hard disks, thereby ensuring the timely provision of data for high-speed computing.

\subsubsection{Communication Pattern}
Communication happens mostly in the Aggregation step and gradient synchronization.
Due to the distributed storage of graph data, it encounters a lot of irregular transmissions during the Aggregation step when collecting the features of its neighboring vertices that are stored in other nodes due to graph partition \cite{eurosys_dgcl}.
Since the features of vertices are vectors or even tensors, the amount of data transmitted in the Aggregation step is large.
Also, the communication is irregular due to the irregularity graph structure \cite{hpca/hygcn/gnn_asic_acc,micro/alleviating_irregularity_in/yan/graphdyns/graph_processing_asic_acc,Tigr}, which brings difficulty in optimizing connection.
In contrast, the communication overhead of gradient synchronization is minuscule due to the small size of model parameters and the regular communication pattern.
As a result, the communication between nodes in the Aggregation step is the main concern of preset-workload-based execution in distributed full-batch training.

%说明为了优化xxx问题，主要有哪些优化方法
%\textit{XXX Optimization Technique.}
%优化的问题
%优化技术的核心
%引用具体工作详细描述优化技术或者优化技术变体
%\subsubsection{Optimization Techniques}
%Next, we introduce the optimization techniques used to \textbf{balance workload, reduce communication, and reduce memory pressure} for preset-workload-based execution of distributed full-batch training in detail.
%We categorize them into four categories, including \blackcircled{1} Graph Pre-partition, \blackcircled{2} Communication Optimization, \blackcircled{3} Delayed Aggregation, and \blackcircled{4} Activation Rematerialization. A summary for these categories is shown in Table \ref{table:summary_of_partition_execution}.

\subsubsection{Optimization Techniques}\label{sec:dft:preset:tech}

Here we introduce the optimization techniques used to balance workload, reduce transmission, and reduce memory pressure for preset-workload-based execution in detail.
We classify them into four categories: \blackcircled{1} Graph Pre-partition, \blackcircled{2} Transmission Optimization, \blackcircled{3} Delayed Aggregation, and \blackcircled{4} Activation Rematerialization.
A summary for these categories is shown in Table \ref{table:summary_of_partition_execution}.

\begin{table*}[hbtp]
    %\vspace{-12pt}
 \caption{Summary of various categories of optimization techniques used in preset-workload-based execution of distributed full-batch training.} \label{table:summary_of_partition_execution}
 %\vspace{-10pt}
 \centering
\renewcommand\arraystretch{1.5} 
\resizebox{1\textwidth}{!}{
\begin{tabular}{|l|cccc|}
\hline
\textbf{Optimization Technique}     & \textbf{Function}  & \textbf{Operation Object} & \textbf{Focus} & \textbf{Related Work}\\ \hline \hline
\multirow{2}{*}{\begin{tabular}[c]{@{}c@{}}\blackcircled{1} Graph Pre-partition\end{tabular}} &\multirow{2}{*}{\begin{tabular}[c]{@{}c@{}}Workload Balance and \\Transmission Reduction\end{tabular}}  &\multirow{2}{*}{\begin{tabular}[c]{@{}c@{}}Entire Graph\end{tabular}} &\multirow{2}{*}{\begin{tabular}[c]{@{}c@{}}Graph Partition Strategy\end{tabular}}  &\multirow{2}{*}{\begin{tabular}[c]{@{}c@{}}DGCL \cite{eurosys_dgcl}, Dorylus \cite{osdi_dorylus}, \\FlexGraph \cite{eurosys_flexgraph}, DistGNN \cite{sc_distgnn}\end{tabular}}  \\
    &   &   &   &   \\ \hline
\multirow{2}{*}{\begin{tabular}[c]{@{}c@{}}\blackcircled{2} Transmission Optimization\end{tabular}} &\multirow{2}{*}{\begin{tabular}[c]{@{}c@{}}Transmission Reduction\end{tabular}}  &\multirow{2}{*}{\begin{tabular}[c]{@{}c@{}}Transmission Data\end{tabular}}  &\multirow{2}{*}{\begin{tabular}[c]{@{}c@{}}Transmission Strategy\end{tabular}} &\multirow{2}{*}{\begin{tabular}[c]{@{}c@{}} DGCL \cite{eurosys_dgcl},  FlexGraph \cite{eurosys_flexgraph}\end{tabular}}  \\
    &   &   &   &   \\ \hline
\multirow{2}{*}{\begin{tabular}[c]{@{}c@{}}\blackcircled{3} Delayed Aggregation\end{tabular}} &\multirow{2}{*}{\begin{tabular}[c]{@{}c@{}}Transmission Reduction\end{tabular}}  &\multirow{2}{*}{\begin{tabular}[c]{@{}c@{}}Aggregation Operation\end{tabular}} &\multirow{2}{*}{\begin{tabular}[c]{@{}c@{}}Aggregation Strategy\end{tabular}}  &\multirow{2}{*}{\begin{tabular}[c]{@{}c@{}}Dorylus \cite{osdi_dorylus}, DistGNN \cite{sc_distgnn}\end{tabular}}  \\
    &   &   &   &   \\ \hline
\multirow{2}{*}{\begin{tabular}[c]{@{}c@{}}\blackcircled{4} Activation Rematerialization\end{tabular}} &\multirow{2}{*}{\begin{tabular}[c]{@{}c@{}}Memory Pressure Reduction\end{tabular}}  &\multirow{2}{*}{\begin{tabular}[c]{@{}c@{}}Intermediate Results\end{tabular}} &\multirow{2}{*}{\begin{tabular}[c]{@{}c@{}}Intermediate Results \\Retrieve Strategy\end{tabular}}  &\multirow{2}{*}{\begin{tabular}[c]{@{}c@{}}SAR \cite{arxiv_sar}\end{tabular}}  \\
    &   &   &   &   \\ \hline

%\blackcircled{1} Graph Pre-partition        &  Workload Balance and Transmission Reduction     & Entire Graph    & Graph Partition Strategy    & DGCL \cite{eurosys_dgcl}, Dorylus \cite{osdi_dorylus}, FlexGraph \cite{eurosys_flexgraph}, DistGNN \cite{sc_distgnn} \\ \hline
%\blackcircled{2} Communication Optimization  &  Transmission Reduction &  Transmission Data & Transmission Strategy      & DGCL \cite{eurosys_dgcl},  FlexGraph \cite{eurosys_flexgraph}  \\ \hline 
%\blackcircled{3} Delayed Aggregation     & Transmission Reduction  & Aggregation Operation & Aggregation Strategy        & Dorylus \cite{osdi_dorylus}, DistGNN \cite{sc_distgnn}  \\ \hline   
%\blackcircled{4} Activation Rematerialization    & Memory Pressure Reduction  & Intermediate Results  & Intermediate Results Retrieve Strategy         & SAR \cite{arxiv_sar}  \\ \hline   
\end{tabular}
    }
\end{table*}

%\textbf{\multirow{2}{*}{\begin{tabular}[c]{@{}c@{}}Distributed\\Mini-batch Training\end{tabular}}}

\textbf{\blackcircled{1} Graph Pre-partition.}
Graph pre-partition refers to partitioning the whole graph into several subgraphs according to the number of computing nodes, mainly to balance workload and reduce transmission \cite{eurosys_dgcl}. This operation is conducted in the preprocessing phase.
The two key principles in designing the partitioning algorithm are listed as follows.

First, in order to pursue workload balance, the subgraphs need to be similar in size.
In preset-workload-based execution, each worker performs the computation of vertices within its own subgraph.
Therefore, the size of the subgraph determines the workload of the worker.
The main reference parameters are the number of vertices and edges in the subgraph.
Since preset-workload-based execution has a gradient synchronization barrier, workload balance is very important for computing nodes to complete a round of computation at a similar time.
Otherwise, some computing nodes will be idle, causing performance loss.

Second, minimizing the number of edge-cuts in the graph pre-partition can reduce communication overhead.
It is inevitable to cut edges in the graph pre-partition, meaning that the source and destination vertices of an edge might be stored in different computing nodes.
When the information of neighboring vertex is required during the computation of the Aggregation step, communication between workers is introduced.
Therefore, reducing the number of edges cut can reduce communication overhead in the Aggregation step.

DGCL \cite{eurosys_dgcl} uses METIS library \cite{metis/dgcl_partition_algorithm} to partition the graph for both the above two targets.
Dorylus \cite{osdi_dorylus} also uses an edge-cut algorithm \cite{osdi/gemini/dorylus_graph_partition} for workload balance.
DistGNN \cite{sc_distgnn} aims at the two targets too. 
However, it uses a vertex-cut based graph partition technique instead.
This means distributing the edges among the partitions. 
Thus, each edge exists in only one partition, while a vertex can reside in multiple partitions. 
Any updates to such vertex must be synchronized to its replicas in other partitions.
FlexGraph \cite{eurosys_flexgraph} partitions the graph in the manner of edge-cut too.
Besides, it learns a cost function to estimate the training cost for the given GNN model.
Using the estimated training cost, FlexGraph migrates the workload from overloaded partitions to underloaded ones to pursue workload balance.
%{\color{red}add p3 in all it need} 
%P3 

\textbf{\blackcircled{2} Transmission Optimization.}
Transmission optimization refers to adjusting the transmission strategy between computing nodes to reduce communication overhead. Due to the irregular nature of communication in preset-workload-based execution, the demand for transmission optimization is even stronger than above.

DGCL \cite{eurosys_dgcl} provides a general and efficient communication library for distributed GNN training.
It tailors the shortest path spanning tree algorithm to transmission planning, which jointly considers fully utilizing fast links, avoiding contention, and balancing workloads on different links.

Compared to traditional transmission planning, FlexGraph \cite{eurosys_flexgraph} takes a different approach to take advantage of the aggregation nature of the Aggregate function.
FlexGraph \cite{eurosys_flexgraph} partially aggregates the features of neighboring vertices that co-locate at the same partition when possible, aiming to reduce the amount of data transmission and overlap partial aggregations and communications.
When each computing node receives a neighbor information request from other nodes, it first partially aggregates the neighbor information locally, and then sends the partial result to the requesting computing node, instead of directly sending the initial neighbor information.
The requesting computing node only needs to aggregate the received partial result with its local nodes to continue the computation. % TODO: Check correctness?
As a result, due to the reduction in data transmission and the overlap of computation and communication, the communication overhead is significantly reduced.

\textbf{\blackcircled{3} Delayed Aggregation.}
By allowing the computing nodes to utilize old transmitted data in the Aggregation step, delayed aggregation can reduce the overhead of data transmission.
Normally, there is no scope for intra-epoch overlap due to the dependence between consecutive phases in an epoch.
Delayed aggregation solves the problem to overlap the computation and communication, allowing the model to use the previously transmitted data.
However, we need to point out that in order to ensure convergence and final accuracy, the delayed aggregation is mainly based on bounded asynchronous training \cite{osdi_dorylus}.

Dorylus \cite{osdi_dorylus} uses bounded asynchronous training on two synchronization points: the update of weight parameters in the backward propagation, and the neighbor aggregation in the Aggregation step.
Bounded asynchronous training of GNNs is based on bounded staleness \cite{atc/exploiting_bounded_staleness/bounded_staleness/1, /nips/hogwild/bounded_staleness/2, icdcs/dynamic_stale_synchronous/bounded_staleness/3, /nips/more_effective_distributed/bounded_staleness/4}, an effective technique for mitigating the convergence problem by employing lightweight synchronization.
The key policy is to restrict the number of iterations between the fastest worker and the slowest worker to not exceed more than a user-specified staleness threshold $s$, where $s$ is a natural number.
As long as the policy is not violated, there is no waiting time among workers.

DistGNN \cite{sc_distgnn} proposes the delayed remote partial aggregates (DRPA) algorithm to overlap the communication with local computation, which is inter-epoch computation-communication overlap.
In the algorithm, the set of vertices that may be queried by other computing nodes is partitioned into $r$ subsets.
For each epoch computation, only the data of one subset is transmitted.
The transmitted data is not required to be received at this epoch, but after $r$ epochs.
This means that the computing nodes do not use the latest global data of vertices, but locally existing data of them.
This algorithm allows communication to overlap with more computational processes, thereby reducing communication overhead.

%\todo{R2 Q1}
\Revise{
Given the efforts described above involving the trade-off between model accuracy and computational efficiency, a detailed analysis of this trade-off is presented below to provide a deeper understanding of these endeavors. 
In the case of Dorylus~\cite{osdi_dorylus}, the use of delayed aggregation demands more epochs to achieve the same level of convergence accuracy. This variation depends on the asynchronous parameter settings, resulting in an increase in the number of convergence epochs by 8\% and 41\% for within-epoch and within-two-epochs delayed aggregation, respectively. The great improvement in computational efficiency stems from the significant reduction in time per epoch due to the asynchronous execution.
%, thereby preserving computational efficiency. 
Regarding DistGNN \cite{sc_distgnn}, under the same number of epochs, the delayed aggregation method manages to maintain the final accuracy decline within 1\% compared to the original synchronous training method.
Hence, it is crucial to recognize that while delayed aggregation methods lead to performance enhancements, they require additional training epochs to compensate for the accuracy loss. 
Therefore, maintaining precise control over hyperparameters in asynchronous training, such as the length of asynchronous strides, is essential to strike a balance between accuracy and computational efficiency.
}

\textbf{\blackcircled{4} Activation Rematerialization.}
Activation rematerialization uses data retransmission and recomputation during the computation process to reduce the memory pressure of computing nodes caused by intermediate results or data.

In preset-workload-based execution, the graph data is stored in each worker in a distributed manner: if there are $n$ workers, then each worker only needs to store $1/n$ of raw data initially.
However, during the computation, each worker needs to store the information received from other workers.
In forward propagation, each worker needs to query other workers to obtain the neighbor information of its local vertices, and the information needs to be stored for later use in backward propagation.
As a result, the actual data stored by each worker during the computation is much larger than its initial size.
In addition, its size is difficult to estimate and may lead to memory overflow problems.

Activation rematerialization, a widely-applied and mature technology in DNN, solves the problem by storing all activations during forward propagation. Activation is the output of each layer of neural networks, which means the representation of each vertex of each layer in GNNs.
Its idea is to recompute or load the activations directly from disks during the backward propagation to reduce the pressure of memory \cite{/arxiv/training_deep_nets/rematerialization/1, /mlsys/checkmate_breaking_the_memory/rematerialization/2}.

%\todo{R1 Q8}

\Revise{The sequential aggregation and rematerialization (SAR) method, as proposed in SAR \cite{arxiv_sar}, builds upon the concept of ``activation rematerialization'' and introduces sequential aggregation and rematerialization for distributed GNN training.}
The specific execution flow is as follows.
In forward propagation, each computing node only receives activation from one other computing node at a time.
After the aggregation operation is completed, the activation is removed immediately.
Then the computing node receives the activation from the next computing node and continues the aggregation operation.
This makes the activation of each vertex only exist in the computing node where it is located, and there is no replicas.
In backward propagation, the computation is also performed sequentially as above.
Each computing node transmits activation sequentially to complete the computation.
Through this method, memory will not overflow as long as the memory capacity of the computing node is larger than the size of two subgraphs.
This allows SAR to scale to arbitrarily large graphs by simply adding more workers.

\section{Distributed Mini-batch Training}\label{sec:dmt}

\begin{figure*}[hbtp]
    %\vspace{-15pt}
    \centering
    \includegraphics[page=4, width=0.99\textwidth]{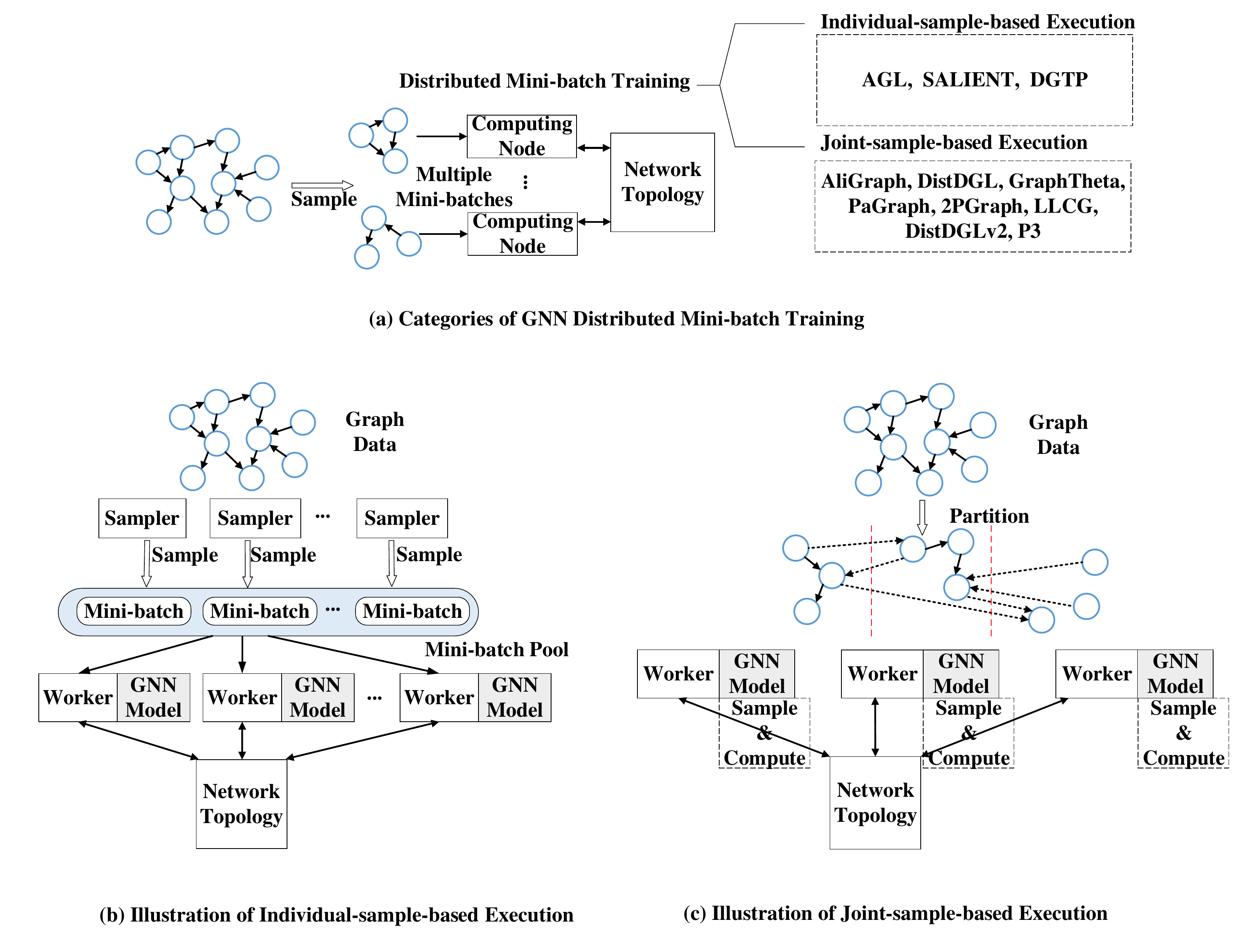}
    %\vspace{-15pt}
    \caption{GNN distributed mini-batch training: (a) Categories of GNN distributed mini-batch training; (b) Illustration of individual-sample-based execution; (c) Illustration of joint-sample-based execution.}
    \label{fig:03_dmt}
    %\vspace{-20pt}
\end{figure*}

This section describes GNN distributed mini-batch training in detail.
Our taxonomy classifies it into two categories according to whether the sampling and
model computation are decoupled, namely individual-sample-based execution and joint-sample-based execution, as illustrated in Fig. \ref{fig:03_dmt} (a).

\subsection{Individual-sample-based Execution} \label{sec:dmt:sample_individual_execution}

The individual-sample-based execution of distributed mini-batch training is illustrated in Fig. \ref{fig:03_dmt} (b).
Its workflow, computational pattern, communication pattern, and optimization techniques are introduced in detail as follows.

\subsubsection{Workflow}
%工作流程和工作分配

%In the individual-sample-based execution of distributed mini-batch training, multiple samplers and workers are used to perform training. \textbf{The characteristic of individual-sample-based execution is that it decouples sampling from the model computation.}
%The sampler first samples the graph data to generate mini-batch, and then send the generated mini-batch to the workers.
%The worker performs the computation of mini-batch and conducts gradient synchronization with other workers to update the model parameters.
%By providing enough computing resources for the samplers to prepare mini-batches for the workers, the computation can be performed without stalls.

The individual-sample-based execution involves multiple samplers and workers so that it decouples sampling from the model computation.
The sampler first samples the graph data to generate a mini-batch, and then sends the generated mini-batch to the workers.
The worker performs the computation of the mini-batch and conducts gradient synchronization with other workers to update the model parameters.
By providing enough computing resources for the samplers to prepare mini-batches for the workers, the computation can be performed without stalls.

A more detailed description of the workflow follows.
First, the samplers generate mini-batches by querying the graph structure.
%This can be done online or offline.
Since each worker requires one mini-batch per round of training, the samplers need to generate enough mini-batches in time.
These mini-batches are transferred to the workers for subsequent computations.
The workers perform forward propagation and backward propagation on their own received mini-batch and generate gradients.
After that, gradient synchronization is conducted between workers to update model parameters.

\subsubsection{Computational Pattern}
%在计算节点内计算上体现的行为，同时描述challenge

The computational pattern in the individual-sample-based execution is dominated by the computation of the sampler and the workers, which are responsible for the sampling phase and other computational phases respectively.
The sampling phase relies on the graph structure, which leads to irregularities in the computation.
These irregularities are reflected in the uncertainty of the data, including its amount and storage address, making it difficult to estimate the computational efficiency of the samplers, as well as the efficiency of mini-batch generation.
After that, the sampler sends the mini-batches to the workers, which perform model computation on the batches.
In contrast to sampling, the computational efficiency of workers is easy to estimate, since the amount of data and computation between mini-batches are similar.
In addition, since the size of the GNN model is small, the gradient synchronization overhead is also small, which makes it generally not a bottleneck \cite{ia3_distdgl, socc_pagraph}.
As a result, the difference in the computational pattern between the samplers and workers makes the generation of mini-batch and the balance of consumption the focus of attention.
The key optimization point is to accelerate the sampling process, so that it provides enough mini-batches in time to avoid stalls.

\subsubsection{Communication Pattern}
%在计算节点间通信上体现的行为，同时描述challenge或者concerns

The main communication of individual-sample-based execution is the mini-batch transmission between samplers and workers, which is characterized by regular but frequent.
%\Dang{which is regular but frequent}.
Since the mini-batch consists of a fixed number of target vertices and their limited neighbors, the amount of data is consistent and small.
In addition, its transmission target is determined.
This makes the transmission regular.
Due to the small amount of data in the mini-batch, the amount of computation is also small and it is easy to estimate the time cost required for computation.
Due to the continuous consumption of mini-batches by workers, frequent mini-batch transfers are required for maintaining a timely supply of mini-batches.

\subsubsection{Optimization Techniques}
%优化的问题
%优化技术的核心
%引用具体工作详细描述优化技术或者优化技术变体

%Here, we introduce the optimization techniques used to parallelize mini-batch generation, balance workload, and reduce transmission overhead for the individual-sample-based execution in detail.
%We categorize them into four categories, including 
%\blackcircled{1} Parallel Mini-batch Generation, \blackcircled{2} Dynamic Mini-batch Allocation, \blackcircled{3} Mini-batch Transmission Pipelining, and \blackcircled{4} Asynchronous Mini-batch Training.
%A summary for these categories is shown in Table \ref{table:summary_of_sample_individual_execution}.

%Here we introduce the optimization techniques used to generate mini-batch, balance workload, reduce transmission, and reduce waiting time for individual-sample-based execution in detail.
%We classify them into four categories: \blackcircled{1} Parallel Mini-batch Generation, \blackcircled{2} Dynamic Mini-batch Allocation, \blackcircled{3} Mini-batch Transmission Pipelining, and \blackcircled{4} Asynchronous Mini-batch Training.
Here we introduce the optimization techniques used to generate mini-batch, balance workload, reduce transmission, and exploit parallelism for individual-sample-based execution in detail.
We classify them into four categories: \blackcircled{1} Parallel Mini-batch Generation, \blackcircled{2} Dynamic Mini-batch Allocation, \blackcircled{3} Mini-batch Transmission Pipelining, and \blackcircled{4} Parallel Aggregation with Edge Partitioning.
A summary for these categories is shown in Table \ref{table:summary_of_sample_individual_execution}.

\begin{table*}[hbtp]
    %\vspace{-12pt}
 \caption{Summary of various categories of optimization techniques used in individual-sample-based execution of distributed mini-batch training.} \label{table:summary_of_sample_individual_execution}
 %\vspace{-10pt}
 \centering
\renewcommand\arraystretch{1.5} 
\resizebox{1\textwidth}{!}{
\begin{tabular}{|l|cccc|}
\hline
\textbf{Optimization Technique}     
& \textbf{Function} & \textbf{Operation Object}  & \textbf{Focus}  & \textbf{Related Work}\\ \hline \hline

\multirow{2}{*}{\begin{tabular}[c]{@{}c@{}}\blackcircled{1} Parallel Mini-batch Generation\end{tabular}} 
&\multirow{2}{*}{\begin{tabular}[c]{@{}c@{}}  Mini-batch Generation   \end{tabular}}  
&\multirow{2}{*}{\begin{tabular}[c]{@{}c@{}}  Generation Operation  \end{tabular}} 
&\multirow{2}{*}{\begin{tabular}[c]{@{}c@{}}  Generation Parallel Strategy  \end{tabular}}  
&\multirow{2}{*}{\begin{tabular}[c]{@{}c@{}}  SALIENT \cite{arxiv_salient}, \\AGL \cite{pvldb_agl}  \end{tabular}}  \\
    &   &   &   &   \\ \hline
\multirow{2}{*}{\begin{tabular}[c]{@{}c@{}}   \blackcircled{2} Dynamic Mini-batch Allocation  \end{tabular}} 
&\multirow{2}{*}{\begin{tabular}[c]{@{}c@{}}   Workload Balance  \end{tabular}}  
&\multirow{2}{*}{\begin{tabular}[c]{@{}c@{}}   Mini-batch  \end{tabular}} 
&\multirow{2}{*}{\begin{tabular}[c]{@{}c@{}}   Mini-batch Allocation Strategy  \end{tabular}}  
&\multirow{2}{*}{\begin{tabular}[c]{@{}c@{}}   SALIENT \cite{arxiv_salient}, \\AGL \cite{pvldb_agl}  \end{tabular}}  \\
    &   &   &   &   \\ \hline
\multirow{2}{*}{\begin{tabular}[c]{@{}c@{}}   \blackcircled{3} Mini-batch Transmission Pipelining  \end{tabular}} 
&\multirow{2}{*}{\begin{tabular}[c]{@{}c@{}}  Transmission Reduction  \end{tabular}}  
&\multirow{2}{*}{\begin{tabular}[c]{@{}c@{}}  Transmission Operation  \end{tabular}} 
&\multirow{2}{*}{\begin{tabular}[c]{@{}c@{}}  Transmission Strategy  \end{tabular}}  
&\multirow{2}{*}{\begin{tabular}[c]{@{}c@{}}  SALIENT \cite{arxiv_salient}, \\AGL \cite{pvldb_agl}  \end{tabular}}  \\
    &   &   &   &   \\ \hline
%\multirow{2}{*}{\begin{tabular}[c]{@{}c@{}}   \blackcircled{4} Asynchronous Mini-batch Training  \end{tabular}} 
%&\multirow{2}{*}{\begin{tabular}[c]{@{}c@{}}  Waiting Time Reduction  \end{tabular}}  
%&\multirow{2}{*}{\begin{tabular}[c]{@{}c@{}}  Model Parameters  \end{tabular}} 
%&\multirow{2}{*}{\begin{tabular}[c]{@{}c@{}}  Model Parameters \\Update Strategy  \end{tabular}}  
%&\multirow{2}{*}{\begin{tabular}[c]{@{}c@{}}  AGL \cite{pvldb_agl}  \end{tabular}}  \\
%    &   &   &   &   \\ \hline
\multirow{2}{*}{\begin{tabular}[c]{@{}c@{}}   \blackcircled{4} Parallel Aggregation with Edge Partitioning  \end{tabular}} 
&\multirow{2}{*}{\begin{tabular}[c]{@{}c@{}}  Parallelism Exploitation  \end{tabular}}  
&\multirow{2}{*}{\begin{tabular}[c]{@{}c@{}}  Aggregation Operation  \end{tabular}} 
&\multirow{2}{*}{\begin{tabular}[c]{@{}c@{}}  Computation Strategy  \end{tabular}}  
&\multirow{2}{*}{\begin{tabular}[c]{@{}c@{}}  AGL \cite{pvldb_agl}  \end{tabular}}  \\
    &   &   &   &   \\ \hline

%\blackcircled{1} Parallel Mini-batch Generation       & Mini-batch Generation & Generation Operation & Generation Parallel Strategy         & SALIENT \cite{arxiv_salient}, AGL \cite{pvldb_agl}   \\ \hline
%\blackcircled{2} Dynamic Mini-batch Allocation       &  Workload Balance  & Mini-batch & Mini-batch Allocation Strategy          & SALIENT \cite{arxiv_salient}, AGL \cite{pvldb_agl}  \\ \hline  
%\blackcircled{3} Mini-batch Transmission Pipelining   &  Transmission Reduction & Transmission Data  & Transmission Strategy      & SALIENT \cite{arxiv_salient}, AGL \cite{pvldb_agl}  \\ \hline   
%\blackcircled{4} Asynchronous Mini-batch Training  & Waiting Time Reduction  & Model Parameters & Model Parameters Update Strategy        & AGL \cite{pvldb_agl} \\ \hline   
\end{tabular}
    }
\end{table*}

\textbf{\blackcircled{1} Parallel Mini-batch Generation.}
Parallel mini-batch Generation refers to parallelizing the generation of mini-batches to reduce the waiting time of the workers for mini-batches. It aims to speed up the sampling to provide the workers with sufficient mini-batches in time.
As mentioned before, the computational pattern of the sampling is irregular, which is hard to be efficiently performed by GPUs.
Therefore, CPUs are generally used to perform sampling \cite{cluster_2pgraph, socc_pagraph}.
%The wide-used parallel mini-batch generation is to use the multi-thread technology of CPUs to parallelize the generation of mini-batches. 
A widely-used solution is to parallelize the generation of mini-batches by using CPUs' multi-thread design.

SALIENT \cite{arxiv_salient} parallelizes mini-batch generation by using the multi-thread technology of CPUs.
It uses C++ threads to end-to-end implement mini-batch generation rather than Python threads to avoid Python's global interpreter lock, and thus improves the performance of mini-batch generation.

AGL \cite{pvldb_agl}, on the other hand, generates $k$-hop neighborhoods of vertices through preprocessing to simplify and speed up the generation of mini-batches.
Here, the $k$-hop neighborhoods are the vertices that can be visited within the maximum $k$ edges by the given vertex.
AGL proposes a distributed pipeline to generate $k$-hop neighborhoods of each vertex based on message passing, and implements it with MapReduce infrastructure \cite{/mapreduce/}.
The generated $k$-hop neighborhoods information is stored in the distributed file system.
Since it is only necessary to collect the $k$-hop neighborhoods information of the target vertices when a mini-batch is required, such a design greatly accelerates the sampling process.
%In order to limit the size of $k$-hop neighborhoods, AGL proposes a novel sampling method to selectively collect neighbors.
In order to limit the size of $k$-hop neighborhoods, AGL implements a set of sampling strategies to selectively collect neighbors.

%Dynamic mini-batch allocation represents that mini-batches are dynamically allocated to the workers to ensure that each computing node has \textbf{mini-batch supply in time and alleviate workload imbalance}.
%Usually, a well-design allocator is used to collect the mini-batch generated by the sampler and then supply the mini-batch in time according to the needs of the workers.
%In this way, the performance of the sampling is determined only by the efficiency of all samplers, not the slowest sampler.
%This can effectively avoid the situation that the computation is stalled due to the long sampling time of some samplers.

\textbf{\blackcircled{2} Dynamic Mini-batch Allocation.}
In dynamic mini-batch allocation, mini-batches are dynamically allocated to the workers to ensure that each computing node has mini-batch supply in time and alleviate workload imbalance.
Usually, a well-designed allocator is used to collect the mini-batch generated by the samplers, and then supply the mini-batch in time according to the needs of the workers.
In this way, the performance of sampling is determined by the efficiency of all samplers, instead of the slowest sampler.
This can effectively avoid the situation that the computation is stalled due to the long sampling time of some slow samplers.

%In contrast, the mini-batch static allocation is that each worker has one or several corresponding samplers, and its sampler is responsible for sampling and transferring mini-batches to it.
%The disadvantage of static allocation is that it is easy to cause workload imbalance.
%Since the computational pattern of the sampling is irregular, its computing time is difficult to estimate. As a result, some samplers cost much time in the sampling than the average sampling time of others. This leads to serious workload imbalance in static allocation as those corresponding workers have to wait for their mini-batches.
%The dynamic mini-batch allocation uniformly manages the workload and dispatch mini-batches to workers in time, thereby achieving better performance.

Meanwhile, dynamic mini-batch allocation helps to alleviate workload imbalance.
Traditionally, the mini-batch static allocation is that each worker has one or several corresponding samplers, and its sampler is responsible for sampling and transferring mini-batches to it.
Due to the irregular computational pattern of sampling, the computing time of a sampler is difficult to estimate, and it may lead to serious workload imbalance in static allocation, as the workers corresponding to some slow samplers have to wait for their mini-batches.
In contrast, dynamic mini-batch allocation uniformly manages the workload and dispatches mini-batches to workers in time, thereby avoiding workload imbalance and achieving better performance.

SALIENT \cite{arxiv_salient} implements dynamic mini-batch allocation by using a lock-free input queue.
CPU responsible for sampling and GPU responsible for model computation are not in a static correspondence.
The number of each worker is sequentially stored in the queue.
The mini-batch generated by each CPU is assigned a destination worker number according to the number stored in the queue during the generation.
After the generation, the mini-batch is sent to the corresponding worker according to the destination number immediately.
%The lock-free input queue is a simple design while effectively avoiding the problem faced by static allocation.
The lock-free input queue is a simple design that effectively avoids the problem faced by static allocation.

\textbf{\blackcircled{3} Mini-batch Transmission Pipelining.}
Mini-batch transmission pipelining refers to pipelining the transmission of mini-batches and the model computation to reduce the overhead of data transmission.

Due to the decoupled nature of sampling and model computation in individual-sample-based execution, mini-batches need to be transferred from samplers to workers.
The overhead of the transmission is high and even occupies up to 35\% of the per-epoch time \cite{arxiv_salient}.
Fortunately, the mini-batch transmission and model computation do not require strict sequential execution.
Thus, the mini-batch transmission and model computation pipelining can be implemented to reduce the mini-batch transmission overhead.
This reduces the occurrence of worker stagnation due to the failure of the mini-batch to be supplied in time.

In SALIENT \cite{arxiv_salient}, the samplers are CPUs and the workers are GPUs, so mini-batches need to be transferred from CPUs to GPUs.
To pipeline mini-batch transmission and model computation, it uses different GPU threads to deal with each of them respectively.
AGL \cite{pvldb_agl} uses the idea of pipelining to reduce the data transmission too.

\textbf{\blackcircled{4} Parallel Aggregation with Edge Partitioning.}
Parallel aggregation with edge partitioning refers to partitioning the edges in mini-batches according to their destination vertices and assigning each partition to an individual thread to accomplish aggregation in parallel. 

%For the Aggregation operation of GNNs, the neighboring vertex's feature needs to be aggregated for each vertex.
For the Aggregation operation of GNNs, each vertex need to aggregate its in-neighbors' features.
It means the computation is actually determined by the edges.
AGL \cite{pvldb_agl} proposes edge partitioning techniques to accomplish parallel aggregation according to this property.
In each mini-batch, AGL partitions the edges into several partitions according to their destination vertices ensuring that the edges with the same destination vertex are in the same partition.
Then these partitions are handled by multiple threads independently.
As there are no data conflicts between these threads, the Aggregation operation can be accomplished effectively in parallel.
In addition, the workload balance between these threads is guaranteed.
It is because the mini-batch generated by sampling ensures that each vertex has a similar number of neighboring vertex, resulting in a similar number of edges per edge partition. 
%resulting in each edge partition has similar number of edges.

\subsection{Joint-sample-based Execution} \label{sec:dmt:sample_joint_execution}

The joint-sample-based execution of distributed mini-batch training is illustrated in Fig. \ref{fig:03_dmt} (c). Its workflow, computational pattern, communication pattern, and optimization techniques are introduced in detail as follows.

\subsubsection{Workflow}
In the joint-sample-based execution, multiple collaborative workers are used to perform training.
The graph is split into subgraphs through the partition in the preprocessing. 
Each worker holds one subgraph and a replica of the model parameters.
The workers sample their own subgraphs to generate mini-batches and perform forward propagation as well as backward propagation on the mini-batches to obtain gradients.
Then they update the model synchronously through communication with other workers.

%A more detailed description of the workflow follows.
%First, in order to make the computation of each worker more independent, a well-design graph partition algorithm is used to split the original graph into multiple subgraphs in the preprocessing.
%Each subgraph is assigned to a worker.
%The worker samples its local subgraph, which means the target vertices of its mini-batch are all in its local subgraph.
%However, it may need to query other workers for the neighborhoods of target vertices in the sampling. 
%Second, the worker performs model computation on the mini-batch generated by itself.
%Through forward and backward propagation, each worker produces gradients and conducts gradient synchronization together to update model parameters.
%Note that the sampling and model computation is conducted on a single worker.
%This feature makes the computation of each computing node more independent, which greatly reduces the demand for data transmission.

A more detailed description of the workflow follows.
First, in order to make the computation of each worker more independent, a well-design graph partition algorithm is used to split the original graph into multiple subgraphs in the preprocessing.
Each subgraph is assigned to a worker, and the worker samples its local subgraph, which means the target vertex chosen for sampling is chosen only from its own subgraph.
However, the worker may need to query other workers for the neighborhoods of target vertices in the sampling. 
Second, the worker performs model computation on the mini-batch generated by itself.
Through forward and backward propagation, each worker produces gradients and conducts gradient synchronization together to update model parameters.
Note that the sampling and model computation is conducted on a single worker.
This feature makes the computation of each computing node more independent, which greatly reduces the demand for data transmission.

\subsubsection{Computational Pattern}
%在计算节点内计算上体现的行为，同时描述challenge

In the joint-sample-based execution, the computation of each computing node is relatively independent.
The computation of each computing node includes two parts: sampling and model computation.
In addition, gradient synchronization between computing nodes is required for updating model parameters.
In the sampling phase, each computing node mainly uses its own local data, and there may be a need to access remote data for neighbor information of vertices.
The mini-batch generated by sampling is directly consumed by itself for model computation.
For such a computing pattern, the main concern is how to make the computation of computing nodes more independent, so as to obtain better performance.

\subsubsection{Communication Pattern}
%在计算节点间通信上体现的行为，同时描述challenge或者concerns

There are three communication requirements in joint-sample-based execution, each with a different communication pattern.
The first exists in the sampling phase, and its communication pattern is irregular.
In the joint-sample-based execution, each computing node selects the target vertices of the mini-batch from its own subgraph.
Then the computing node needs to sample the neighbors of these target vertices.
Since the graph is partitioned, it may be necessary to query other computing nodes to obtain the neighbor information of these target vertices \cite{ia3_distdgl}.
As the communication is determined by the graph structure and the graph itself has an irregular connection pattern, its communication pattern is irregular.

The second communication requirement is in the gradient synchronization phase.
After sampling and model computation, the computing nodes need to synchronize to update the model parameters.
Unlike DNNs, the size of model parameters of GNNs is much smaller, since GNNs have few layers and share weights across all vertices.
As a result, it is expected that the gradient synchronization phase occupies a small part of the execution time \cite{socc_pagraph, ia3_distdgl}.
However, synchronization with the nature of a fence requires all computing nodes to complete the computation before it can start, so it is highly sensitive to workload imbalance.
The low performance of a single computing node causes other computing nodes to wait, which results in a large time overhead for this part of the communication \cite{/cal22/linhaiyang/characterizing_and_gnn/}.

The third communication requirement only arises when the sampling and model computation of a single worker are performed on different computing components, such as CPUs and GPUs on a single server \cite{socc_pagraph, cluster_2pgraph}.
Its communication pattern is regular but with high redundancy.
The existence of the same vertex in different mini-batches is called inter-mini-batch redundancy, which leads to wasted transmission.
This redundancy is exacerbated by the fact that each computing node in the joint-sample-based execution mainly samples on its own subgraph: compared to the sampling on the entire graph, the sampling on the same subgraph exhibits a higher probability of sampling the same vertices for different mini-batches \cite{socc_pagraph}. 
As a result, reducing transmission overhead by eliminating this redundancy is an effective optimization point.

\subsubsection{Optimization Techniques}\label{sec:dmt:joint:tech}
%优化的问题
%优化技术的核心
%引用具体工作详细描述优化技术或者优化技术变体

%\subsubsection{Computational Pattern} \label{sec:sample_joint_execution:computational_pattern}

%\subsubsection{Optimization Techniques} \label{sec:sample_joint_execution:optimziation_techniques}
Here we introduce the optimization techniques used to reduce transmission overhead and waiting overhead for joint-sample-based execution in detail.
We categorize them into four categories, including 
\blackcircled{1} Locality-aware Partition, \blackcircled{2} Partition with Overlap, \blackcircled{3} Independent Execution with Refinement, and \blackcircled{4} Frequently-used Data Caching.
A summary for these categories is shown in Table \ref{table:summary_of_sample_joint_execution}.

\begin{table*}[hbtp]
    %\vspace{-12pt}
 \caption{Summary of various categories of optimization techniques used in joint-sample-based execution of distributed mini-batch training.} \label{table:summary_of_sample_joint_execution}
 %\vspace{-10pt}
 \centering
\renewcommand\arraystretch{1.5} 
\resizebox{1\textwidth}{!}{
\begin{tabular}{|l|cccc|}
\hline
\textbf{Optimization Technique}     
& \textbf{Function}  & \textbf{Operation Object} & \textbf{Focus} & \textbf{Related Work}\\ \hline \hline
\multirow{2}{*}{\begin{tabular}[c]{@{}c@{}}   \blackcircled{1} Locality-aware Partition  \end{tabular}} 
&\multirow{2}{*}{\begin{tabular}[c]{@{}c@{}}   Transmission Reduction \end{tabular}}  
&\multirow{2}{*}{\begin{tabular}[c]{@{}c@{}}   Entire Graph \end{tabular}} 
&\multirow{2}{*}{\begin{tabular}[c]{@{}c@{}}   Graph Partition Strategy \end{tabular}}  
&\multirow{2}{*}{\begin{tabular}[c]{@{}c@{}}   PaGraph \cite{socc_pagraph}, 2PGraph \cite{cluster_2pgraph}, \\DistDGL \cite{ia3_distdgl}, AliGraph \cite{pvldb_aligraph}  \end{tabular}}  \\
    &   &   &   &   \\ \hline
\multirow{2}{*}{\begin{tabular}[c]{@{}c@{}}   \blackcircled{2} Partition with Overlap  \end{tabular}} 
&\multirow{2}{*}{\begin{tabular}[c]{@{}c@{}}  Transmission Reduction  \end{tabular}}  
&\multirow{2}{*}{\begin{tabular}[c]{@{}c@{}}  Neighboring Vertices  \end{tabular}} 
&\multirow{2}{*}{\begin{tabular}[c]{@{}c@{}}  Duplicate Strategy  \end{tabular}}  
&\multirow{2}{*}{\begin{tabular}[c]{@{}c@{}}  PaGraph \cite{socc_pagraph}, 2PGraph \cite{cluster_2pgraph}  \end{tabular}}  \\
    &   &   &   &   \\ \hline
\multirow{2}{*}{\begin{tabular}[c]{@{}c@{}}   \blackcircled{3} Independent Execution with Refinement  \end{tabular}} 
&\multirow{2}{*}{\begin{tabular}[c]{@{}c@{}}  Waiting Time Reduction  \end{tabular}}  
&\multirow{2}{*}{\begin{tabular}[c]{@{}c@{}}  Computing Fashion  \end{tabular}} 
&\multirow{2}{*}{\begin{tabular}[c]{@{}c@{}}  Computing Strategy  \end{tabular}}  
&\multirow{2}{*}{\begin{tabular}[c]{@{}c@{}}  LLCG \cite{arxiv_llcg}  \end{tabular}}  \\
    &   &   &   &   \\ \hline
\multirow{2}{*}{\begin{tabular}[c]{@{}c@{}}   \blackcircled{4} Frequently-used Data Caching  \end{tabular}} 
&\multirow{2}{*}{\begin{tabular}[c]{@{}c@{}}  Transmission Reduction  \end{tabular}}  
&\multirow{2}{*}{\begin{tabular}[c]{@{}c@{}}  Frequently-used Data  \end{tabular}} 
&\multirow{2}{*}{\begin{tabular}[c]{@{}c@{}}  Caching Strategy  \end{tabular}}  
&\multirow{2}{*}{\begin{tabular}[c]{@{}c@{}}  AliGraph \cite{pvldb_aligraph}, PaGraph \cite{socc_pagraph}, \\2PGraph \cite{cluster_2pgraph}  \end{tabular}}  \\
    &   &   &   &   \\ \hline

%\blackcircled{1} Locality-aware Partition        & Transmission Reduction    & Entire Graph   & Graph Partition Strategy   & PaGraph \cite{socc_pagraph}, 2PGraph \cite{cluster_2pgraph}, DistDGL \cite{ia3_distdgl}, AliGraph \cite{pvldb_aligraph}   \\ \hline
%\blackcircled{2} Partition with Overlap  & Transmission Reduction  & Neighboring Vertices   & Duplicate Strategy       & PaGraph \cite{socc_pagraph}, 2PGraph \cite{cluster_2pgraph}  \\ \hline  
%\blackcircled{3} Independent Execution with Refinement     & Waiting Time Reduction  & Compute Fashion & Computation Strategy         & LLCG \cite{arxiv_llcg}  \\ \hline   
%\blackcircled{4} Caching Strategy \todo{need revise}    &  Transmission Reduction & Frequently-used Data & Caching Strategy          & AliGraph \cite{pvldb_aligraph}, PaGraph \cite{socc_pagraph}, 2PGraph \cite{cluster_2pgraph}  \\ \hline   
\end{tabular}
    }
\end{table*}

\textbf{ \blackcircled{1} Locality-aware Partition.}
Locality-aware partition refers to partitioning the graph into subgraphs with good locality, that is, vertices and their neighbors have a high probability to be in the same subgraph, so most of the data required for sampling is local to the computing node.
This partition is conducted in the preprocessing phase.

Locality-aware partition aims to make the computation of the nodes more independent by reducing the communication between them.
According to the workflow of mini-batch training, the model computation is restricted in using the mini-batch data and does not involve access to the entire raw graph data.
That means the access to the raw graph data is almost completely in the sampling phase.
By focusing on the locality of the graph, each vertex and its neighbors are clustered into a subgraph as much as possible.
This makes the workers mainly access their own subgraph during the sampling phase, reducing the remote queries of neighboring vertices, and thus improving the independence of each worker's computation.

The graph partition here is very different from the one of distributed full-batch training mentioned earlier in the paper, which is mainly reflected in its purpose.
The graph partition in the preset-workload-based execution of distributed full-batch training focuses more on workload balance due to its significant impact on performance.
Because a large amount of information of the subgraph is related to the execution time, it is necessary to take various factors into account to guide the graph partition, such as the numbers of vertices and edges.
In contrast, workload balance is not the critical issue for graph partition in distributed mini-batch training.
Because the computation amount of the subgraph is mainly determined by the number of target vertices contained in it, it is only necessary to ensure that the number of target vertices in each subgraph tends to be consistent, which is simple but effective.
Such characteristics cause the two graph partitions to be quite different, which in turn leads to different techniques used.

%\todo{
%这个章节不是说locality-aware partition吗？目的是减少通信吧。怎么后续两个段落的描述又回到了workload balance呢？
%就是目前的描述，反而均衡比重更强，locality都没有被描述的感觉，除了第一段。你看看要不要改改描述？
%}
%\Revise{revising}

To implement locality-aware partition, PaGraph \cite{socc_pagraph} proposes a formula to guide the graph partition.
Its partition algorithm scans the whole target vertex set, and iteratively assigns the scanned vertex to one of $P$ partitions according to the score computed from the formula:
\begin{equation}\label{eq4_pagraph_score} 
    score_{v_t}^{(i)} = |TV_i \cap IN(V_t)|\cdot\frac{(TV_{avg}-|TV_i|)}{|PV_i|}
\end{equation}
where $TV_i$ denotes the train vertex set already assigned to the $i$-th partition, $IN(V_t)$ denotes the $k$-hop ($k$ is the layer number of GNN model) in-neighbors of the target vertex $v_t$, and $PV_i$ denotes the total number of vertices in the $i$-th partition.
%in-neighbor set
$TV_{avg}$ denotes the expected number of target vertices in the final $i$-th partition, which is typically set as the total number of vertices $|TV_i|$ divided by the partition number $P$ for workload balance. 
This formula fully considers the distribution of the vertices' neighbors when splitting them into each subgraph, thus making great use of the locality of the graph.

Other researches mainly use existing graph partition algorithms to harvest the locality of the graph.
2PGraph \cite{cluster_2pgraph} uses cluster-based graph partition algorithm \cite{/siamsc/a_fast_and_high_quality/distdgl_metis/2pgraph_partition_algorithm1/, /pami/weighted_graph_cuts/2pgraph_partition_algorithm2/} in its graph partition.
DistDGL \cite{ia3_distdgl} takes advantage of the multi-constraint mechanism in METIS \cite{/siamsc/a_fast_and_high_quality/distdgl_metis/2pgraph_partition_algorithm1/} for graph partition.
%to balance workloads in each partition.
%\Revise{
AliGraph \cite{pvldb_aligraph} implements four graph partition algorithms in its system and suggests that the four algorithms are suitable for different circumstances.
They are METIS \cite{metis/dgcl_partition_algorithm} for processing sparse graphs, vertex and edge cut partitions \cite{/osdi/powergraph/aligraph_graph_partition_2/} for processing dense graphs, 2-D partition \cite{/sc/scalable_matrix_computations/aligraph_graph_partition_3/} for systems with a fixed number of workers, and streaming-style partition strategy \cite{/kdd/streaming_graph_partitioning/aligraph_graph_partition_4/} for graphs with frequently edge updates.
%}
%, including METIS \cite{metis/dgcl_partition_algorithm}, Vertex cut and edge cut partitions \cite{/osdi/powergraph/aligraph_graph_partition_2/}, 2-D partition \cite{/sc/scalable_matrix_computations/aligraph_graph_partition_3/}, and Streaming-style partition strategy \cite{/kdd/streaming_graph_partitioning/aligraph_graph_partition_4/}.
%It suggests that they are suitable for different circumstances.
%Respectively, METIS method for processing sparse graphs, Vertex and edge cut method for processing dense graphs, 2-D partition for systems with a fixed number of workers, and Streaming-style partitioning method for graphs with frequently edge updates.

\begin{figure*}[hbtp]
    \vspace{5pt}
    \centering
    \includegraphics[page=7, width=0.6\textwidth]{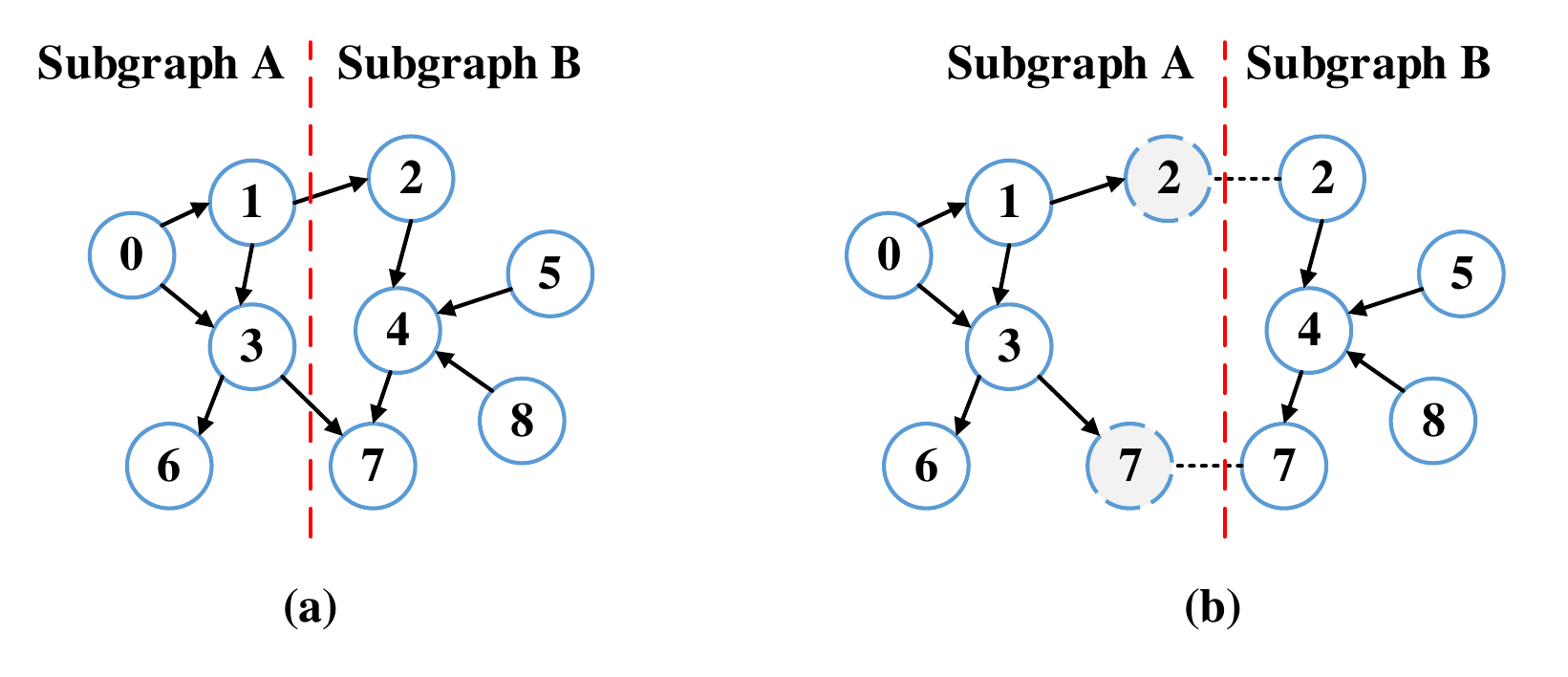}
    %\vspace{5pt}
    \caption{Illustration of graph partition with overlap: (a) Graph partition with non-overlap where no duplicate method is used; (b) Graph partition with overlap. The duplicate range is 1-hop neighbors. 
    The vertices in white are original target vertices, while the vertices in grey are duplicated, referred to as mirror vertices.
    }
    \label{fig:06_overlap}
    \vspace{5pt}
\end{figure*}

\textbf{\blackcircled{2} Partition with Overlap.}
Partition with overlap refers to duplicating the neighboring vertices of the target vertex when partitioning the graph, thereby reducing or even completely eliminating the data transmission between computing nodes in the sampling phase.

%{\color{red} modify the number from (a) to (a) (b)}

For comparison, we first review the general graph partitioning method used in traditional graph analytics, i.e., partition with non-overlap which means no duplicate method is used.
%The way to partition graph is generally partition with non-overlap, which means no duplicate method is used.
An example of partition with non-overlap is demonstrated in Fig. \ref{fig:06_overlap} (a).
The original graph is split into two subgraphs, denoted as A and B.
The vertices held by each subgraph are disjoint.
This has the benefit of linearly reducing the requirement of memory capacity when training large graphs.
In the case of $n$ workers, each worker only needs to hold $1/n$ of the original graph data.
However, partition with non-overlap causes some vertices' neighbors to be stored in remote workers, resulting in the need for remote queries during the sampling phase.
Intensive transmissions would lead to inefficiencies in the parallel execution of distributed training.

To reduce the amount of transmission in the sampling phase, partition with overlap is proposed.
An example of partition with overlap is illustrated in Fig. \ref{fig:06_overlap} (b), whose duplicate range is 1-hop neighbors.
The original graph is split into two subgraphs with duplicate vertices.
Vertex 2 and 7 are duplicated as they are the 1-hop neighboring vertex of target vertices in subgraph A.
This method takes advantage of the characteristics of GNN mini-batch training.
When the mini-batch is sampled, only the $L$-hop neighboring vertices of the target vertices are needed, where $L$ is the number of GNN layers.
Therefore, by duplicating the features of its neighboring vertices, the transmission overhead in the sampling phase can be reduced.
These duplicate vertices are regarded as mirror vertices, which do not participate in training as target vertices \cite{/osdi/powergraph/aligraph_graph_partition_2/}. 
If the $L$-hop neighbors are fully duplicated, the transmission requirements during the sampling phase can be completely eliminated.
Note that duplicating more vertices causes an increase in the memory capacity requirements of the computing nodes, so there is a trade-off in how many neighboring vertices are duplicated.

Both PaGraph \cite{socc_pagraph} and 2PGraph \cite{cluster_2pgraph} use this method to make the computation of each worker more independent. 
DGCL \cite{eurosys_dgcl} defines a replication factor, which is computed from the total number of vertices (including the original and mirror vertices) kept by all workers divided by the number of vertices in the original graph.
In its experiments, for the dense Reddit dataset \cite{nips_inductive_representation_ley_graphsage_minibatch} with 16 GPU workers and 2-hop duplicate range, the replication factor reach 15.
In contrast, for the sparser Web-Google dataset \cite{/im/community_structure_in_large/web_google/dgcl_dataset} with the same setting, the replication factor is much smaller which is 2.5. 
It suggests that sparse graphs are more suitable for the duplicate method as they have less memory capacity overhead.

%\todo{
%只优化参数同步阶段，需要细化描述和标题内容
%}
%\Revise{awswer
%并不是只是优化同步阶段
%其方法就是 1.各节点在自己子图上sampling和GNN model computation  （sampling完全只考虑自己的子图，不去访问其他节点的图数据）
%然后 2.周期性收集大家的weights， 平均得到统一的weight
%然后 3. 有一台机器（leader) sampling全图，然后训练update weights。 （称为refine) 然后发回给各节点

%也是就  独立计算阶段，各worker只会用自己subgraph数据，完全独立计算。
% refine阶段： 收集平均，利用全图sampling，训几轮，发回去继续。
%}

\textbf{\blackcircled{3} Independent Execution with Refinement.}
Independent execution with refinement refers to letting each worker compute and update the model parameters independently, and then periodically averages the model parameters using additional refine operation to release the potential of parallel computation while ensuring high accuracy.

Although the workers of joint-sample-based execution mainly perform computations on their own local subgraph, there are still interactions between them to advance the training process, including the sampling phase and the gradient synchronization phase.
The limitations of these two phases cause each worker's execution to be affected by the state of other workers, resulting in idle wait time.
%Although the duplicate method mentioned before can reduce the amount of communication in the sampling phase, its additional memory capacity requirement makes it difficult to scale to large graphs.
%上一句话和上一段以及下一段都没有联系，删掉了。

%\todo{R1 Q8}
\Revise{
%In order to achieve more independent computation of each computing node,  a method of independent execution with refinement building on the concept of learn locally correct globally (LLCG) is proposed in LLCG \cite{arxiv_llcg}. 
To enable more independent computations for each computing node, a method called independent execution with refinement is proposed, building upon the concept of ``learn locally, correct globally'' (LLCG) as outlined in LLCG \cite{arxiv_llcg}.
}
%\Revise{In order to achieve more independent computation of each computing node, \todo{R1 Q8} Learn Locally, Correct Globally (LLCG)}\cite{arxiv_llcg} proposes a method of independent execution with refinement.
%It adopts an architecture similar to parameter server architecture, which is divided into parameter server and workers. 
In its implementation, each worker first performs sampling and corresponding model computation independently on their own subgraphs.
%That is to say, the data of remote workers is not accessed at all and each worker updates their own model parameters.
That is to say, the data of remote workers is not accessed at all, and each worker only updates its own model parameters locally.
A parameter server then periodically collects the model parameters from each worker, and performs the average operation of the parameters.
Finally, the parameter server refines the average result, that is, it samples the mini-batch on the whole graph and conducts model computation to update the model parameters of every worker. 
This is due to the fact that each worker lacks the structural information of the whole graph, which may lead to low accuracy. 
Through this method, the computation of each worker is more independent while the refinement on model parameters ensures the high accuracy of GNN model.

%\todo{R2 Q1}
\Revise{
Considering the aforementioned effort involving the balance between model accuracy and computational efficiency, an analysis of this trade-off is presented below to offer a more detailed understanding of this effort. LLCG's methodology has been validated for convergence through rigorous algorithmic theory and practical experiments, as described in its paper~\cite{arxiv_llcg}. However, given its focus on subgraph computations, the method results in a marginal decrease in final model accuracy, usually within a 1\% range. Simultaneously, it achieves a remarkable reduction in communication volume by almost 100$\times$, leading to a substantial enhancement in computational efficiency.
}

%\Revise{
%\textit{Tradeoff between Model Accuracy and Computational Efficiency:}
%The trade-off between model accuracy and computational efficiency is investigated as follows. LLCG's approach has been validated for convergence through both algorithmic theory and practical experiments in its paper~\cite{arxiv_llcg}. Nevertheless, since the method predominantly involves subgraph computations, it leads to a marginal decrease in final model accuracy, typically within a 1\% range. Concurrently, it achieves a remarkable reduction in communication volume by nearly 100$\times$, thereby significantly improving computational efficiency.
%}

\textbf{\blackcircled{4} Frequently-used Data Caching.}
%Caching strategy refers to caching the feature of frequently used vertices to reduce the overhead of data transmission.
Frequently-used data caching refers to caching the information of frequently used vertices from other computing nodes, including neighbors or features, to reduce the overhead of data transmission.

To reduce the transmission overhead of obtaining the neighboring vertices in the sampling phase, AliGraph \cite{pvldb_aligraph} proposes to cache the outgoing neighbors of important vertices wherever important vertices are located.
Important vertices are identified by the following formula.
\begin{equation}\label{eq_important_vertice}
    Imp^k(v) = \frac{Nin^k(v)}{Nout^k(v)} > U^k
    %\mathcal{L} = \frac{1}{|\mathcal{V}_s|}\sum_{v_i \in \mathcal{V}_s} \nabla l(y_i, z_i)
\end{equation}
where $Imp^k(v)$ denotes the importance score of the vertex v, $Nin^k(v)$ and $Nout^k(v)$ represents the number of $k$-hop incoming and outgoing neighbors of the vertex $v$, and $U^k$ is a user-specified threshold.
When the $Imp^k(v)$ exceeds $U^k$, the vertex $v$ is considered to be an important vertex.
By caching the outgoing neighbors of important vertex $v$, the transmission overhead derived from the queries of other vertices via $v$ can be reduced.

Similarly, the frequently-used data caching technique can also reduce the transmission of mini-batches between CPUs and GPUs.
It is the third communication requirement of joint-sample-based execution which occurs when CPUs and GPUs are responsible for sampling and model computation respectively.
Due to the benefit of mini-batch training, each GPU only needs to load one mini-batch of data per round of computation, and it is possible to load cached data into the free memory during model computation.
Thus, by caching vertex features and combining them into mini-batches on GPUs, the amount of mini-batch transmission can be reduced.

PaGraph \cite{socc_pagraph} uses a very simple caching strategy.
It sorts vertices according to their degree from largest to smallest and then caches them in order, where the degree is the number of incoming edges of a vertex.
The higher the degree, the higher the cache priority.
%This is easy to understand because when a vertex has a larger degree, this vertex is more likely to be a neighbor of other target vertices.
This is a simple but effective method: when a vertex has a larger degree, this vertex is more likely to be a neighbor of other target vertices.
As for the proportion of cached vertices, it is determined by measuring the free memory capacity of GPUs during the pre-run.
%This simple method works very well, reducing data transmission up to 96.8\%.
2PGraph \cite{cluster_2pgraph} proposes GNN Layer-aware Caching.
It adopts a sampling method that uses a fixed order of target vertices, which allows it to schedule the vertices to train next.
By pre-caching all of the target vertices' $L$-hop neighbors into GPU memory before the computation, the time overhead of data transmission can be almost completely eliminated.
However, fixing the order of target vertices may compromise the model accuracy, so it proposes to periodically permute the order to reduce this effect.

\section{Software Frameworks for \\ Distributed GNN Training}\label{sec:software_framework}
%\section{Software Framework}\label{FRAMEWORK}

%\begin{table}[!t]
\begin{table*}[hbtp]
 \caption{Software frameworks for distributed GNN training.} \label{table:framework}
  %\caption{Software Frameworks of distributed GNN training. \textit{DMT means distributed mini-batch training and DFT means distributed full-batch training.}} \label{table:framework}
 \centering
 \renewcommand\arraystretch{1.8} 
    \resizebox{1\textwidth}{!}{
\begin{tabular}{|c|cccccccc|}
\hline
%\textbf{Names}    &\textbf{Multi-CPU}  &\textbf{Multi-GPU}   &\textbf{DMT} &\textbf{DFT}  &\textbf{Homogeneous Graph} &\textbf{Heterogeneous Graph} &\textbf{Static Graph} &\textbf{Dynamic Graph}\\ \hline \hline
%\textbf{Names}    &\textbf{Multi-CPU}  &\textbf{Multi-GPU}   &\textbf{\multirow{2}{*}{\begin{tabular}[c]{@{}c@{}}Distributed \\ Mini-batch\end{tabular}}} &\textbf{Distributed Full-batch}  &\textbf{Homogeneous Graph} &\textbf{Heterogeneous Graph} &\textbf{Static Graph} &\textbf{Dynamic Graph}\\ 
%& & & & & & & & \\ \hline \hline
\multicolumn{1}{|c|}{\multirow{2}{*}{\textbf{Name}}}    &\multirow{2}{*}{\textbf{Multi-CPU}}  &\multirow{2}{*}{\textbf{Multi-GPU}}   &\textbf{\multirow{2}{*}{\begin{tabular}[c]{@{}c@{}}Distributed\\Mini-batch Training\end{tabular}}} &\textbf{\multirow{2}{*}{\begin{tabular}[c]{@{}c@{}}Distributed\\Full-batch Training\end{tabular}}}  &\textbf{\multirow{2}{*}{\begin{tabular}[c]{@{}c@{}}Homogeneous\\Graph\end{tabular}}} &\textbf{\multirow{2}{*}{\begin{tabular}[c]{@{}c@{}}Heterogeneous\\Graph\end{tabular}}} &\textbf{\multirow{2}{*}{\begin{tabular}[c]{@{}c@{}}Static\\Graph\end{tabular}}} &\textbf{\multirow{2}{*}{\begin{tabular}[c]{@{}c@{}}Dynamic\\Graph\end{tabular}}}\\
\multicolumn{1}{|c|}{} & & & & & & & & \\ \hline \hline
             PyG \cite{arxiv_pyg}       &\checkmark   &\checkmark   &\checkmark   &\checkmark  &\checkmark &\checkmark  &\checkmark  &\checkmark   \\\hline
             DGL \cite{arxiv_dgl}      &\checkmark   &\checkmark   &\checkmark   &\checkmark  &\checkmark &\checkmark   &\checkmark &\checkmark   \\\hline
             NeuGraph \cite{atc_neugraph}      &   &\checkmark   &   &\checkmark &\checkmark  &   &\checkmark &   \\\hline
             Roc \cite{mlsys_roc}      &   &\checkmark   &   &\checkmark  &\checkmark &   &\checkmark &   \\\hline
             %DGCL \cite{eurosys_dgcl}      &   &\checkmark   &   &\checkmark   &   &   \\\hline
             FlexGraph \cite{eurosys_flexgraph}      &\checkmark   &   &   &\checkmark  &\checkmark &\checkmark   &\checkmark &   \\\hline
             MG-GCN \cite{arxiv_mg_gcn}      &   &\checkmark   &   &\checkmark &\checkmark  &   &\checkmark &   \\\hline
             Dorylus \cite{osdi_dorylus}       &\checkmark   &   &   &\checkmark  &\checkmark &   &\checkmark &   \\\hline
             AliGraph \cite{pvldb_aligraph}     &\checkmark   &   &\checkmark   & &\checkmark  &\checkmark   &\checkmark &\checkmark   \\\hline
             AGL \cite{pvldb_agl}      &\checkmark   &   &\checkmark   & &\checkmark  &   &\checkmark &   \\\hline
             %DistDGL \cite{ia3_distdgl}       &\checkmark   &   &\checkmark   &   &   &   \\\hline
             GraphTheta \cite{arxiv_graphtheta}      &\checkmark   &   &\checkmark   &\checkmark &\checkmark  &   &\checkmark &   \\
\hline 
\end{tabular}
    }
\end{table*}

%\begin{tabular}[c]{@{}c@{}}Limited Memory Resource, Resource Contention, \\ and High Transmission Overhead\end{tabular}

In this section, we introduce the software frameworks of distributed GNN training. 
Table \ref{table:framework} provides a summary of these frameworks and their supported attributes.

\textbf{Basic Software Frameworks.} PyG \cite{arxiv_pyg} and DGL \cite{arxiv_dgl} are the two most popular software frameworks in the GNN community.
PyG \cite{arxiv_pyg} is a geometric deep learning extension library for PyTorch to enable deep learning on irregular structure data such as graphs.
It supports both CPU and GPU computing, providing convenience for using GPU to accelerate the computing process.
Through the message-passing application programming interface (API), it is easy to express various GNN models, as neighbor aggregation is a kind of message propagation.
DGL \cite{arxiv_dgl} is a framework specialized for deep learning models on graphs.
It abstracts the computation of GNNs into a few user-configurable message passing primitives, thus helping users express GNNs more conveniently.
It achieves good performance by exploring a wide range of parallelization strategies.
It also supports both CPU and GPU computing.

Due to their open source and easy-to-use nature, more and more software frameworks are building on top of them and propose optimizations for distributed GNN training, such as DistGNN \cite{sc_distgnn}, SAR \cite{arxiv_sar}, DistDGL \cite{ia3_distdgl}, PaGraph \cite{socc_pagraph}, LLCG \cite{arxiv_llcg}, DistDGLv2 \cite{arxiv_distdglv2}, SALIENT \cite{arxiv_salient}, P3 \cite{pvldb_g3}, and so on.
Next, we introduce the software frameworks dedicated to distributed full-batch training and distributed mini-batch training.
%For the supported computing platforms and training methods, each software framework mainly focuses on the optimization of one of them.

\textbf{Software Frameworks Dedicated to Distributed Full-batch Training.}
The software frameworks dedicated to distributed full-batch training includes NeuGraph \cite{atc_neugraph}, Roc \cite{mlsys_roc}, FlexGraph \cite{eurosys_flexgraph}, MG-GCN \cite{arxiv_mg_gcn}, and Dorylus \cite{osdi_dorylus}.

NeuGraph \cite{atc_neugraph} is a distributed GNN training software framework proposed in 2019, using multi-GPU hardware platform.
It is categorized as the dispatch-workload-based execution of distributed full-batch training.
%It proposes an abstract model for the programming of GNN operations, called SAGA-NN (Scatter-ApplyEdge-Gather-ApplyVertex with Neural Networks).
%SAGA-NN splits each layer of model computation into four stages: Scatter, ApplyEdge, Gather, and ApplyVertex.
It proposes SAGA-NN, an abstract model for the programming of GNN operations which splits each layer of model computation into four stages: Scatter, ApplyEdge, Gather, and ApplyVertex.
SAGA-NN is named after these stages.
The Scatter stage means the vertices scatter their features to their output edges, while the Gather stage means the vertices gather the value from their input edges.
There are two user-defined functions used in the ApplyEdge stage and ApplyVertex stage, for users to declare neural network computations on edges and vertices respectively.
%It is illustrated in Fig. {\color{red} [~]}.
Through the abstraction of SAGA-NN, users can easily express various GNN models and execute them in a parallelized way.
NeuGraph optimizes the training process based on this abstract model, using techniques including vertex-centric workload partition, transmission planning, and feature-dimension workload partition.

Roc \cite{mlsys_roc} is a distributed multi-GPU software framework for fast GNN training and inference, proposed in 2020.
It is categorized as the dispatch-workload-based execution of distributed full-batch training.
Its optimization of distributed training mainly focuses on balanced workload generation and transmission planning.
In terms of balanced workload generation, an online linear regression cost model is proposed to achieve efficient graph partition. The cost model is being tuned by collecting runtime data.
According to this cost model, the training resources and time required for the subgraph can be estimated to guide the workload partition for workload balance.
In terms of transmission planning, a recursive dynamic programming algorithm is introduced to find the global optimal solution to decide which part of the data should be cached in GPU memory for reuse.

FlexGraph \cite{eurosys_flexgraph} is a distributed multi-CPU training software framework proposed in 2021.
It is categorized as the preset-workload-based execution of distributed full-batch training.
In order to express more kinds of GNNs including GNNs for heterogeneous graphs, it proposes a novel programming abstraction, namely NAU.
NAU splits the computation of one GNN layer into three stages including NeighborSelection, Aggregation, and Update stages, each with a user-defined function.
Based on NAU, FlexGraph optimizes the training process using techniques including graph pre-partition and partial aggregation.
In terms of graph pre-partition, it introduces a cost model to estimate the runtime overhead of the workload, so as to guide the workload partition for workload balance.
In terms of partial aggregation, FlexGraph partially aggregates the features of vertices collocated at the same partition when possible.
In addition, it overlaps partial aggregations and communication to reduce the transmission overhead.

MG-GCN \cite{arxiv_mg_gcn} is a distributed multi-GPU training software framework proposed in 2021.
%It is categorized as the dispatch-workload-based execution of distributed full-batch training.
It is categorized as the preset-workload-based execution of distributed full-batch training.
It focuses on the efficient parallelization of the sparse matrix-matrix multiplication (SpMM) kernel on multi-GPU hardware platform.
It uses matrix partitioning method to distribute raw data to multiple GPUs, and each GPU is responsible for completing the workload of its own local matrix.
%To partition workload, it adopts the matrix partition.
It involves the efficient reuse of memory buffers to reduce the memory footprint of training GNN models, and overlaps communication and computation to reduce communication overhead.
Specifically, the memory buffer in the computing node is used to cache the data reused by the forward propagation and backward propagation, thereby reducing data transmission.
As for the communication and computation overlap, it uses two GPU streams for computation and communication, respectively.

Dorylus \cite{osdi_dorylus} is a distributed multi-CPU training software framework proposed in 2021.
It is categorized as the preset-workload-based execution of distributed full-batch training.
Its main focus is on how to train GNNs at a low cost, so it adopts serverless computing.
Serverless computing refers to ``cloud function''  threads, such as AWS Lambda and Google Cloud Functions, that can be used massively in parallel at an extremely low price.
The hardware platform of Dorylus consists of CPUs and serverless threads.
CPUs mainly perform the Aggregation operation, while the serverless threads are used for the Combination operation due to more regular computation and simpler workload partition in the Combination operations.
It adopts a fine-grained workload partition to adapt to the situation that the available hardware resources of serverless threads are quite limited.
In addition, asynchronous training is used to make full use of computing resources and reduce stagnation.

\textbf{Software Frameworks Dedicated to Distributed Mini-batch Training.}
The software frameworks dedicated to distributed mini-batch training includes AliGraph \cite{pvldb_aligraph} and AGL \cite{pvldb_agl}.

AliGraph \cite{pvldb_aligraph} is a distributed multi-CPU training framework proposed in 2019.
It is categorized as the joint-sample-based execution of distributed mini-batch training.
It supports not only GNNs for homogeneous graphs and static graphs, but also GNNs for heterogeneous graphs and dynamic graphs.
In terms of storage, it adopts a graph partitioning method to store graph data in a distributed manner.
The structure and features of the subgraph in each computing node are stored separately. In addition, two caches are added for the features of vertices and edges.
Furthermore, it proposes a caching strategy to reduce the communication overhead between computing nodes, that is, each computing node caches the outgoing neighbors of frequently-used vertices.

AGL \cite{pvldb_agl} is a distributed multi-CPU training software framework proposed in 2020.
It is categorized as the individual-sample-based execution of distributed mini-batch training.
To speed up the sampling process, it introduces a distributed pipeline to generate $k$-hop neighborhood in the spirit of message passing, which is implemented with MapReduce infrastructure \cite{/mapreduce/}.
In this way, in the sampling phase, mini-batch data can be rapidly generated by collecting the $k$-hop neighbors of the target vertices.
In the training phase, the computing nodes are partitioned into workers and parameter servers.
The workers perform the model computation on the mini-batches, while the parameter servers maintain the current version of the model parameters.
It also uses the commonly used optimization techniques for better efficiency, such as the transmission pipeline.

\textbf{Software Frameworks Dedicated to Both Distributed Full-batch Training and Mini-batch Training.}
There exists a software framework dedicated to both distributed full-batch training and distributed mini-batch training.

GraphTheta \cite{arxiv_graphtheta} is a distributed multi-CPU training software framework proposed in 2021.
It supports three training methods: mini-batch, full-batch, and cluster-batch training.
The cluster-batch training methods is proposed by Chiang \cite{kdd/cluster_gcn/subgraph_based1} in 2019.
It first partitions a large graph into a set of smaller clusters.
Then, it generates a batch of data either based on one cluster or a combination of multiple clusters.
Apparently, cluster-batch restricts the neighbors of a target vertex into only one cluster, which is equivalent to conducting a globalized convolution on a cluster of vertices.
There is a parameter server in GraphTheta, which is responsible for managing multi-version model parameters.
The worker obtains the model parameters from the parameter server and transfers the generated gradient back to the parameter server for the update of model parameters.
Multi-version parameter management makes it possible to train GNNs asynchronously as well as synchronously.

%Instead of the straightforward training method which is to load the subgraph structure and the related data into memory and perform matrix operations on the subgraph located at the same machine, GraphTheta completes the training procedure with distributed computing by only spanning the structure of the distributed subgraph. 
%To construct a subgraph, it introduces a breadth-first-search traverse operation. For each target node, this operation initializes a minimal number of layers per node, which are involved in the computation, in order to reduce unnecessary propagation of graph computing.

\section{Hardware Platforms for \\ Distributed GNN Training}\label{sec:hardware_platform}

In this section, we introduce the hardware platforms for distributed GNN training.
At present, it is mainly categorized into two types: multi-CPU hardware platform and multi-GPU hardware platform.
A summary of the hardware platform for distributed GNN training is shown in Table \ref{table:summary_of_hardware_platform}.
Next, we introduce them separately.

%\todo{
%我发现你前面用了很多multi-CPU和multi-GPU，这个章节是否要提前到Background？不然就有点还没介绍就用的感觉
%}

%\todo{
%你这里加一个表格吧，就是对你描述内容的概括，只描述平台的特色和不足，不对比
%}
%\Revise{done}

\begin{figure*}[hbtp]
    %\vspace{-15pt}
    \centering
    \includegraphics[page=5, width=0.8\textwidth]{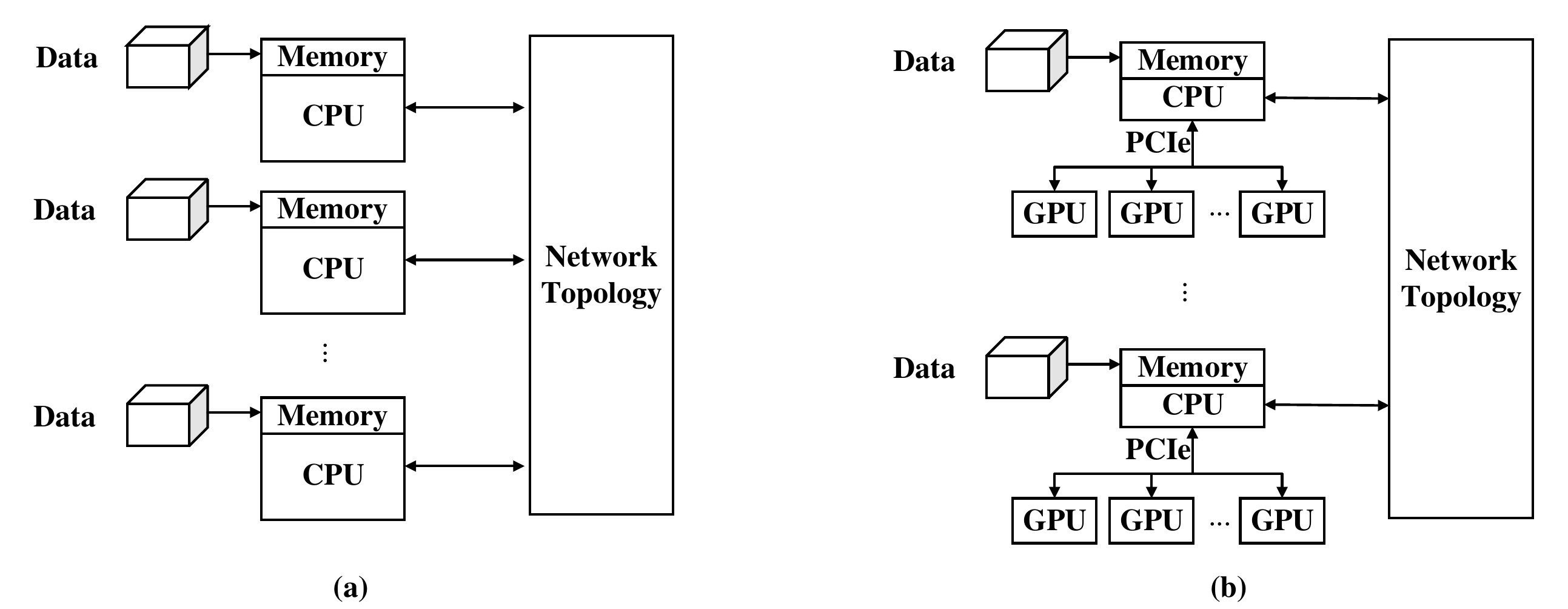}
    %\vspace{-15pt}
    \caption{Hardware platforms for distributed GNN training: (a) Multi-CPU hardware platform; (b) Multi-GPU hardware platform.}
    \label{fig:04_hardware_platform}
    %\vspace{-20pt}
\end{figure*}

\begin{table*}[hbtp]
    %\vspace{-12pt}
 \caption{Summary of hardware platforms for distributed GNN training.} \label{table:summary_of_hardware_platform}
 %\vspace{-10pt}
 \centering
\renewcommand\arraystretch{1.5} 
\resizebox{0.98\textwidth}{!}{
\begin{tabular}{|c|ccc|}
\hline
\multicolumn{1}{|c|}{\textbf{Hardware Platform}} & \textbf{Main Advantages}                                                              & \textbf{Main Disadvantages}                                                                             & \multicolumn{1}{c|}{\textbf{Suitable Training Method}} \\ \hline
\multicolumn{1}{|c|}{\multirow{2}{*}{Multi-CPU}}                  & \multirow{2}{*}{\begin{tabular}[c]{@{}c@{}}Rich Memory Resource \\ and Good Scalability\end{tabular}} & \multirow{2}{*}{Limited Computing Resource}                                                                             & \multicolumn{1}{c|}{\multirow{2}{*}{Distributed Full-batch Training}}   \\
\multicolumn{1}{|c|}{}                                            &                                                                                                       &                                                                                                                         & \multicolumn{1}{c|}{}                                                   \\ \hline
\multicolumn{1}{|c|}{Multi-GPU}                                   & Rich Computing Resource                                                                               & \begin{tabular}[c]{@{}c@{}}Limited Memory Resource, Resource Contention, \\ and High Transmission Overhead\end{tabular} & \multicolumn{1}{c|}{Distributed Mini-batch Training}                    \\ \hline
\multicolumn{1}{l}{}                                              & \multicolumn{1}{l}{}                                                                                  & \multicolumn{1}{l}{}                                                                                                    & \multicolumn{1}{l}{}                                                   
\end{tabular}
    }
\end{table*}

\subsection{Multi-CPU Hardware Platform}\label{sec:hp:multi_cpu_hardware_platform}

The structure diagram of multi-CPU hardware platform is illustrated in Fig. \ref{fig:04_hardware_platform} (a). It is mainly composed of multiple computing nodes. Each computing node has one or multiple CPUs. CPUs in each computing node communicate with each other through the network.
The graph data is generally stored in each computing node in a distributed storage manner.
Each computing node performs computations mainly on its own local data. 
When remote data is required, it queries other computing nodes through the network to obtain the required data.

The main advantage of multi-CPU hardware platform is the good scalability due to the large memory \cite{sc_distgnn, ia3_distdgl}.
By increasing the number of computing nodes, large graphs can be maintained completely in memory.
This makes it possible to avoid data access to high-latency storage components, such as hard disks, thereby ensuring the timely supply of data.
This characteristic makes it ideal for preset-workload-based execution of distributed full-batch training \cite{sc_distgnn, eurosys_flexgraph} and joint-sample-based execution of distributed mini-batch training \cite{pvldb_aligraph, ia3_distdgl, arxiv_graphtheta}.
The graph is partitioned and distributed to different computing nodes for storage.
Each computing node is responsible for the workload of its own subgraph.
This results in computing nodes using mostly local data for computation, although additional communication is also involved.
By further optimizing the communication overhead, multi-CPU platforms can achieve good scalability.

The main disadvantage of multi-CPU hardware platform is the limited computing resource in each computing node.
CPU is a latency-oriented architecture with limited computing resources \cite{throughput_oriented_architectures}.
This makes it less friendly to neural network operations that incur a lot of computation \cite{NVIDIA_Tesla, GPU_Computing}.
This problem is even more pronounced when using distributed mini-batch training.
In this case, CPUs perform both the sampling and model computation, leading to a shortage of computing resources.

\subsection{Multi-GPU Hardware Platform}\label{sec:hp:multi_gpu_hardware_platform}
The structure diagram of multi-GPU hardware platform is illustrated in Fig. \ref{fig:04_hardware_platform} (b).
Graph data is generally stored in a distributed manner on multiple computing nodes.
Each computing node consists of CPUs and GPUs.
Computing nodes are connected through high-speed network interfaces.
For example, GPUs communicate with each other over PCIe under the same computing node and over InfiniBand across different computing nodes. GPUs can also be interconnected through NVLink and NVSwitch for higher-speed inter-GPU communication.

The main advantage of multi-GPU hardware platform is the rich computing resources.
Compared with CPUs, GPUs have more computing resources that can be used for large-scale parallel computing. 
This is because GPU is a throughput-oriented architecture which focuses on improving the throughput by using massive computing resources \cite{throughput_oriented_architectures, NVIDIA_Tesla, GPU_Computing}. 
In view of the development of DNN, both sparse matrix multiplication and dense matrix multiplication in GPU have been fully developed, which facilitates the model computation of GNN.
Due to its limited memory capacity, it is not suitable for distributed full-batch training.
Otherwise, it will result in a large amount of data transmission.
Currently, distributed mini-batch training is usually performed on multi-GPU hardware platform.
The most popular way to perform distributed mini-batch training on multi-GPU hardware platform is that CPUs are responsible for the sampling phase and GPUs are responsible for the model computation.
This is because the computational pattern of the sampling phase is irregular and CPUs are more suitable for the irregular computational pattern than GPUs \cite{Tigr, micro/alleviating_irregularity_in/yan/graphdyns/graph_processing_asic_acc}.

The main disadvantages of multi-GPU hardware platform are resource contention issues and expensive data transmission overhead.
Since there are generally multiple GPUs on a computing node, these GPUs share the computing node's resources, including CPU sampling threads and network interface \cite{/cal22/linhaiyang/characterizing_and_gnn/, atc_neugraph}.
The competition for these resources makes multiple GPUs hard to obtain sufficient input data in time.
This causes the occurrence of computational stagnation, resulting in poor scalability, manifested by a large gap between the actual speedup and the ideal \cite{/cal22/linhaiyang/characterizing_and_gnn/}.
The expensive data transmission overhead is due to the tight memory resources of GPUs.
When using a multi-GPU hardware platform for distributed full-batch training, it needs to continuously transmit input data and output data \cite{atc_neugraph}.
When performing distributed mini-batch training, it is also necessary to continuously transmit the mini-batch data sampled by CPUs \cite{socc_pagraph}.
This problem is further exacerbated when the graph is large.

%\todo{
%你把两个平台对比的内容抽出来，放到这个章节中，同时配个表
%}
%\Revise{ unable
%跟前面相同问题，没有什么内容主要是。。。
%}

\subsection{Multi-CPU Hardware Platform V.S. Multi-GPU Hardware Platform}\label{sec:hp:comparison_cpu_or_cpu}

The summary of the comparison between multi-CPU hardware platform and multi-GPU hardware platform is shown in Table \ref{table:summary_of_hardware_platform}.
Multi-CPU and multi-GPU hardware platforms have their own advantages and are suitable for different scenarios.
The different characteristics of the two also lead to different optimization directions.
Multi-CPU hardware platform is more suitable for distributed full-batch training.
The entire graph can be cached in the memory, which leads to good scalability.
Multi-GPU hardware platform is more suitable for distributed mini-batch training given its limited memory resources.
However, when the graph size is small and there are high-speed interconnections between GPUs, such as NVLink, distributed full-batch training using multi-GPU hardware platform can achieve good performance \cite{mlsys_roc}.

\section{Comparison to Distributed DNN Training}\label{sec:comparison_to_dnn}

In this section, we highlight the characteristics of distributed computing of GNNs by comparing the distributed training of GNNs with that of DNNs.
We first introduce the two categories of distributed DNN training: data parallelism and model parallelism.
Among them, for GNN distributed full-batch training, its operation mode is similar to the model parallelism of DNNs.
GNN distributed mini-batch training is similar to the data parallelism of DNNs.
The two pairs are compared separately below.
A summary of the comparison between distributed GNN training and distributed DNN training is shown in Table \ref{table:summary_of_DNN_GNN_comparison}.
Next, we introduce them separately.

%\todo{ley
%补充这个表格
%修改正文中相关描述
%}
\begin{table*}[hbtp]
    %\vspace{-12pt}
 \caption{Summary of the comparison between distributed GNN training and distributed DNN training.}\label{table:summary_of_DNN_GNN_comparison}
 %\vspace{-10pt}
 \centering
\renewcommand\arraystretch{1.5} 
\resizebox{0.95\textwidth}{!}{
\begin{tabular}{|c|c|cc|}
\hline
& \textbf{Comparison Object}  & \textbf{Distributed Full-batch Training}  & \textbf{DNN Model Parallelism} \\ \hline \hline
%本质应该是从模型角度出发，将一个模型的Workload进行划分，从而挖掘并行性，利用多节点加速
\multirow{2}{*}{\textbf{Similarity}}  
                     % & \Revise{Object of Workload Partitioning}       &  \multicolumn{2}{c|}{Model} \\  \\ \cline{2-4}
                      & Perspective of Parallelism Exploitation   &  \multicolumn{2}{c|}{Model} \\ \cline{2-4}
                      & Fashion of Parallelism Exploitation   &  \multicolumn{2}{c|}{Partitioning Workload of Model Computation} \\ \hline
\multirow{4}{*}{\textbf{Difference}}    
                      & Communication Uncertainty                 & High  & Low \\  \cline{2-4}
                      & Transmission Planing Complexity          & High  & Low \\  \cline{2-4}
                      & Probability of Communication Congestion  & High  & Low \\  \cline{2-4}
                      & Probability of Computation Stagnation    & High  & Low \\  \hline
%\textbf{Difference} &Irregular Data Transmission   &Strong Execution Dependency between Computing Nodes  \\
\hline
\hline
& \textbf{Comparison Object}  & \textbf{Distributed Mini-batch Training}  & \textbf{DNN Data Parallelism} \\ \hline \hline
%本质应该是从数据角度出发，将一个模型的数据进行划分，从而挖掘并行性，利用多节点加速
\multirow{2}{*}{\textbf{Similarity}}  
                     %& Training Optimizer & \multicolumn{2}{c|}{Stochastic Gradient Descent} \\ \cline{2-4}
                    % & \Revise{Object of Workload Partitioning}       &  \multicolumn{2}{c|}{Input Data} \\  \\ \cline{2-4}
                     & Perspective of Parallelism Exploitation & \multicolumn{2}{c|}{Input Data} \\ \cline{2-4}
                     & Fashion of Parallelism Exploitation    & \multicolumn{2}{c|}{Partitioning Raw Data} \\ \hline
\multirow{6}{*}{\textbf{Difference}}   
                     & Source of Input Data & Generated by Sampling Phase & Directly Loaded From Raw Data  \\ \cline{2-4}
                     %& Main Phase of Inter-node Communication & Sampling & Model Parameters Update \\ \cline{2-4}
                     & Main Phase of Inter-node Communication & Sampling & Gradient Synchronization \\ \cline{2-4}
                     & Communication Uncertainty                 & High  & Low \\  \cline{2-4}
                     & Transmission Planing Complexity          & High  & Low \\  \cline{2-4}
                     & Probability of Communication Congestion  & High  & Low \\  \cline{2-4}
                     & Probability of Computation Stagnation    & High  & Low \\  \hline
\end{tabular}
    }
\end{table*}

\subsection{Brief Introduction to Distributed DNN Training}\label{sec:comparison_to_dnn:brief_intro_dnn} 

Distributed DNN training is categorized into model parallelism and data parallelism \cite{jpdc/a_hitchhiker/dnn_review}.
\textit{In model parallelism, the model is split into several parts and different computing nodes compute each part.}
AlexNet \cite{nips/imagnenet_classification/alexnet} is the first one to use model parallelism, because its model size could not fit in one GPU at that time. 
%Therefore, this model is split into two parts and then is computed by two GPUs together.
Therefore, the model is split into two parts, which are then computed by two GPUs together.

%However, due to its complex partitioning work and poor efficiency in systems with high communication overhead, it is seldom used in practical applications. 
%Data parallelism is common nowadays. %So this paper focuses on data parallelism.

\textit{In data parallelism, each computing node holds a copy of the model parameters, and the input dataset is distributed to each computing node.
Each computing node conducts the training process completely using its own local data and uses the generated gradients to update the model parameters with other computing nodes together.}
Data parallelism is classified into synchronous training and asynchronous training. 
However, compared with synchronous distributed training, asynchronous distributed training has lower final accuracy, and sometimes non-convergence may occur \cite{jpdc/a_hitchhiker/dnn_review}. 
In synchronous distributed training, after each computing node completes one round of training on a small piece of data, the system starts to collect gradients, which are used to update the model parameters uniformly. 
In this way, each round of training in each computing node is carried out with the same model parameters, which is equivalent to the case of one computing node.

\subsection{Distributed Full-batch Training V.S. DNN Model Parallelism}\label{sec:comparison_to_dnn:full_vs_model}

%\todo{
%%一张大图醒目看出相同点和不同点，左边 GNN DFT， 右边 model parallelism
%}
%\Revise{answer
%不会画
%model parallelism 不会画。 我没有见人画过。而且想一想画出我文中表述的区别 区别就更夸张了。
%sorry
%这个想法我也有过，能画出相同不同的，当然最好。
%如果您有具体怎么画的思路，请告诉我。
%}

\textbf{Similarities.} Both of them distribute the workload of model computation to different computing nodes from the perspective of the model.
Each computing node needs to transfer the intermediate results of model computation to each other and cooperate to complete a round of training.
In DNN model parallelism, DNN model computation is partitioned into different parts and distributed to different computing nodes \cite{nips/imagnenet_classification/alexnet}.
Each computing node may be responsible for executing several layers of DNN, or a set of operations within each layer.
Similarly, GNN distributed full-batch training also distributes the workload of model computation to different computing nodes from the perspective of the model \cite{atc_neugraph, mlsys_roc, arxiv_mg_gcn, sc_reducint_communication_in_graph_cagnet}. 
Since the operation of GNN model computation is to operate on vertices, including the Aggregation and Combination operations, the workload is generally distributed to different computing nodes by partitioning the graph.
Each computing node is responsible for completing the Aggregation and Combination operation of its assigned vertices and communicating with other computing nodes to jointly advance the computation \cite{eurosys_dgcl, sc_distgnn, eurosys_flexgraph}.

\textbf{Differences.} DNN model parallelism has low data communication uncertainty, while GNN distributed full-batch training exhibits high data communication uncertainty.
In DNN model parallelism, most of the transmitted data is intermediate data generated in the computation.
Since the transmission of intermediate data directly affects the execution of other computing nodes, there is a strong execution dependence among computing nodes.
Due to the high certainty of data transmission in DNN model parallelism, which is reflected in the fixed data volume and transmission target, the computation and data transmission can be overlapped by reasonable planning to reduce the communication overhead.
However, in GNN distributed full-batch training, the data communication uncertainty is high.
This is because the data transmitted between each computing node is mainly the features of the neighboring vertex needed in the Aggregation step.
Due to the irregular connection pattern of the graph itself and the fluctuation of the computational efficiency of each computing node, there are great uncertainties in data transmission \cite{eurosys_dgcl}.
Therefore, the congestion of the communication network can easily occur, which in turn leads to stagnation of computation \cite{MultiGCN}.

\subsection{Distributed Mini-batch Training V.S. DNN Data Parallelism} \label{sec:comparison_to_dnn:mini_vs_data}

\textbf{Similarities.} Both of them use small pieces of input data per round of training to update the model parameters.
The computing nodes use different small pieces of input data for computation, and then jointly update the model parameters.
In DNN data parallelism, each computing node holds a portion of the original data, typically stored in its memory.
In each round of computation, a small piece of data is taken from its own local data and computed.
Then, using the generated gradients, the model parameters are updated asynchronously or synchronously.
In GNN distributed mini-batch training, the training of each computing node also uses independent input data, that is, the mini-batch generated by sampling \cite{socc_pagraph, cluster_2pgraph}.

\textbf{Differences.} The input data required for each round of GNN distributed mini-batch training need to be generated by the sampling, while DNN data parallelism is loaded directly from memory.
In DNN data parallelism, each computing node performs computations on its own local data \cite{jpdc/a_hitchhiker/dnn_review}.
It does not need to communicate with other computing nodes except for the stage of updating model parameters.
In GNN distributed mini-batch training, a large number of communication requirements are required in the sampling phase, including querying graph data and transmitting mini-batch data.
As graphs generally exhibit an irregular connection pattern, the communication in the sampling phase is highly irregular, resulting in high communication uncertainty and planning complexity.
This leads to the fact that distributed GNN training may not be able to provide input data in time due to the inefficiency of the sampling, resulting in computational stagnation.
This makes it necessary to optimize the sampling stage to reduce the communication overhead when conducting GNN distributed mini-batch training \cite{socc_pagraph, cluster_2pgraph}.

\section{Summary and Discussion}\label{sec:summary}

This section summarizes the aforementioned details of distributed GNN training, and discusses several interesting issues and opportunities in this field.

The main focus of optimizing distributed full-batch training is to reduce its communication overhead\cite{atc_neugraph, mlsys_roc, sc_reducint_communication_in_graph_cagnet, eurosys_dgcl, sc_distgnn, eurosys_flexgraph, osdi_dorylus}.
In this training method, each computing node inevitably needs to communicate with other nodes for information about the entire graph structure, so that it can finish operations which require graph data that are not stored in the local split, such as aggregating information of neighboring vertices which are distributed in other computing components.
As a result, reducing communication overhead, mainly by reasonably splitting the graph and planning the data transmission, can greatly improve computational efficiency.

By contrast, the main focuses of optimizing distributed mini-batch training are to speed up the sampling phase and to reduce the mini-batch transmission overhead \cite{arxiv_salient, pvldb_agl, ia3_distdgl, pvldb_aligraph, socc_pagraph, cluster_2pgraph}.
Currently, it is popular to perform distributed mini-batch training in which CPUs are responsible for sampling and GPUs are responsible for GNN model computation.
The sampling capability of CPUs is insufficient, producing a lot of idleness and limiting the utilization of GPUs due to insufficient data supply.
By accelerating sampling and optimizing data transmission, the utilization of computing components can be improved, thereby improving efficiency.

The two methods show their advantages in different aspects respectively. 
In current designs of distributed training of GNNs, the adoption of the mini-batch method is gradually increasing due to its advantages of a simpler implementation and less memory pressure.
Distributed full-batch training, on the other hand, requires programmers to put more effort into programming and design, so it is more difficult to optimize.
However, some research shows that the final precision of distributed full-batch training can be higher \cite{mlsys_roc}.

Next, we discuss several interesting issues and opportunities in this field. 

\subsection{Quantitative Analysis of Performance Bottleneck}\label{sec:summary:quantitative}
%(0) 性能瓶颈和执行行为的分析
%先说一下单节点上的characterization 工作，然后再说一下你的IEEE CAL工作，再说目前的分析不够深入和完善，需要更多的人参与。
%

As mentioned before, the performance of distributed GNN training is primarily hindered by the workload imbalance across different computing nodes, irregular transmission between computing nodes, etc.
Although these performance bottlenecks are well-known in the GNN community, the quantitative analysis of these bottlenecks, which is important to guide the optimizations for distributed GNN training, is still obscure.

Recently, a lot of characterization efforts for GNN training or inference have been conducted on a single computing node or a single GPU~\cite{GCN_Characterization_CAL,GNN_Architectural_Implication,Characterizing_HGNNs,character_Distributed_GNN,Understanding_GNN,/mlsys/understanding_gnn_computational/gat_intermediate_data}. 
For example, Yan et al. \cite{GCN_Characterization_CAL} quantitatively disclose the computational pattern of the inference on a single GPU.
Zhang et al. \cite{GNN_Architectural_Implication} quantitatively characterize the training of a large portion of GNN variants concerning general-purpose and domain-specific hardware architectures.
However, there are few characterization efforts on distributed GNN training \cite{character_Distributed_GNN}. 

Diagnosing performance issues without a quantitative understanding of the bottlenecks can cause misdiagnosis: A potentially expensive effort that involves optimizing something that is not the real problem. 
Although Lin et al. \cite{character_Distributed_GNN} quantitatively characterize the end-to-end execution of distributed GNN training, it is not enough for the long-term development of distributed GNN training. We believe that the quantitative analysis of performance bottleneck will be one of the main directions in the field of distributed GNN training. 
%\todo{R1 Q9}
\Revise{
Here are some potential directions for further exploration:
1) Performance Modeling and Measurement: GNN distributed training methods can be subjected to mathematical modeling and analysis, allowing us to gain insights into their theoretical potential and identify primary bottlenecks. To enhance model modeling, it is beneficial to measure the performance metrics of the distributed system in real-world operation, thus improving the precision of performance evaluation.
%2) Scalability Testing: Evaluating the scalability of distributed training frameworks and techniques across a range of workload and resource conditions is crucial. This assessment should consider factors like performance growth and variations in workload balancing to ensure robust scalability.
2) Scalability Testing: Evaluating the scalability of distributed training frameworks and techniques across a range of workload and resource conditions is crucial. This assessment should consider factors like computing capability growth and variations in workload balancing to ensure robust scalability.
3) Cost-Benefit Analysis: Allocating additional computational resources for acceleration introduces extra computational costs. It is imperative to measure the input-output ratio, assess the cost of hardware resources, and optimize resource utilization for cost-effective scaling.
4) Automated Evaluation Tools: Developing automated evaluation tools that can perform relevant tests automatically based on provided distributed configuration scenarios streamlines the testing process. 

Moreover, there are opportunities to delve into the reliability and security aspects of GNN distributed systems. In summary, quantitative analysis of GNN distributed training systems plays a pivotal role in identifying and addressing performance bottlenecks.

}

\subsection{Performance Benchmark} \label{sec:summary:performance_benchmark}
Despite the large quantity of performance benchmarks presented to fairly evaluate and compare the performance of software frameworks and hardware platforms for DNNs, there are few performance benchmarks for distributed GNN training. 
Most of the recent benchmarks for GNNs mainly focus on evaluating the prediction accuracy of trained GNN models.
For example, Open Graph Benchmark (OGB)~\cite{nips_open_graph_benchmark_ogbdataset} provides a set of large-scale real-world benchmark datasets to facilitate the graph machine learning research.
However, current benchmarks focus mainly on the accuracy, instead of performance.
As a result, standardized performance benchmark suites with typical, representative, and widely-acknowledged workloads are needed for the industry and academia to comprehensively compare the performance of distributed GNN training systems.

In the field of DNNs, MLperf~\cite{MLPerf_IEEE_Micro}, an industry standard benchmark suite for machine learning performance, has made a great contribution to the development of software frameworks and systems.
Therefore, we believe that performance benchmark suites for GNNs similar to MLperf are vital to drive the rapid development of GNNs.

\subsection{Distributed Training on Extreme-scale Hardware Platform}\label{sec:summary:distributed_training_on_extreme_scale}

For distributed GNN training on large-scale graphs or even extreme-scale graphs, the performance can be greatly improved by using the extreme-scale hardware platform.
The scale of the graph data has reached a staggering magnitude and is growing at an exaggerated rate.
For example, Sogou \cite{DBLP:conf/sc/LinZYTXCZHMLZX18/shentu} graph dataset has 12 trillion edges and Kronecker2 \cite{DBLP:conf/pkdd/LeskovecCKF05/kronecker} graph dataset has 70 trillion edges.
The training on graphs in this order of magnitude puts a very high requirement on the computing resources and memory resources of the computing system.
Only by using the extreme-scale hardware platform can the training task be completed in a reasonable time.
In this respect, the research of graph processing and distributed DNN training on the extreme-scale hardware platform can be used for reference.
ShenTu \cite{DBLP:conf/sc/LinZYTXCZHMLZX18/shentu} realizes processing graph with trillion edges in seconds by using petascale computing system.
It suggests that the extreme-scale hardware platform can be used to accelerate distributed GNN training, which needs to explore and establish more hardware platform support for distributed GNN training.

\subsection{Domain-specific Distributed Hardware Platform}\label{sec:summary:domain_specific}
%类似DNN那样，目前已经有TPU-Pod的专用集群设计，设计单节点的加速器已经证明效率非常高，其次也涉及了MultiGCN的集群设计以及谢老师ISCA 2022 的sampling加速工作，但是目前的工作仍然减少，但是研究的必要性是非常强的。

There are many single-node domain-specific hardware platforms designed for GNNs. 
%However, very little efforts are conducted on distributed hardware platform.
However, very few efforts are conducted on distributed hardware platforms.
Recently, many single-node domain-specific hardware accelerators designed for GNNs have achieved significant improvement in performance and efficiency compared with the single-GPU platform. 
For example, HyGCN~\cite{hpca/hygcn/gnn_asic_acc} proposes a hybrid architecture consisting of the Aggregation engine and Combination engine for the acceleration of the Aggregation operation and Combination operation respectively. Compared with NVIDIA V100 GPU, HyGCN achieves an average 6.5$\times$ speedup with 10$\times$ energy reduction, respectively.
However, the single-node platform is not sufficient to handle the rapid development of GNNs since the ever-growing scale of graphs dramatically increases the training time, generating the demand for distributed GNN hardware platform. In response to that, MultiGCN~\cite{MultiGCN} proposes a distributed hardware platform for the acceleration of GNN inference, which consists of multiple computing nodes to perform the inference in parallel. 
MultiGCN achieves 2.5$\sim$8$\times$ speedup over the state-of-the-art multi-GPU software framework. 
But only a single work is not enough.   
We have witnessed that TPU-Pod~\cite{TPU-Pod}, a domain-specific supercomputer for training DNNs, greatly accelerates the deployment of large-scale DNN models.
Therefore, we believe that a domain-specific distributed hardware platform for GNNs similar to TPU-Pod is time to present, which is vital to drive the rapid development and deployment of GNNs.

\subsection{General Communication Library for Distributed GNN Training}\label{sec:summary:general_communication}

The variety of hardware platforms and irregular communication patterns of distributed GNN training lead to strong demand for a general communication library to ease the programming effort on transmission planning while achieving high efficiency.
The communication characteristics of hardware platforms are various due to the differences in the number of computing nodes, network topology, and communication bandwidth as well as communication latency of interconnection interface.  
In order to better adapt to the various transmission requirements of distributed GNN training on various hardware platforms, it is urgent to develop a general communication library which adapts to different network topologies, as well as various interconnection interfaces of different communication bandwidths and latency, so as to ease the implementation of novel communication patterns while ensuring efficiency.

%\todo{R2 Q3}

\Revise{
%\subsection{Distributed Training towards Temporally-varying GNNs}
\subsection{Distributed Training for Dynamic GNNs}\label{sec:summary:dynamic_gnn}

Dynamic GNNs are a specific variant of GNNs designed for dynamic graphs, which inherently evolve over time. These dynamic graphs fall into two main categories: continuous-time dynamic graphs (CTDGs) and discrete-time dynamic graphs (DTDGs)~\cite{DBLP:conf/ppopp/WangSB23/pipad, DBLP:conf/hpdc/XiaZWYZC23/sven}. CTDGs capture continuous dynamics of dynamic graphs, while DTDGs consist of a sequence of graph snapshots sampled at regular intervals.

Dynamic GNNs differ from GNNs focusing on static graphs in two crucial ways: Firstly, they operate on dynamic graphs with changing input, and secondly, they incorporate temporal neural networks like LSTM~\cite{DBLP:journals/neco/HochreiterS97/lstm} to capture temporal information. 
As a result, dynamic GNNs pose a higher computational complexity compared to static GNNs, making distributed training for them more challenging. Applying the distributed training optimization techniques discussed earlier to dynamic GNNs necessitates specific adjustments. For instance, optimization techniques related to graph partitioning must consider how to update graphs dynamically.

To conclude, developing distributed training systems for dynamic GNNs presents several challenges. 
1) Dynamic Input Data: Swift propagation of changes in graph topology and vertex features to computing nodes is crucial. Efficient network transmission support and strategies to minimize transmission overhead and reduce waiting times for computing nodes are necessary~\cite{DBLP:conf/ppopp/WangSB23/pipad,DBLP:conf/sc/ChakaravarthyPR21/esdg}.
2) Real-Time Requirements: Dynamic GNNs are designed for rapidly changing graph data. As a result, dynamic GNNs may require real-time output generation to satisfy predefined time constraints. 
3) Workload Balancing: The dynamic nature of input data in dynamic GNNs makes traditional static workload partitioning methods impractical, necessitating innovative workload balancing approaches.
4) Parallelism Exploitation: Dynamic GNNs, especially in the case of processing DTDGs, perform computations across multiple graphs. Exploring parallel computations across these graphs offers opportunities for parallelism.

Current research focuses on reducing redundant data transmission by leveraging similarities between graphs~\cite{DBLP:conf/ppopp/WangSB23/pipad,DBLP:conf/sc/ChakaravarthyPR21/esdg}. Two frameworks, PyTorch Geometric Temporal~\cite{DBLP:conf/cikm/RozemberczkiSHP21/pygt} and TGL~\cite{DBLP:journals/pvldb/ZhouZNISK22/tgl} (based on PyG~\cite{arxiv_pyg} and DGL~\cite{arxiv_dgl} respectively), support distributed training of dynamic GNNs. Sven~\cite{DBLP:conf/hpdc/XiaZWYZC23/sven}, an extension of TGL, has significantly accelerated distributed training by eliminating data transmission redundancy and utilizing pipeline parallelism.

While still in its early stages, research on optimizing distributed training for dynamic GNNs is poised to become a central focus for future investigations.

}

\Revise{
\subsection{Distributed Training for Deep GNNs
%Impact for Performance Deterioration on Distributed Training
}\label{sec:summary:performance_deterioration}
%Distributed Training for Large-scale GNN Model
%放到最后一个章节
%论文中的Distributed Training的motivation是啥？
%集中在模型并行性的挖掘
%\todo{R2 Q2: Under Revising}
The issue of performance deterioration has garnered significant attention in the field of GNNs~\cite{DBLP:conf/iclr/KlicperaBG19/predict_then_propagate/pd01,DBLP:conf/iclr/RongHXH20/dropedge/pd02,DBLP:journals/corr/abs-2006-07739/deepergcn/pd03,DBLP:conf/icml/ChenWHDL20/simple_and_deep_gcn/pd04,DBLP:journals/corr/abs-2106-15810/edgeproposal/pd05, DBLP:conf/aistats/LiuQWL23/EEGNN/pd07_review,DBLP:conf/iclr/0002Y21/bottleneck_of_gnn/pd08}, which significantly affects the number of GNNs' layers. Specifically, it has been observed that using deeper layer stacks can lead to a partial loss in performance, which is in contrast to the behavior of  DNNs~\cite{DBLP:conf/nips/VaswaniSPUJGKP17/attentoin_is_all_you_need/transformer}. Consequently, the best-performing GNN models currently tend to have 2-4 layers~\cite{DBLP:conf/aistats/LiuQWL23/EEGNN/pd07_review}. This phenomenon can primarily be attributed to two factors: 1) over-smoothing and 2) under-reaching~\cite{DBLP:conf/aistats/LiuQWL23/EEGNN/pd07_review, DBLP:conf/iclr/0002Y21/bottleneck_of_gnn/pd08}.
Over-smoothing refers to the distinctiveness of vertex features diminishing as information is propagated through multiple layers. Targeted solutions for this issue include techniques like DropEdge~\cite{DBLP:conf/iclr/RongHXH20/dropedge/pd02} and DropNode~\cite{DBLP:journals/corr/abs-2008-09864/pd09/dropnode}, which involve selectively removing edges and vertices. Under-reaching, on the other hand, signifies the challenge of propagating information across long-range paths. Specific strategies to address this problem include adding virtual edges~\cite{DBLP:conf/icml/GilmerSRVD17/virtual_edge/pd10} and super vertices~\cite{DBLP:journals/tnn/ScarselliGTHM09/super_node/pd11}. Additionally, it's worth noting that EEGNN~\cite{DBLP:conf/aistats/LiuQWL23/EEGNN/pd07_review} introduces another contributing factor called mis-simplification, where excessive simplification occurs during graph building, neglecting the importance of edges and self-loops.

The impact of performance deterioration on distributed GNN training significantly influences the overall optimization direction in this domain. 
Because of the performance deterioration issue, current GNN models tend to have a relatively small number of layers~\cite{DBLP:conf/aistats/LiuQWL23/EEGNN/pd07_review}, resulting in the smaller scale of model parameters. 
This discrepancy becomes especially apparent when comparing them to large-scale graph datasets~\cite{nips_open_graph_benchmark_ogbdataset}. The stark disparity between the vast volume of input data and the modest model parameters makes GNN execution memory-intensive or communication-intensive. Consequently, there is a strong emphasis on researching methods to minimize the preparation overhead of input data, particularly concerning data transmission costs. This pursuit has led to techniques such as activation rematerialization in Section~\ref{sec:dft:partition_execution}, mini-batch transmission pipelining in Section~\ref{sec:dmt:sample_individual_execution}, and frequently-used data caching in Section~\ref{sec:dmt:sample_joint_execution}. This also explains the popularity of current distributed mini-batch training, which is similar to data parallelism in DNNs, as partitioning GNN models into smaller pieces doesn't hold much significance.

Currently, there is a continuous emergence of techniques addressing performance deterioration in GNNs~\cite{DBLP:conf/iclr/RongHXH20/dropedge/pd02,DBLP:journals/corr/abs-2008-09864/pd09/dropnode,DBLP:conf/icml/GilmerSRVD17/virtual_edge/pd10, DBLP:journals/tnn/ScarselliGTHM09/super_node/pd11, DBLP:conf/aistats/LiuQWL23/EEGNN/pd07_review}, encouraging the exploration of deeper layer structures in GNN architectures. This evolving trend is shaping the future optimization strategies for distributed GNN training. 
%One key aspect is the necessity to accommodate the rise of larger GNN models, which may favor approaches akin to model parallelism, such as pipeline parallelism~\cite{DBLP:conf/nips/HuangCBFCCLNLWC19/gpipe}. 
One key aspect is the necessity to accommodate the rise of deep GNN models, which may favor approaches akin to model parallelism, such as pipeline parallelism~\cite{DBLP:conf/nips/HuangCBFCCLNLWC19/gpipe}. 
Research efforts are likely to focus on implementing the execution of deep GNNs on distributed training platforms.
}

\section{Conclusion}\label{sec:conclusion}
In this paper, we comprehensively review distributed training of GNNs.
After investigating recent efforts on distributed GNN training, we classify them in more detail according to their workflow, including the dispatch-workload-based execution as well as preset-workload-based execution of distributed full-batch training, and the individual-sample-based execution as well as joint-sample-based execution of distributed mini-batch training.
Each taxonomy's workflow, computational pattern, and communication pattern are summarized, and various optimization techniques are further introduced to facilitate the understanding of recent research status, and to enable readers to quickly build up a big picture of distributed training of GNNs.
In addition, this paper introduces the training software and hardware platforms, and then contrasts distributed GNN training with distributed DNN training.
In the end, we provide several discussions about interesting issues and opportunities of distributed GNN training.

The emergence of distributed GNN training has successfully expanded the usage of GNNs on large-scale graphs, making it a powerful tool when learning from real-world associative relations.
We optimistically look forward to new research which can further optimize this powerful tool on large-scale graph data and bring its application into the next level.

\section*{Acknowledgments}
This work was supported by the National Natural Science Foundation of China (Grant No. 61732018, 61872335, and 62202451), Austrian-Chinese Cooperative R\&D Project (FFG and CAS) (Grant No. 171111KYSB20200002), CAS Project for Young Scientists in Basic Research (Grant No. YSBR-029), and CAS Project for Youth Innovation Promotion Association.

%\begin{thebibliography}{1}
\bibliographystyle{IEEEtran}
\bibliography{gnndt_ref}

%\bibitem{ref1}
%{\it{Mathematics Into Type}}. American Mathematical Society. [Online]. Available: https://www.ams.org/arc/styleguide/mit-2.pdf

%\end{thebibliography}

\newpage

%\todo{
%最后与Proposal同步
%}
%\Revise{answer
%已把cover_letter的复制过来了
%}
% \Dang{% 其实可以在一个专门的文件里写这些IEEEbiography，然后在proposal和这里分别\input{xxxx.tex}，然后只修改这个专门的文件就可以保持两边统一了
% }

\section{Biography Section}

\begin{IEEEbiography}[{\includegraphics[width=1in,height=1.25in,clip,keepaspectratio]{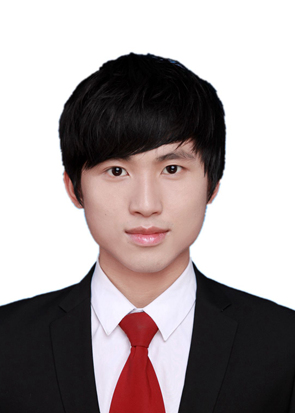}}]{Haiyang Lin} received the B.S. degree from Shanghai Jiaotong University, Shanghai, China, in 2018, and currently working toward the Ph.D. degree at Institute of Computing Technology, Chinese Academy of Sciences, Beijing, China.
His current research interests include distributed training, graph-based hardware accelerator, and high-throughput computer architecture.

\end{IEEEbiography}

\begin{IEEEbiography}[{\includegraphics[width=1in,height=1.25in,clip,keepaspectratio]{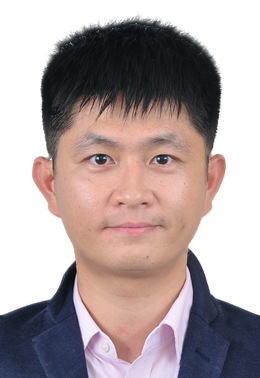}}]{Mingyu Yan} received the Ph.D. degree from University of Chinese Academy of Sciences, Beijing, China in 2020. 
He is currently an Associate Professor at Institute of Computing Technology, Chinese Academy of Sciences, Beijing, China. 
His current research interests include graph-based machine learning, domain-specific hardware architecture for graph-based machine learning, domain-specific distributed system for graph-based machine learning, and high-throughput computer architecture.
To date, Dr. Yan has published over 20 research papers in top-tier journals and conferences, including the MICRO, HPCA, IJCAI, DAC, ICCAD, IEEE Transactions on Computers (TC), IEEE Transactions on Computer-Aided Design of Integrated Circuits and Systems (TCAD), IEEE/CAA Journal of Automatica Sinica (IEEE/CAA JAS), and IEEE Journal of Selected Topics in Signal Processing (IEEE J-STSP). 

\end{IEEEbiography}

\begin{IEEEbiography}[{\includegraphics[width=1in,height=1.25in,clip,keepaspectratio]{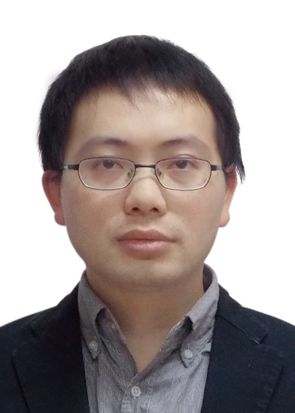}}]{Xiaochun Ye} received the Ph.D. degree in computer architecture from Institute of Computing Technology, Chinese Academy of Sciences, Beijing, in 2010. 
He is currently a Professor at Institute of Computing Technology, Chinese Academy of
Sciences, Beijing. 
His main research interest is domain-specific hardware architecture for graph-based machine learning and high-throughput computer architecture.
To date, Dr. Ye has published over 90 research papers in top-tier journals and conferences, 
the MICRO, HPCA, IJCAI, PACT, IPDPS, DAC, ICCAD, DATE, HOT CHIPS, ICCV, IEEE Transactions on Computers (TC), IEEE/CAA Journal of Automatica Sinica (IEEE/CAA JAS), and IEEE Micro.

\end{IEEEbiography}

\begin{IEEEbiography}[{\includegraphics[width=1in,height=1.25in,clip,keepaspectratio]{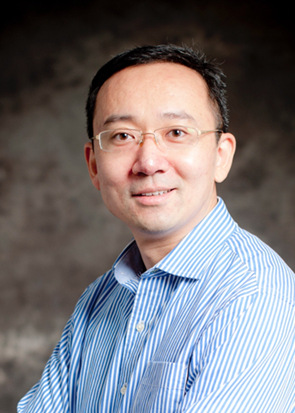}}]{Dongrui Fan} received the Ph.D. degree in computer architecture from Institute of Computing Technology, Chinese Academy of Sciences, Beijing, in 2005. 
He is currently a Professor and Ph.D. Supervisor at Institute of Computing Technology, Chinese Academy of Sciences, Beijing. 
His main research interests include high-throughput computer architecture, high-performance computer architecture, and low-power design.
To date, Dr. Fan has published over 140 research papers in top-tier journals and conferences, including the IEEE Transactions on Parallel and Distributed Systems (TPDS), IEEE Transactions on Computers (TC), IEEE micro, Journal of Parallel and Distributed Computing (JPDC), MICRO, HPCA, PPoPP, IJCAI, DAC, ICCAD, DATE, ICCV, PACT, HOT CHIPS. He has been a member of the program committee of many important academic conferences, including MICRO, HPCA, etc.

\end{IEEEbiography}

\begin{IEEEbiography}[{\includegraphics[width=1in,height=1.25in,clip,keepaspectratio]{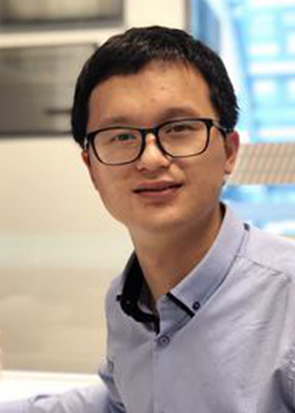}}]{Shirui Pan} received a Ph.D. in computer science from the University of Technology Sydney (UTS), Ultimo, NSW, Australia. He is a Professor at the School of Information and Communication Technology, Griffith University, Australia. Prior to this, he was a Senior Lecturer with the Faculty of IT, Monash University, Australia. His research interests include data mining and machine learning. To date, Dr. Pan has published over 150 research papers in top-tier journals and conferences, including the IEEE Transactions on Pattern Analysis and Machine Intelligence (TPAMI), IEEE Transactions on Knowledge and Data Engineering (TKDE), IEEE Transactions on Neural Networks and Learning Systems (TNNLS), ICML, NeurIPS, KDD, AAAI, IJCAI, WWW, CVPR, ICDE, and ICDM. His research has attracted over 10,000 citations. 

Dr. Pan is an expert in graph neural networks (GNNs). He is recognized as one of the AI 2000 AAAI/IJCAI Most Influential Scholars in Australia (2021, 2022), and one of the World’s Top 2\% Scientists (2021). His research on GNNs received the IEEE ICDM Best Student Paper Award (2020), and the JCDL Best Paper Honorable Mention Award (2020). He has seven papers on GNNs recognized as the Most Influential Papers in KDD (x1), IJCAI (x5), and CIKM (x1) (Feb 2022). In particular, his survey paper on \textit{``A Comprehensive Survey on Graph Neural Networks"} in TNNLS-21 has been cited over 4,100 times; his research on ``Graph WaveNet''  (582 citations) is the most cited research at IJCAI-2019 as of September 2022. He serves as an Area Chair/Senior Program Committee member (SPC) for ICDM, AAAI, and IJCAI, and a PC member for NeurIPS, KDD, WWW, etc. He is an Australian Research Council (ARC) Future Fellow.

\end{IEEEbiography}

\begin{IEEEbiography}[{\includegraphics[width=1in,height=1.25in,clip,keepaspectratio]{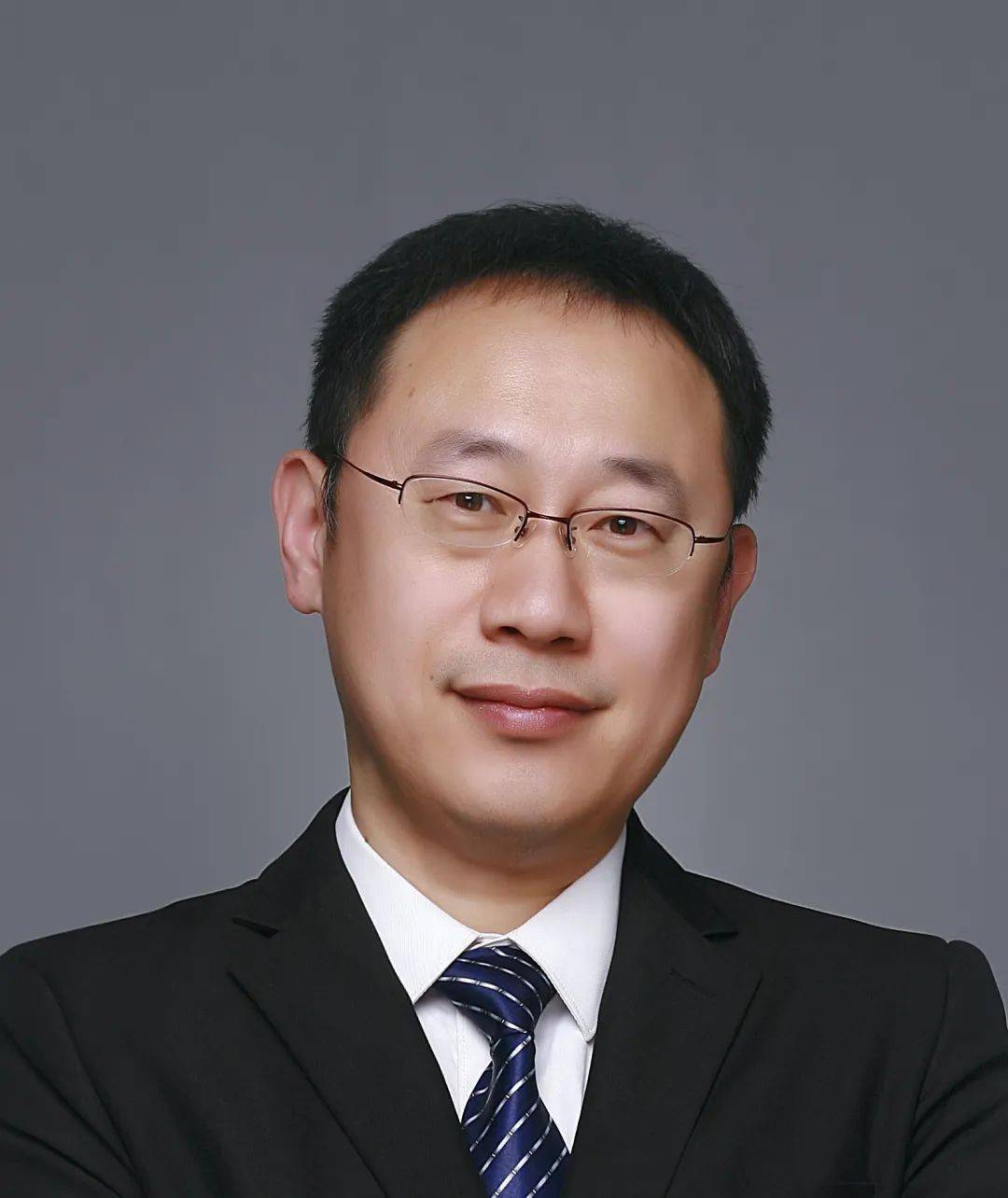}}]{Wenguang Chen} received the B.S. and Ph.D. degrees in computer science from Tsinghua University, in 1995 and 2000, respectively. He was the CTO with Opportunity International Inc. from 2000-2002. Since January 2003, he joined Tsinghua University. He is currently a professor and associate head with the Department of Computer Science and Technology, Tsinghua University. He is chair of ACM China Council, chief editor of ACM Communications Chinese Edition. 
He joined Ant Group in 2020 as the head of graph processing technology. 
Since 2022, he has served as the President of Ant Technology Research Institute. 

Dr. Chen is an expert in parallel and distributed systems for extreme-scale graph processing.
To date, Dr. Chen has published over 100 research papers in top-tier conferences and journals, including the OSDI, PPoPP, SC, ASPLOS, ICS, EuroSys, IPDPS, MICRO, ATC, CGO, VLDB, PACT, Nature, IEEE Transactions on Parallel and Distributed Systems (TPDS), and IEEE Transactions on Computers (TC).
Their work in extreme-scale graph processing, i.e., ``ShenTu: Processing Multi-trillion Edge Graphs on Millions of Cores in Seconds", was a finalist for the ACM Gordon Bell Prize in 2018. The Gordon Bell Prize is commonly referred to as the Nobel Prize of Supercomputing.
He has been a member of the program committee of many important academic conferences, including OSDI, SOSP, PLDI, PPoPP, SC, ASPLOS, TPDS, TC, ATC, CGO, IPDPS, CCGrid, ICPP, PACT, EuroSys, ICS, etc.

\end{IEEEbiography}

\begin{IEEEbiography}[{\includegraphics[width=1in,height=1.25in,clip,keepaspectratio]{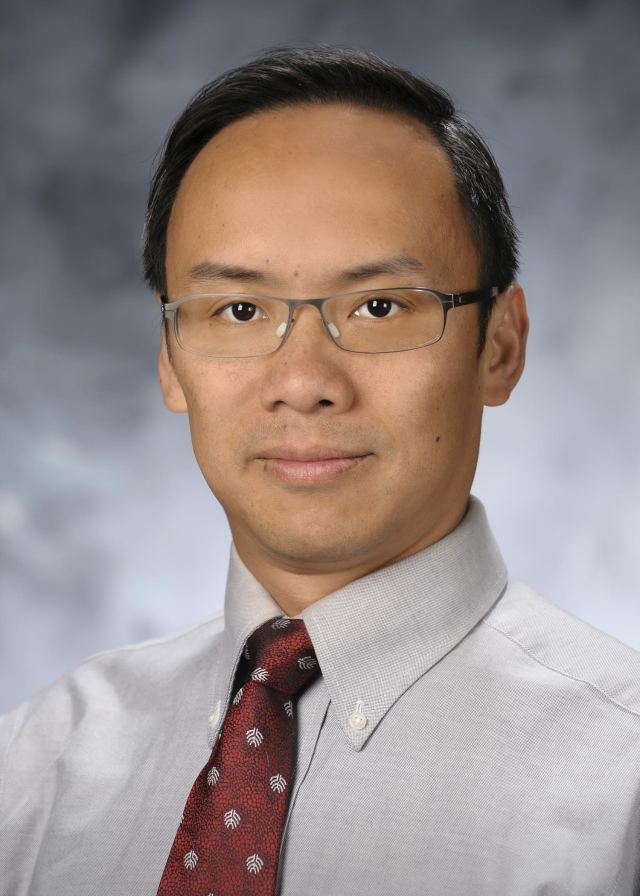}}]{Yuan Xie} (Fellow, IEEE) received the B.S. degree in electronic engineering from Tsinghua University, Beijing, China, in 1997, and the M.S. and Ph.D. degrees in electrical engineering from Princeton University, Princeton, NJ, USA, in 1999 and 2002, respectively.
He was an Advisory Engineer with the IBM Microelectronic Division, Burlington, NJ, USA, from 2002 to 2003. 
He was a Full Professor with Pennsylvania State University, State College, PA, USA, from 2003 to 2014. 
He was a Visiting Researcher with the Interuniversity Microelectronics Center (IMEC), Louvain, Belgium, from 2005 to 2007 and in 2010. 
He was a Senior Manager and a Principal Researcher with the AMD Research China Laboratory, Beijing, from 2012 to 2013. 
%He is currently a Professor with the Department of Electrical and Computer Engineering, University of California at Santa Barbara, Santa Barbara, CA, USA. 
He is currently a Chair Professor with the Department of Electronic and Computer Engineering, The Hong Kong University of Science and Technology.
His research interests include VLSI design, electronics design automation (EDA), computer architecture, and embedded systems.

Dr. Xie serves as a Committee Member for the IEEE Design Automation Technical Committee (DATC). 
He is a Fellow of AAAS/ACM and an expert in computer architecture who has been inducted into ISCA/MICRO/HPCA Hall of Fame. 
He was a recipient of the Best Paper Awards at HPCA 2015, ICCAD 2014, GLSVLSI 2014, ISVLSI 2012, ISLPED 2011, ASPDAC 2008, and ASICON 2001; the Best Paper Nominations at ASPDAC 2014, MICRO 2013, DATE 2013, ASPDAC 2010–2009, and ICCAD 2006; the 2016 IEEE Micro Top Picks Award at the 2008 IBM Faculty Award, and the 2006 NSF CAREER Award. 
He has served as the TPC Chair for ICCAD 2019, HPCA 2018, ASPDAC 2013, ISLPED 2013, and MPSOC 2011.
He serves as the Editor-in-Chief for ACM Journal on Emerging Technologies in Computing Systems and as an Associate Editor for the ACM Transactions on Design Automation for Electronics Systems, the IEEE Transactions on Computers, the IEEE Transactions on Computer-Aided Design of Integrated Circuits and Systems, the IEEE Transactions on VLSI, IEEE Design and Test, and IET Computers and Design Techniques. 
Through extensive collaboration with industry partners, AMD, HP, Honda, IBM, Intel, Google, Samsung, IMEC, Qualcomm, Alibaba, Seagate, Toyota, and so on, he has helped the transition of research ideas to industry.

\end{IEEEbiography}

\vfill
\end{document}